\documentclass[10pt]{book}
\usepackage{times}

\usepackage[english]{babel}

\usepackage[utf8]{inputenc}

\usepackage{lmodern}

\usepackage{indentfirst}

\usepackage[plmath]{polski}

\usepackage{icomma}

\usepackage{amsmath}

\usepackage{amssymb}

\usepackage{mathrsfs}

\usepackage{bbold}

\usepackage{mathtools}
\usepackage{float}

\usepackage{booktabs}
\usepackage{subcaption}
\usepackage{adjustbox}
\usepackage{multirow}

\usepackage[table]{xcolor}

\usepackage{tabu}

\usepackage{xcolor}

\definecolor{lgray}{HTML}{9F9F9F}
\definecolor{dgray}{HTML}{5F5F5F}

\usepackage[ruled,vlined,linesnumbered,longend,algochapter]{algorithm2e}

\SetCommentSty{algcomment}

\usepackage{textcomp}%

\usepackage{listings} 

\usepackage[headheight=18pt, top=25mm, bottom=25mm, left=30mm, right=25mm]{geometry}


\usepackage{titling}

\usepackage{tocloft}	

\tocloftpagestyle{tableOfContentStyle}

\usepackage{fancyhdr}

\fancypagestyle{custom}{
	\fancyhf{}									
	\fancyhead[RO]{								
		\hrulefill \hspace{16pt} \large \ifnum \thechapter>0 {Chapter \thechapter} \else{Introduction}\fi
		\put(-472.1, 12.1){%
			\makebox(0,0)[l]{%
				\includegraphics[width=0.05\textwidth]{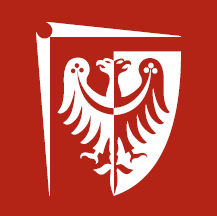}
			}
		}
		\put(-443,5.5){%
			\makebox(0,0)[l]{%
				\small Wrocław University of Science and Technology
			}
		}
	}
	\fancyhead[LE]{								
		\large \ifnum \thechapter>0 {Chapter \thechapter} \else{Introduction}\fi \hspace{16pt} \hrulefill  
		\put(-22, 12.1){%
			\makebox(0,0)[l]{%
				\includegraphics[width=0.045\textwidth]{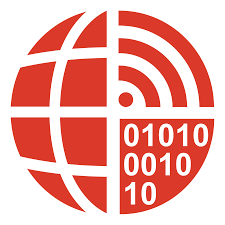}
			}
		}
		\put(-250,5.5){%
			\makebox(0,0)[l]{%
				\small Faculty of Information and Communication Technology
			}
		}
	}
	\fancyfoot[LE,RO]{							
		\thepage
	}
}

\fancypagestyle{bibliographyStyle}{
	\fancyhf{}									
	\fancyhead[RO]{								
		\hrulefill \hspace{16pt} \large Supplementary \thechapter
		\put(-472.1, 12.1){%
			\makebox(0,0)[l]{%
				\includegraphics[width=0.05\textwidth]{resources/pwr-logo.png}
			}
		}
		\put(-443,5.5){%
			\makebox(0,0)[l]{%
				\small Wrocław University of Science and Technology
			}
		}
	}
	\fancyhead[LE]{								
		\large Bibliography \hspace{16pt} \hrulefill 
		\put(-22, 12.1){%
			\makebox(0,0)[l]{%
				\includegraphics[width=0.05\textwidth]{resources/wit-logo.png}
			}
		}
		\put(-250,5.5){%
			\makebox(0,0)[l]{%
				\small Faculty of Information and Communication Technology
			}
		}
	}
	\fancyfoot[LE,RO]{							
		\thepage
	}
}

\fancypagestyle{listOfPlotsStyle}{
	\fancyhf{}									
	\fancyhead[RO]{								
		\hrulefill \hspace{16pt} \large List of figures
		\put(-472.1, 12.1){%
			\makebox(0,0)[l]{%
			\includegraphics[width=0.05\textwidth]{pwr-logo}
			}
		}
		\put(-443,5.5){%
			\makebox(0,0)[l]{%
				\small Wrocław University of Science and Technology
			}
		}
	}
	\fancyhead[LE]{								
		\large List of figures \hspace{16pt} \hrulefill 
		\put(-22, 12.1){%
			\makebox(0,0)[l]{%
\includegraphics[width=0.05\textwidth]{wit-logo}
			}
		}
		\put(-210,5.5){%
			\makebox(0,0)[l]{%
				\small Faculty of Information and Communication Technology
			}
		}
	}
	\fancyfoot[LE,RO]{							
		\thepage
	}
}

\fancypagestyle{appendixStyle}{
	\fancyhf{}									
	\fancyhead[RO]{								
		\hrulefill \hspace{16pt} \large Supplementary \thechapter
		\put(-472.1, 12.1){%
			\makebox(0,0)[l]{%
				\includegraphics[width=0.05\textwidth]{resources/pwr-logo.png}
			}
		}
		\put(-443,5.5){%
			\makebox(0,0)[l]{%
				\small Wrocław University of Science and Technology
			}
		}
	}
	\fancyhead[LE]{								
		\large Supplementary \thechapter \hspace{16pt} \hrulefill 
		\put(-22, 12.1){%
			\makebox(0,0)[l]{%
				\includegraphics[width=0.05\textwidth]{resources/wit-logo.png}
			}
		}
		\put(-250,5.5){%
			\makebox(0,0)[l]{%
				\small Faculty of Information and Communication Technology
			}
		}
	}
	\fancyfoot[LE,RO]{							
		\thepage
	}
}

\fancypagestyle{chapterBeginStyle}{
	\fancyhf{}%
	\fancyfoot[LE,RO]{
		\thepage
	}

}

\fancypagestyle{tableOfContentStyle}{
	\fancyhf{}%
	\fancyfoot[LE,RO]{
		\thepage
	}

}

\usepackage{titlesec}

\titleformat{\chapter}[hang]					
{
	\vspace{-10ex}
	\Huge
	\bfseries
}												
{}												
{
	10pt
}												
{
	\Huge
	\bfseries
}												
[
\vspace{2ex}
]												

\titleformat{\section}[hang]					
{
	\vspace{2ex}
	\Large\bfseries
}												
{
	\thesection									
}
{
	0pt
}												
{
	\Large
	\bfseries
}												

\usepackage{hyperref}							

\hypersetup{
	colorlinks	=	true,					
	linkcolor		=	blue,					
        citecolor		=	blue,					
	urlcolor		=	blue					
}

\usepackage{enumitem}

\newlist{myitemize}{itemize}{3}

\setlist[myitemize,1]{
	label		=	\textbullet,
	leftmargin	=	4mm}

\setlist[myitemize,2]{
	label		=	$\diamond$,
	leftmargin	=	8mm}

\setlist[myitemize,3]{
	label		=	$\diamond$,
	leftmargin	=	12mm
}

\usepackage{graphicx}

\usepackage{caption}
\usepackage{subcaption}

\usepackage{footnote}

\usepackage{dirtree}

\usepackage{multirow}

\usepackage{calc}

\newenvironment{streszczenie}{
	\par\noindent {\large \textbf{Streszczenie}\\[14pt]\indent}
}{}
\newenvironment{abstract}{
	\par\noindent {\large \textbf{Abstract}\\[14pt]\indent}
}{}

\frontmatter

\begin{document}

    \pagestyle{empty}
	\begin{titlingpage}
		\vspace*{\fill}
		\begin{center}
			\begin{picture}(430,500)
				\put(-25,590){\makebox(0,0)[l]{\huge \textbf{Wrocław University of Science and Technology}}}
				\put(20,565){\makebox(0,0)[l]{\Large \textbf{Faculty of Information and Communication Technology}}}
                \put(0,550){\line(1,0){430}}
                \put(0,510){\makebox(0,0)[l]{\large Field of study: \textbf{SZT}}}
                \put(0,490){\makebox(0,0)[l]{\large Speciality: \textbf{-}}}                
				\put(0,370){\begin{minipage}{0.9\textwidth}
				\centering
				\Huge \textsc{MASTER THESIS}
                \end{minipage}
				}
				\put(0,230){\begin{minipage}{0.9\textwidth}
				\centering
				\LARGE \textbf{Estimation of FFR in coronary arteries with deep learning}
                \end{minipage}
				}
				\put(0,170){\begin{minipage}{0.9\textwidth}
				\centering
				\Large {
				Patryk Rygiel
				}
				\end{minipage}
				}
				\put(0,90){\begin{minipage}{0.9\textwidth}
				\centering
				\large{
				Supervisor\\
				\textbf{dr hab. inż. Maciej Zięba}
				}
				\end{minipage}
				}
				\put(0,-30){
				\begin{minipage}{0.9\textwidth}
				\normalsize{
				deep learning, geometric deep learning, point clouds, CAD, FFR, hemodynamics}
				\end{minipage}
				}
                \put(0,-80){\line(1,0){430}}
				\put(155,-100){\makebox(0,0)[bl]{\large \textsc{Wrocław 2023}}}
			\end{picture}
		\end{center}	
		\vspace*{\fill}
	\end{titlingpage}
	
    \cleardoublepage
	\begin{abstract}

\noindent
Coronary artery disease (CAD) is one of the most common causes of death in the European Union and the USA.
The crucial biomarker in its diagnosis is called Fractional Flow Reserve (FFR) and its in-vivo measurement is obtained via an invasive diagnostic technique in the form of coronagraphy. 
In order to address the invasive drawbacks associated with a procedure, a new approach virtual FFR (vFFR) measurement has emerged in recent years. This technique involves using computed tomography angiography (CTA) to obtain virtual measurements of FFR.
By utilizing Computational Fluid Dynamics (CFD), vFFR estimates can be derived from CTA data, providing a promising in-silico alternative to traditional methods.
However, the widespread adoption of vFFR from CTA as a diagnostic technique is hindered by two main challenges: time and computational requirements.
In this work, we explore the usage of deep learning techniques as surrogate CFD engine models in the task of vFFR estimation in coronary arteries to drastically limit the required time and computational costs without a major drop in quality.
We propose a novel approach to vFFR estimation by representing the input vessel geometry as a point cloud and utilizing the hybrid neural network that learns geometry representation based on both explicitly and implicitly given features.
We evaluate the method from the clinical point of view and showcase that it can serve as a compelling replacement for commonly utilized CFD-based approaches.

\end{abstract}

	\vspace*{1cm}	
    \begin{streszczenie}

\noindent
Choroba niedokrwienna serca (CAD) jest jedną z najczęstszych przyczyn zgonów w Unii Europejskiej i  USA. Kluczowym biomarkerem w jej diagnozowaniu jest tzw. Wskaźnik Rezerwy Przepływu Wieńcowego (FFR), a jego pomiar in vivo uzyskuje się za pomocą inwazyjnej techniki diagnostycznej w postaci koronarografii.
Aby poradzić sobie z wadami związanymi z inwazyjnym zabiegiem, w ostatnich latach pojawiło się nowe podejście, jakim jest pomiar wirtualnego FFR (vFFR). 
Ta technika polega na wykorzystaniu Komputerowej Mechaniki Płynów (CFD), w celu pozyskania wartości vFFR na podstawie danych z tomografii komputerowej (CTA), co stanowi obiecującą in-silico alternatywę dla tradycyjnych metod.
Jednak powszechne stosowanie vFFR z CTA jako techniki diagnostycznej napotyka dwa główne wyzwania: złożoność czasową i obliczeniową.
W niniejszej pracy badamy zastosowanie technik głębokiego uczenia jako surogatu silnika CFD w zadaniu estymacji vFFR w tętnicach wieńcowych, aby znacznie ograniczyć wymagany czas i koszty obliczeniowe bez większego spadku w jakości.
Proponujemy nowatorskie podejście do szacowania vFFR, polegające na reprezentowaniu geometrii naczynia wejściowego jako chmury punktów i wykorzystaniu hybrydowej sieci neuronowej, która uczy się reprezentacji geometrii na podstawie cech podanych wprost jak i wyuczonych nie wprost.
Oceniamy tę metodę z punktu widzenia klinicznego i pokazujemy, że może ona stanowić przekonującą alternatywę dla powszechnie stosowanych podejść opartych na CFD.

\end{streszczenie}

    \cleardoublepage
	\pagenumbering{Roman}
	\pagestyle{tableOfContentStyle}
	\tableofcontents
	
	\clearpage
	\pagestyle{listOfPlotsStyle}
        \listoffigures
        \addcontentsline{toc}{chapter}{List of figures}
		
	
	\pagestyle{custom}
	\mainmatter
	

	\addcontentsline{toc}{chapter}{Introduction}
\chapter*{Introduction}

\thispagestyle{chapterBeginStyle}

\noindent
The goal of this thesis is to design, develop and implement a novel approach to the estimation of virtual Fractional Flow Reserve (vFFR) in coronary arteries with the usage of deep learning methods.
The scope of this work includes discussion and analysis of approaches to the task of vFFR estimation based on Computational Fluid Dynamics (CFD) and Artificial intelligence (AI).
We primarily focus on AI-based methods which allow us to drastically limit the time-consumption of CFD methods while not suffering great loss in performance. 
We propose a novel approach to the vFFR estimations by leveraging the point cloud representation and designing an adequate learning framework.
We pair implicit feature learning performed by a point cloud based neural network with hand-crafted features to construct a hybrid approach that achieves promising results from the clinical point of view. \\ 

\noindent
This work consists of $5$ chapters: \\

\noindent
\textbf{Chapter} \ref{chapter1} highlights the problem background by introducing how the \textit{in-vivo} FFR measurement is performed and why there is a need for in-silico measurement.
The concept of virtual FFR is introduced in the light of CFD and AI-based methods. \\

\noindent
\textbf{Chapter} \ref{chapter2} showcases related works which previously tackled the problem of vFFR estimation with AI. 
We group the methods under explicit and implicit feature learning categories and discuss its advantages and disadvantages. \\

\noindent
\textbf{Chapter} \ref{chapter3} introduces concepts of learning on 3D shapes.
Common 3D data representations are discussed, and the blueprint of Geometric Deep Learning (GDL) is introduced.
Through the lens of the GDL the architectures used for learning on point clouds, which are the main focus of this work, used in this work are described. \\

\noindent
\textbf{Chapter} \ref{chapter4} showcases the proposed approach to the estimation of vFFR based on point cloud representation.
We discuss the construction of the synthetic dataset used in this work, by describing the process of vessel geometry generation and computation of CFD simulations which serve as the ground truth labels.
Utilized deep learning architectures are described together with the training and inference pipelines. \\

\noindent
\textbf{Chapter} \ref{chapter5} showcases our experiments and results performed to evaluate the proposed approach.
We report various scores that prove the robustness, generalization and clinical viability of the proposed approach.
Finally, we also discuss the limitations and future works. \\

	\cleardoublepage

 	\addcontentsline{toc}{chapter}{Acknowledgments}
\chapter*{Acknowledgments}

\thispagestyle{chapterBeginStyle}

\noindent
This work has been performed together with the company Hemolens Diagnostics sp. z o.o.~\footnote{\url{https://hemolens.eu/}}. \\

\noindent
I would like to thank my supervisor dr hab. inż. Maciej Zięba for the provided support and suggestions in the design of a proposed deep learning approach, dr. Tomasz Konopczyński for providing invaluable insights and guidance during the development of this work from both medical and technical points of view, and dr. Paweł Płuszka for helping with the intricacies of CFD and physics aspects of this work.
	\cleardoublepage

	\chapter{Problem background}
\thispagestyle{chapterBeginStyle}
\label{chapter1}

\noindent
In this chapter, we state and describe the problem of Fractional Flow Reserve (FFR) estimation by introducing the medical background.
We discuss the \textit{in-vivo} coronarography procedure used to obtain FFR and common \textit{in-silico} replacement in the form of CFD simulation.
At last, we point out the common issues with the aforementioned techniques and introduce the concept of AI-based CFD engine surrogates.

\section{Coronarography-based FFR estimation}

\begin{figure}[ht]

\begin{subfigure}{.4\textwidth}
  \centering
  \includegraphics[width=\textwidth]{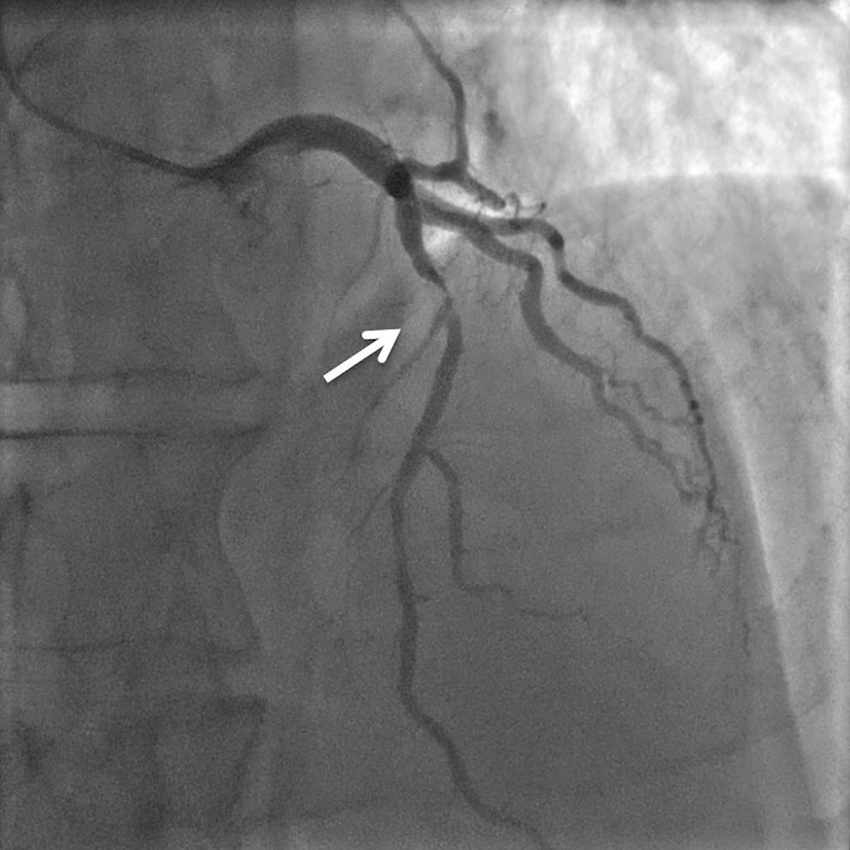}
  \caption{Coronary Angiography imaging}
  \label{fig:ch1:coronarography:angiography}
    \footnotesize{Source: \cite{ch1:coronarography-image}}
\end{subfigure}%
\begin{subfigure}{.6\textwidth}
  \centering
  \includegraphics[width=\textwidth]{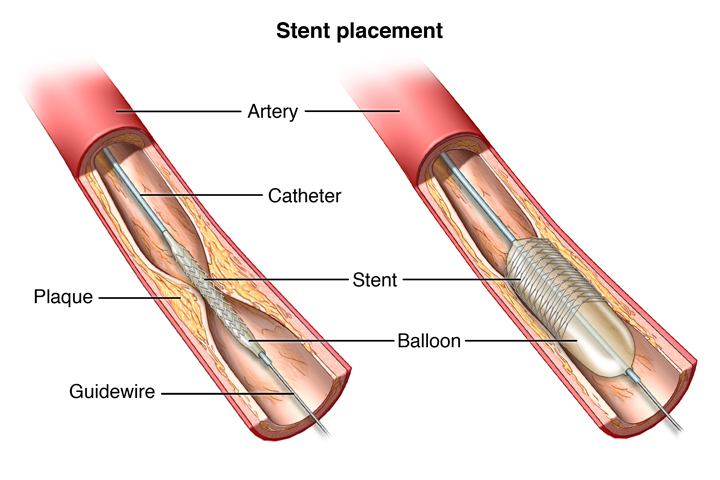}
  \caption{Process of implanting a stent}
  \footnotesize{Source: \cite{ch1:stent-image}}
  \label{fig:ch1:coronarography:stent}
\end{subfigure}

\caption[Coronarography procedure]{Coronarography procedure. The procedure is supervised via coronary angiography imaging and if required, the angioplasty treatment is performed by placing a stent in the stenotic region. The white arrow marks the severe stenosis seen on the coronary angiography imaging.}
\label{fig:ch1:coronarography}
\end{figure}

\noindent
The \textit{coronarography} is an invasive medical procedure performed on the patient's coronary arteries.
It is performed by bringing a catheter to the patient's heart through the radial, femoral, or brachial artery, and injecting contrast dye into the coronary arteries~\cite{ch1:coronarography}.
To control the process the patient is under continuous supervision via coronary angiography imaging (see Fig.~\ref{fig:ch1:coronarography:angiography}). 
The main goal of performing the procedure is an assessment of abnormalities in the patient's coronary arteries.
The procedure is often paired with the angioplasty treatment in the form of stent implantation in places where the vessel lumen is highly occluded by the atherosclerotic plaques or blood clots (the process of stent implanting is showcased in Fig.~\ref{fig:ch1:coronarography:stent}).
Such areas are called \textit{stenoses}~\cite{ch1:stenosis} and are the main reason for Coronary Artery Disease (CAD).
In stenotic areas, the blood flow is reduced and the myocardium (heart muscle) is not sufficiently supplied with blood thus oxygen delivery is impeded causing myocardial ischemia. 
The vessel's cross-sectional area is radically reduced and the role of the stent is to ensure proper blood flow by keeping it from narrowing or closing again~\cite{ch1:stents}.
Accurate assessment of the stent placement is crucial to ensure the correct and safe treatment.
The main biomarker, utilized to determine whether the stenosis impedes the blood flow, such the treatment is required, is called \textit{Fractional Flow Reserve} (FFR). \\

\noindent
The FFR measures pressure differences across coronary artery stenosis to compare the blood flow on either side of a blockage and determine how severe the impact of the stenosis on oxygen delivery to the myocardium is~\cite{ch1:wiki-ffr,ch1:cleveland-ffr}.
It is defined as the ratio of maximal coronary blood flow in a diseased artery to maximal coronary blood flow in the same artery without stenosis.
In reality, the ratio is computed as pressure after (distal to) stenosis relative to the pressure before (proximal to) the stenosis:

\begin{equation}
    FFR=\frac{p_d}{p_a},
\end{equation}
where $p_d$ is the pressure distal to the lesion, and $p_a$ is the pressure proximal to the lesion~\cite{ch1:wiki-ffr}. 

\noindent
The resulting value of FFR is an absolute value between $0$ and $1$, and when interpreting measurements, higher values indicate a non-significant stenosis, whereas lower values indicate a significant one~\cite{ch1:wiki-ffr,ch1:physiology-ffr}.
According to the medical literature and clinical trials~\cite{ch1:clinical-ffr-1,ch1:clinical-ffr-2}, the cut-off of $0.8$ is considered to be a practical determinant (with greater than $90\%$ accuracy~\cite{ch1:procedure-ffr}) of whether the stenosis is significant or not. \\

\noindent
During the coronarography procedure, to correctly estimate FFR, the pressure measurement needs to be performed at a steady state (no pressure fluctuation with time) during maximal coronary blood flow (\textit{hyperemia}). 
According to \cite{ch1:procedure-ffr}, hyperemia is \textit{"a physiologic state where coronary microvascular resistance is minimized"} and it \textit{"is typically achieved by administration of intravenous continuous infusion (140 mcg/kg/min) or intracoronary bolus (right coronary artery 50–100 mcg, left coronary artery 100–200 mcg) adenosine"}. \\

\noindent
Even though the FFR is a minimally invasive diagnostic technique, it still poses risks for patients, especially ones with CAD.
The risks include mainly radiation exposure in form of additional contrast injection (e.g the aforementioned adenosine) and increased risk of coronary arterial dissection with catherer~\cite{ch1:risks-ffr}.
Additionally, since the qualification for the procedure is often based on the non-invasive medical imaging technique in the form of Computed Tomography Angiography (CTA), coronarography is rarely used as a prophylaxis but rather as a target form of treatment.
Due to this fact, in recent years solutions estimating so-called virtual FFR (vFFR) based on CTA have been proposed as an alternative.

\section{Virtual FFR estimation}
\label{sec:1.2}

\noindent
The virtual FFR (vFFR) is an \textit{in-silico} measurement of the FFR, that is most often based on CTA imaging and denoted in literature as $\text{FFR}_{\text{CT}}$~\cite{ch1:ffr-ct}.
It offers a compelling replacement for the classic FFR measurement by omitting its invasive aspect and allowing for wider screening of patients.
The process of computing vFFR based on CTA requires a prior segmentation of coronary arteries.
It can be done threeway: manually by the expert, semi-automatically with an algorithm of choice under expert continuous supervision or fully automatically (most often with the usage of AI-based solutions) with the latter potential refinement by the expert.
The segmentation, obtained in either of the aforementioned ways, is a prerequisite for the \textit{in-silico} estimation of hemodynamic features with the usage of Computer Fluid Dynamics (CFD) or AI.

\subsection{CFD-based vFFR estimation}
\label{sec:1.2.2}

\begin{figure}[ht]
    \centering
    \includegraphics[width=.9\textwidth]{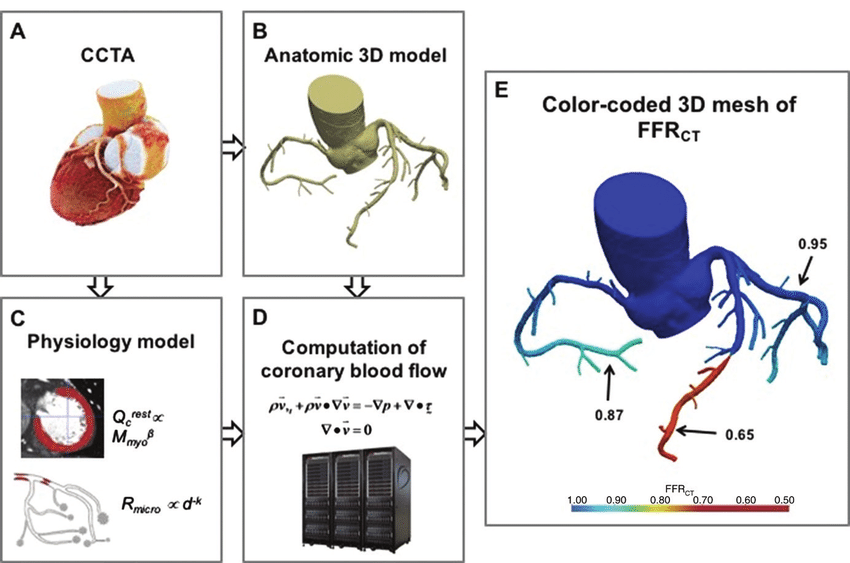}
    \caption[CFD-based vFFR estimation]{CFD-based vFFR estimation methods from CCTA are often derived from the common pipeline. Based on the CCTA image (A) the fine-grained anatomical 3D model (B) and physiology model defining boundary conditions (C), are constructed. Both of them serve as prerequisites of CFD numerical solver (D). The results are often visualised for the physician in the form of colour-coded 3D mesh (E).}
    \label{fig:ch1:ffr-cfd}
    \footnotesize{Source: \cite{ch1:ffr-cfd}}
\end{figure}

\noindent
Estimation of vFFR with a CFD engine is the most common and recognized approach to non-invasive diagnosis.
There are already solutions that have achieved FDA (U.S. Food and Drug Administration) clearance and can provide the product to hospitals and medical facilities~\cite{ch1:cardiac-flow-overview}.
The CFD-based approaches most often are derived from a common pipeline showcased in Fig.~\ref{fig:ch1:ffr-cfd} which consists of $4$ steps: (A) Image acquisition (CCTA), (B) Construction of anatomic 3D model, (C) Definition of physiology model and (D) Computation of coronary blood flow. \\

\noindent
Steps A and B have been already described in general terms in section~\ref{sec:1.2}, however for the correct CFD simulation some further criteria must be met. 
The anatomic 3D model needs to be represented as a surface mesh of a fine-grained quality for the simulation to be stable~\cite{ch1:cfd-meshing}.
Automatic generation of such mesh is not a trivial task and it often requires some postprocessing steps in the form of remeshing and manifold corrections~\cite{ch1:remeshing}.
Additionally, the level of mesh discretization, which can be understood as the maximal edge length, has a crucial influence on simulation quality and accuracy.
One may consider utilizing a very fine discretization at all times to avoid the issue of stability, however, the aspect of time consumption becomes the major drawback.
Thus the trade-off between time consumption and accuracy is an ever-looming issue which needs to be acknowledged and addressed in one way or the other to suit the task at hand. \\

\noindent
In step C the physiology model is constructed based on the CCTA volume.
The model consists of patient-specific boundary conditions (BC) estimated from rest state conditions including e.g. systolic and diastolic blood pressure, heart rate, and ventricular mass.
To model the state of maximal hyperemia, induced by the administration of adenosine, the conditions are appropriately modified by modelling the decrease in microvascular resistance~\cite{ch1:ffr-cfd}.
Moreover, the physiological model might be further enhanced towards specific patients by incorporating continuously recorded blood pressure waveforms obtained in a non-invasive measurement~\cite{ch1:hemolens-cfd-patent}.
The correct definition of the physiology model is crucial for the CFD simulation to well reflect the real blood flow and be accurate. \\

\noindent
The last step (D) is a computation of vFFR with a CFD simulation.
It is based on solving the Navier-Stokes equations, the physical laws that govern fluid dynamics, to compute the blood flow and pressure fields in the coronary vasculature.
However, because these equations are complicated nonlinear partial differential equations, numerical solving needs to be done in an iterative manner and thus is a computationally demanding task~\cite{ch1:ffr-cfd}.
The most common approach derives vFFR from a 3D anatomical model (obtained in step B), however computations for this approach, due to its complexity, need to be implemented off-site on supercomputers or cloud services~\cite{ch1:ffr-cfd-full-model-1,ch1:ffr-cfd-full-model-2}.
This model is called \textit{full-model} and so-called \textit{reduced-order} models have been proposed as alternative, less computationally demanding, approaches.
These models are based on averaging Navier-Stokes equations over vessel-cross sections and generalizing coronary microvasculature resistance parameteres~\cite{ch1:ffr-cfd,ch1:cfd-reduced-order-1,ch1:cfd-reduced-order-2}.
This approach allows for on-site computation of vFFR in less than 1 hour~\cite{ch1:cfd-reduced-order-time}, however, the accuracy in small, stenotic and bifurcating segments is decreased. \\

\noindent
Although CFD-based vFFR estimation is a compelling \textit{in-silico} alternative to invasive FFR measurement, its use as prophylaxis is still limited due to the high computational and time demands.
With the recent rapid advances in machine learning (ML) and deep learning (DL), the interest in utilizing these techniques in the estimation of vFFR has been sparked.
Due to its nature, deep learning allows to drastically limit the computational and time demand in inference settings which is the main drawback of CFD-based approaches.

\subsection{AI-based vFFR estimation}

\noindent
The AI-based approaches utilize a similar pipeline as CFD ones (showcased in Fig.~\ref{fig:ch1:ffr-cfd}), by surrogating steps C and D with an ML model which estimates hemodynamic features directly.
The ML models are most often trained in a supervised matter by considering CFD-based simulation results as the ground truth (GT), thus they can be only as good as the underlying simulation. \\

\noindent
One may consider training the model on invasive FFR measurement, however, this approach is limited by the nature of the measurement which is done only locally on the suspected stenotic segment but not continuously along the whole vessel.
There exists a locally-continuous technique of pressure measurement along the vessel called \textit{intracoronary pressure pullback}.
This technique is performed by bringing the measurement device in the form of a pressure wire to the lesion of interest and then performing a pullback manoeuvre whereby the pressure wire is pulled back at a fixed rate through the vessel towards the aorta, and the pressure along the way is recorded continuously~\cite{ch1:ffr-pullback-1,ch1:ffr-pullback-2}.
However, such a procedure is not commonly performed and only yields pressure values in the local segment of interest. \\

\noindent
The main gain of utilizing the AI-based approaches over CFD ones lies in the computational and time demands in the inference setting. 
As mentioned in Section~\ref{sec:1.2.2}, the iterative solving of numerical equations can take up to hours for a single geometry.
When using AI methods, the time and computational demands are shifted from the inference onto the training process, which is done only once in the non-production environment before deployment.
The already trained model, depending on the exact architecture and approach, drastically reduces the time demand of the vFFR estimation to seconds and can often be run locally on-site.
Due to this fact, AI-based approaches may serve as a compelling surrogate of the CFD engine and could bring vFFR estimation procedures to be a common prophylaxis non-invasive diagnostic tool. 

\section{Summary}

\noindent
In this chapter, we introduced and described the medical background of FFR estimation.
We described in detail a process of obtaining an in-vivo FFR measurement and highlighted its issues, risks for the patient, and obstacles that stand in the way of utilizing it as a common diagnostic technique. 
Next, the concept of in-silico measurement in the form of virtual FFR estimation has been introduced as a compelling non-invasive alternative that allows for a broader patient screening.
The most common technique based on the CFD engine was described and the main drawbacks in the form of time and computational demand were highlighted.
As the solution, we introduced the approach of surrogating a CFD engine with the learned AI model.
In the next chapter, we will delve deeper into AI-based methods and techniques by discussing the related works and the most commonly used approaches up to date.

	\cleardoublepage

	\chapter{Related work}
\thispagestyle{chapterBeginStyle}
\label{chapter2}

\noindent
In this chapter, we will present and discuss the related works that tackle the problem of estimating hemodynamic features (not only in the form of vFFR) with AI.
To solve this task a few different approaches have been proposed in recent years. 
The most basic ones perform extraction of hand-crafted features from the underlying geometry and use them as the geometry representation for the latter regressor of the hemodynamic feature of choice~\cite{ch2:siemens,ch2:siemens-sklet,ch2:keya,ch2:fossan}.
In more advanced approaches, the geometry is encoded implicitly by the dedicated encoder working directly on the input mesh~\cite{ch2:point-clouds-aorta,ch2:gem-wss,ch2:gem-velocity}.
Such methods can be further enhanced by incorporating physics-inspired losses to form so-called \textit{physics-informed neural networks} (PINNs), whereby the underlying differential equation is exploited to guide the network~\cite{ch2:pinns-1,ch2:pinns-2,ch2:pinns-3}. 
However, in this work, we will not delve into PINNs but rather only mention them for the sake of the potential future ways of research.

\section{Learning on explicit hand-crafted features}
\label{sec:2.1}

\noindent
The first models introduced to estimate vFFR via AI proposed to utilize hand-crafted features.
As such features, we consider ones that need to be designed a priori by the domain expert and can be automatically extracted or computed based on the vessel geometry.
When designing such features, expert knowledge is required since the features need to correlate with the estimated hemodynamic feature for the model to be even learned.
This is not a trivial task, since not all correlating and important features are obvious, and can be missed.
Since no other information besides these features is given to the model, its performance relies almost only on the design choice of the expert.
On the other hand, including such features allows for greater explainability.
The features are tangible for the expert and their influence on the result can be measured. \\

\subsection{Siemens framework (Itu et al.)}
\label{sec:2.1.1}
\begin{figure}[ht]
    \centering
    \includegraphics[width=.85\textwidth]{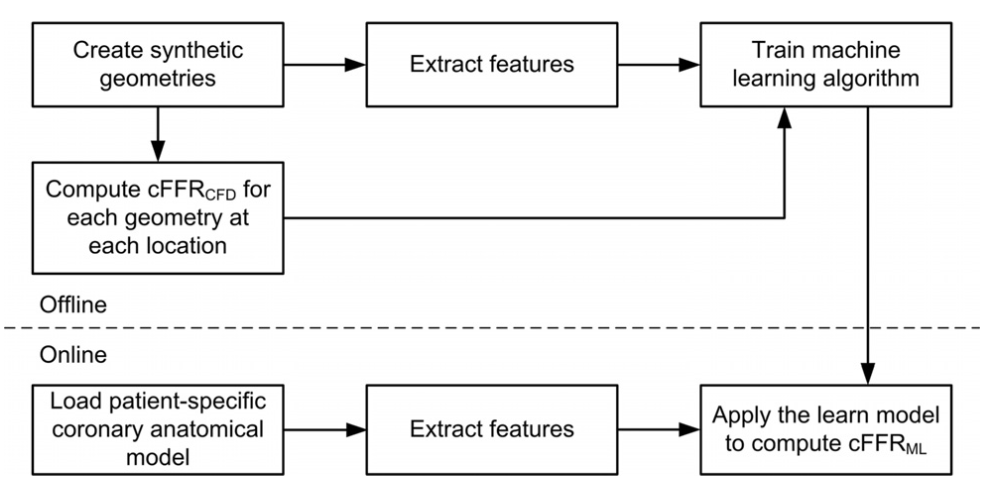}
    \caption[Workflow of Siemens framework]{Workflow of Siemens framework. The model is trained on synthetic data offline and then deployed online to perform vFFR regression on real patients.}
    \label{fig:ch2:siemens-workflow}
    \footnotesize{Source: \cite{ch2:siemens}}
\end{figure}

\noindent
We will start by introducing the work by Itu et al.~\cite{ch2:siemens}, which together with Siemens, proposed the first-ever AI-based vFFR estimation framework.
The steps of the proposed framework are showcased in Fig.~\ref{fig:ch2:siemens-workflow}.
In the first step, the dataset of $12,000$ coronary anatomies is generated to form a rich sample of the different morphologies of coronary blockage.
The parameters of generated geometries are sampled in prespecified ranges based on the published medical literature~\cite{ch2:coronary-anatomy}.
One may raise a question of why the model is not trained on the dataset of real patients' anatomies. 
The authors claim, that it is known that DNNs require lots of data to be properly trained, and datasets of real geometries are often not large enough to be able to cover the vast variability between the patients.
To prove this claim, the authors showcase that in their framework training on synthetic geometries does not hinder the performance on the real anatomies on which the model is validated. \\

\noindent
The framework utilizes a supervised learning scheme and the ground truth labels are generated via CFD simulation (this notion was introduced in Section~\ref{sec:1.2.2}).
The vessel geometry is split into segments that are considered stenotic and non-stenotic.
From the stenotic segments, the hand-crafted features are extracted, and together with CFD ground truths, they form a training dataset for the ML model.
In the inference setting (online section in Fig~\ref{fig:ch2:siemens-workflow}), the ML model trained on synthetic anatomies is utilized to infer vFFR for the patient-specific coronary anatomical models with latter feature extraction in the same manner as for the synthetic geometries. \\

\begin{figure}[ht]
    \begin{subfigure}{.6\textwidth}
      \centering
      \includegraphics[width=\textwidth]{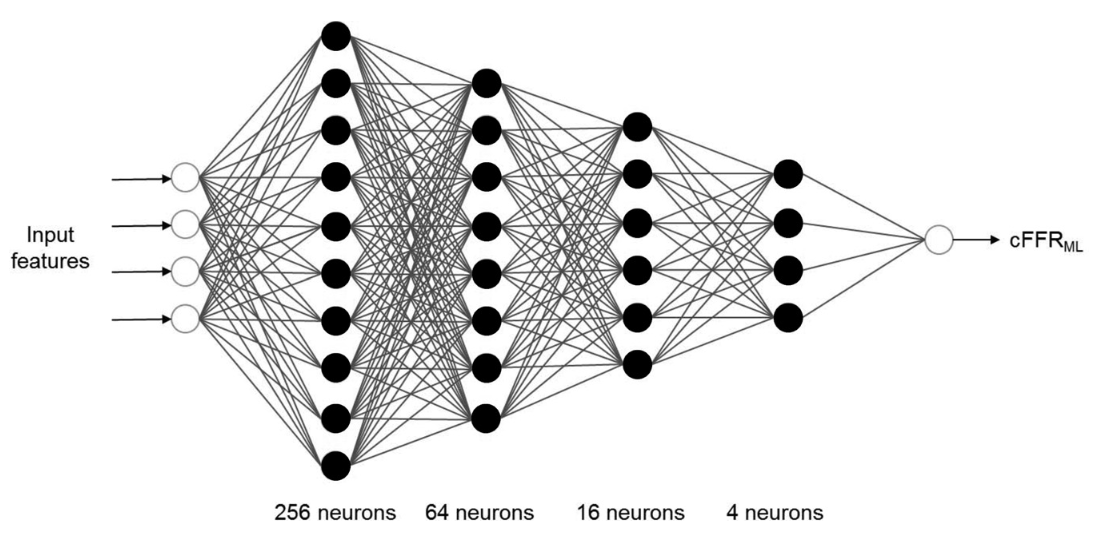}
      \caption{DNN architecture}
      \label{fig:ch2:siemens-ml:architecture}
    \end{subfigure}%
    \begin{subfigure}{.4\textwidth}
      \centering
      \includegraphics[width=\textwidth]{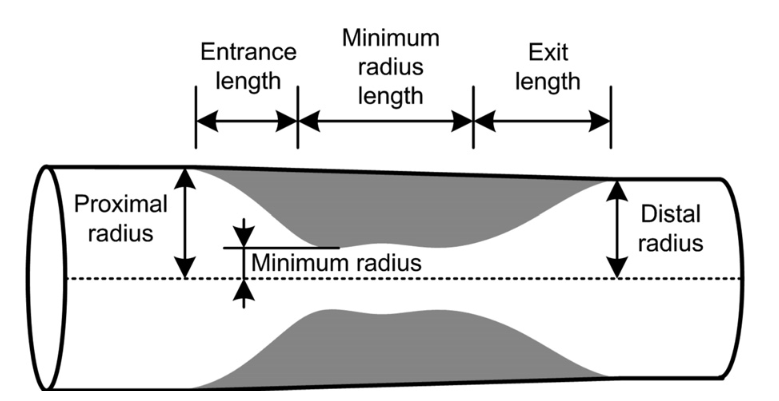}
      \caption{Hand-crafted features}
      \label{fig:ch2:siemens-ml:features}
    \end{subfigure}

    \centering
    \caption[Deep learning approach by Siemens]{Deep learning approach by Siemens. The simple $4$-layer MLP is utilized to regress local vFFR from hand-crafted features in the form of local and global vessel characteristics.}
    \label{fig:ch2:siemens-ml}
    \footnotesize{Source: \cite{ch2:siemens}}
\end{figure}

\noindent
As the ML model, the authors propose a simple multi-layer perception (MLP) built out of $4$ hidden layers (see Fig.~\ref{fig:ch2:siemens-ml:architecture}).
The model performs vFFR regression for the given stenotic segment based on a set of extracted features.
The authors utilize a total of $28$ hand-crafted features which encode local, upstream and downstream  information (examples of local features are presented in Fig.~\ref{fig:ch2:siemens-ml:features}).
The features, however, are not described in detail enough for the approach to be reproducible, which was also reported by Sklet et al.~\cite{ch2:siemens-sklet}. \\

\noindent
The Siemens framework was validated on the database of $87$ patient-specific anatomies resulting in a total of $125$ stenotic segments.
They report a vFFR correlation of $0.9994$ between the CFD and AI approaches.
However, for the invasive FFR, the reported correlation is only $0.729$.
As we mentioned in Section~\ref{sec:1.2.2}, the performance of the ML model can be only as good as the CFD ground truth.
And since, the reduced-order CFD model utilized in this work, achieves a correlation of $0.725$, we cannot expect the ML model to be substantially better.

\subsection{DEEPVESSEL-FFR (Wang et al.)}

\begin{figure}[ht]
    \centering
    \includegraphics[width=.9\textwidth]{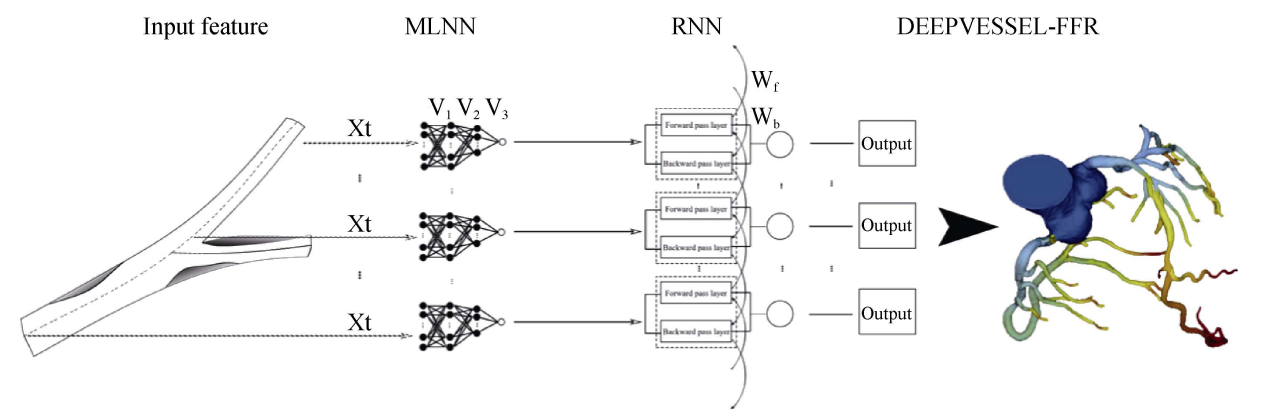}
    \caption[DEEPVESSEL-FFR framework]{DEEPVESSEL-FFR framework. This approach utilizes two phases, a simple feature encoder, similar to Siemens framework~(\ref{sec:2.1.1}), and a latter RNN for more robust vFFR regression that includes upstream and downstream vessel characteristics.}
    \label{fig:ch2:deepvessel-ffr}
    \footnotesize{Source: \cite{ch2:keya}}
\end{figure}

\noindent 
The platform DEEPVESSEL-FFR proposed by Wang et al.~\cite{ch2:keya} follows a similar approach as previously described Siemens framework~(\ref{sec:2.1.1}) in utilizing hand-crafted features.
However, instead of a simple MLP, the DNN model is arranged out of two components, a feature encoder in the form of a multilevel neural network (MLNN) and a bi-directional recursive neural network (BRNN) (see Fig.~\ref{fig:ch2:deepvessel-ffr}).
The first phase encodes hand-crafted features describing vessel geometry in the form of lesion characteristics and proximal/distal markers subsequently defined for each lesion, into feature vectors.
In the second phase, recurrent architecture is utilized to include in local vFFR regression upstream and downstream information in the form of surrounding centerline node embeddings. 
As for the Siemens approach~(\ref{sec:2.1.1}), the utilized ground truths are obtained via classic CFD simulation. \\

\noindent
The authors do not disclose the setup of the training database and only provide the information that the evaluation study has been performed on $68$ patients.
They report the correlation with invasive FFR measurement to be $0.686$.
However, as for the Siemens framework~(\ref{sec:2.1.1}), the proposed approach shares the same issues regarding reproducibility, due to the lacking description of exactly used hand-crafted features.

\subsection{Fossan et al.}

\begin{figure}[ht]
    \begin{subfigure}{.48\textwidth}
      \centering
      \includegraphics[width=\textwidth]{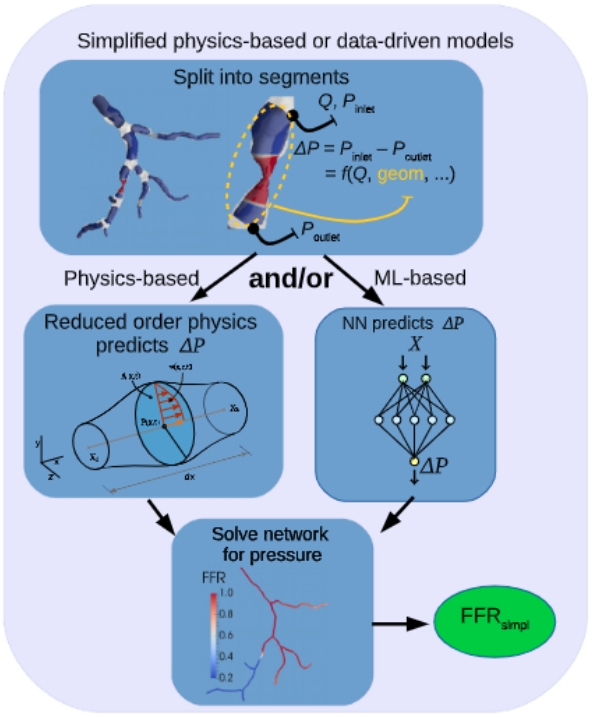}
      \caption{Proposed framework}
      \label{fig:ch2:fossan-ml:framework}
    \end{subfigure}%
    \begin{subfigure}{.52\textwidth}
      \centering
      \includegraphics[width=\textwidth]{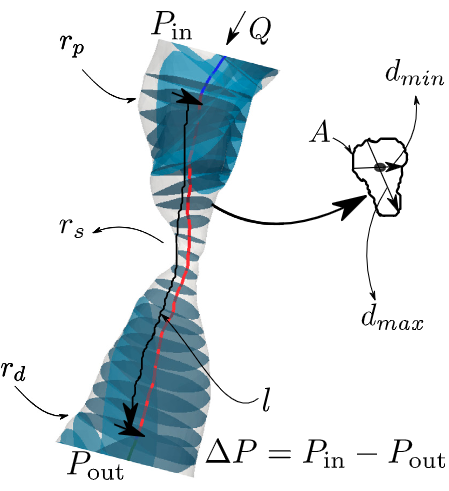}
      \caption{Hand-crafted features}
      \label{fig:ch2:fossan-ml:features}
    \end{subfigure}

    \centering
    \caption[Deep learning approach by Fossan et al.]{Deep learning approach by Fossan et al. In this approach, the pure MLP used in Siemens framework~(\ref{sec:2.1.1}) is paired with a reduced-order physics model to incorporate physics-based knowledge into the training. The hand-crafted features used to encode the local segment are derived both from geometry: radii along the stenosis ($r_p, r_s, r_d$), length of the stenosis ($l$), cross-sectional area ($A$), minimal cross-sectional diameter ($d_{min}$), and maximal cross-sectional diameter ($d_{max}$); and fluid motion: input pressure ($P_{in}$), output pressure ($P_{out}$), inflow ($Q$), pressure loss ($\Delta P_{0D}$) and flow separations changes on upstream pressure ($\Delta P_{sep}$).}
    \label{fig:ch2:fossan-ml}
    \footnotesize{Source: \cite{ch2:fossan}}
\end{figure}

\noindent
The next approach that utilizes the hand-crafted features but also incorporates some physics-based knowledge was proposed by Fossan et al.~\cite{ch2:fossan}.
In this work, the authors utilize a reduced-order physics model and the fully-connected neural network to regress pressure drops on the local vessel segment and based on the predicted results solve the exact vFFR. 
The local segment is represented by the set of hand-crafted features describing the vessel geometry (see Fig~\ref{fig:ch2:fossan-ml:features}) e.g. radii along the stenosis ($r_d, r_p, r_s$), length of the stenosis ($l$) and cross-sectional area information ($A, d_{min}, d_{max}$).
The features are further enhanced by incorporating the physics knowledge derived from the reduced-order physics model in the form of blood flow ($Q$), pressure loss ($\Delta P_{0D}$) and dynamic and flow separation changes/losses on upstream pressure ($\Delta P_{sep}$). \\

\noindent
The model was trained and validated on the dataset collected from $64$ patients with stable CAD and ground truths were generated via the CFD simulation.
The authors report an error standard deviation of $0.021$ between CFD GT and their proposed approach.

\subsection{Summary}

\noindent
To conclude, the models utilizing hand-crafted features were the first ones to be proposed to solve the problem of vFFR regression. 
They offer some form of explainability by utilizing explicit features and can be further enhanced by incorporating more geometry or physics-based features.
However, since the features need to be defined explicitly by the expert for specific tasks, the models are not simply transferable to estimate different hemodynamic features and due to some important features being potentially missed in the design process, the DNN model might not be able to be trained properly.
Moreover, the common lack of in-detail description of utilized hand-crafted features poses a major limitation in reproducing and validating the proposed approaches.

\section{Learning on implicit features}
\label{sec:2.2}

\noindent
The alternative approach to encoding geometry based on hand-crafted features is to train the model to perform implicit feature extraction from the raw geometry.
These approaches utilize DNNs that are tailored towards the processing of 3D structures such as point clouds or meshes, which will be described in further detail in Chapter~\ref{chapter3}.
By encoding raw geometry, the requirement of defining a set of carefully designed hand-crafted features is no longer needed, and thus the models are not limited by the encoding scheme proposed by the expert.
The embeddings of the geometries are learnt implicitly for the task at hand and thus these architectures can be utilized to estimate a variety of hemodynamic features.
In this approach, however, we lose some form of explainability by sacrificing known features for the implicit representation, which does not provide much interpretability.

\subsection{Li et al.}

\begin{figure}[ht]
    \begin{subfigure}{.48\textwidth}
      \centering
      \includegraphics[width=\textwidth]{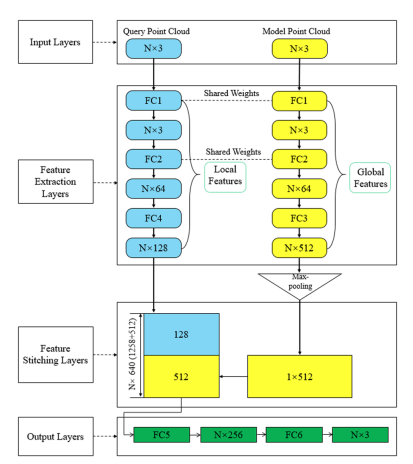}
      \caption{Proposed point cloud DNN architecture}
      \label{fig:ch2:aorta:dnn}
    \end{subfigure}%
    \begin{subfigure}{.52\textwidth}
      \centering
      \includegraphics[width=\textwidth]{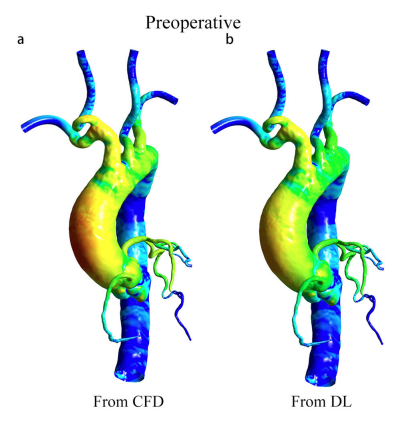}
      \caption{Comparison of pressure fields from CFD and AI}
      \label{fig:ch2:aorta:results}
    \end{subfigure}

    \centering
    \caption[Point cloud based approach by Li et al.]{Point cloud based approach by Li et al.}
    \label{fig:ch2:aorta}
    \footnotesize{Source: \cite{ch2:point-clouds-aorta}}
\end{figure}

\noindent
In Li et al~\cite{ch2:point-clouds-aorta}, the authors propose to represent input geometry as the 3D point cloud to estimate velocity and pressure fields in the coronary arteries and aorta before and after bypass surgery.
The authors propose a novel PointNet-based~\cite{ch3:pointnet} architecture (showcased in Fig.~\ref{fig:ch2:aorta:dnn}) built out of two separate encoding branches.
The input point cloud is split into \textit{model point cloud} which contains boundary points of the geometry, and the \textit{query point cloud} which contains points in the vessel lumen.
According to the authors, the split was performed in this manner to encode local and global features separately to allow for better spatial relationship encoding.
From both point clouds the same amount of input points are sampled and obtained encodings are stitched together to form a global representation.
The obtained representation is further decoded with fully-connected layers to yield the final hemodynamic feature.
When we described the learning on hand-crafted features, the issue of being unable to utilize the network for various hemodynamic features, due to different requirements for input features, has been highlighted.
However, when we deal with implicit feature learning, this is no longer an issue, since the network can learn relevant features for the task on its own.
This case is proved by Li et al.~\cite{ch2:point-clouds-aorta} by showcasing that the same architecture can be utilized with the same success for both 3D velocity and pressure field estimations (qualitative example of obtained pressure field is showcased in Fig.~\ref{fig:ch2:aorta:results}). \\

\noindent
The study has been performed on $110$ real patient data and the ground truths have been obtained via CFD simulation.
They report the correlation of vFFR with CFD-obtained FFR to be $0.9580$ and highlight that the performance must be further evaluated due to the lack of clinical data in the form of invasive FFR to compare with.

\subsection{Suk et al.}

\begin{figure}[ht]
    \centering
    \includegraphics[width=.9\textwidth]{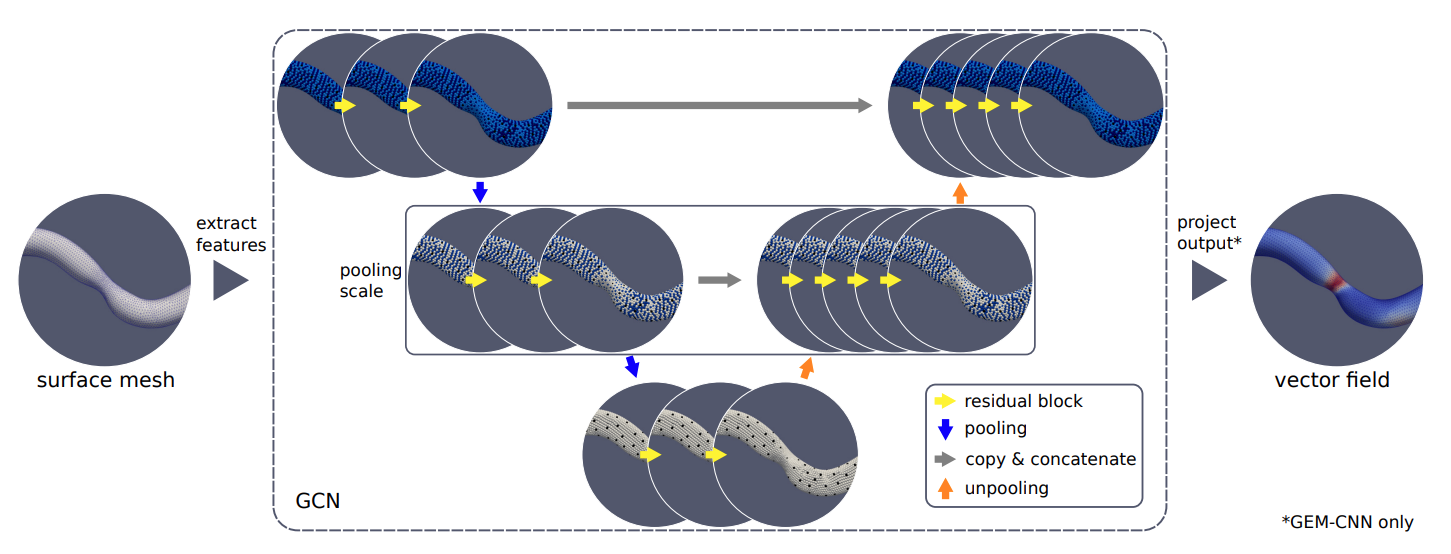}
    \caption[Mesh-based DNN architecture by Suk et al.]{Mesh-based DNN architecture by Suk et al.}
    \label{fig:ch2:gem-dnn}
    \footnotesize{Source: \cite{ch2:gem-wss}}
\end{figure}

\noindent
The works by Suk et al.~\cite{ch2:gem-wss,ch2:gem-velocity} delve into utilizing mesh-based neural networks for the hemodynamics estimation in the form of Wall Shear Stress (WSS) and velocity field.
The input geometry is represented as the surface or volumetric mesh and then processed with UNet-like~\cite{ch3:unet} architecture tailored towards processing raw meshes (see Fig.~\ref{fig:ch2:gem-dnn}).
The nodes of the input mesh, are further enhanced beyond spatial coordinates features only, to include simple hand-crafted features in the form of geodesic distance from the vessel inlet.
This feature is easily computed and serves as the prior which informs the network of fluid flow direction.
As for Li et al.~\cite{ch2:point-clouds-aorta}, the advantage of the proposed approach lies in the notion that the architecture can be utilized for various hemodynamic feature estimations. \\

\noindent
The approach is trained and evaluated on synthetically generated datasets containing simple geometries with stenoses and bifurcating geometries, both containing $2000$ in unique geometries.
Again, as for previously discussed approaches, the ground truths are obtained via CFD simulation.
The authors report the x$180$ speedup by utilising the AI approach in comparison to CFD one and plan on further extending their work in the future to include real geometries.

\subsection{Summary}

\noindent
To sum up, learning on features that are implicitly extracted by the model to the task at hand, is a compelling alternative to approaches utilising hand-crafted features.
The implicit models are far more generic and can be utilized to estimate various hemodynamic features, whereas hand-crafted ones are limited towards the specific hemodynamic feature and cannot be extended to other problems.
The main drawbacks lie in decreased explainability and higher computational demands since more advanced architectures, tailored towards the processing of raw 3D data, must be utilized.
The hand-crafted representation takes the form of a low-dimensional numerical feature vector which can be sufficiently processed with simple MLP which has low computational overload.
However, with recent advances in deep learning for 3D shapes, efficient architectures have been developed to process raw 3D point clouds and meshes, making implicit learning a viable choice.

\section{Summary}

\noindent
In this chapter, we reviewed the most relevant works that tackle the problem of vFFR and other hemodynamic feature estimations.
We grouped the methods into the explicit and implicit feature learning categories and highlighted their advantages and disadvantages.
We discussed the proposed workflows of learning on synthetic and real data and noted that all the methods utilise CFD simulation to generate ground truth labels for the supervised task of vFFR estimation.
In the next chapter, we will discuss in detail learning on 3D shapes (implicit feature learning).
	\cleardoublepage

 	\chapter{Learning on 3D shapes}
\thispagestyle{chapterBeginStyle}
\label{chapter3}

\noindent
In this work, we focus mainly on implicit feature learning (see Section \ref{sec:2.2}) on the raw representations of 3D objects in the form of point clouds.
Thus, in this chapter, we will describe the concepts behind learning on 3D shapes.
We will start by introducing various 3D data representations and discussing their strengths and weaknesses.
Next, the blueprint of the so-called geometric deep learning (GDL)~\cite{ch3:gdl} will be discussed as the set of concepts and guidelines for constructing architectures towards learning robust representations of 3D shapes.
Through the lens of GDL, we will introduce specific architectures for learning on point clouds and meshes that are utilized in this work.

\section{Representations of data in 3D}

\begin{figure}[ht]
    \centering
    \includegraphics[width=\textwidth]{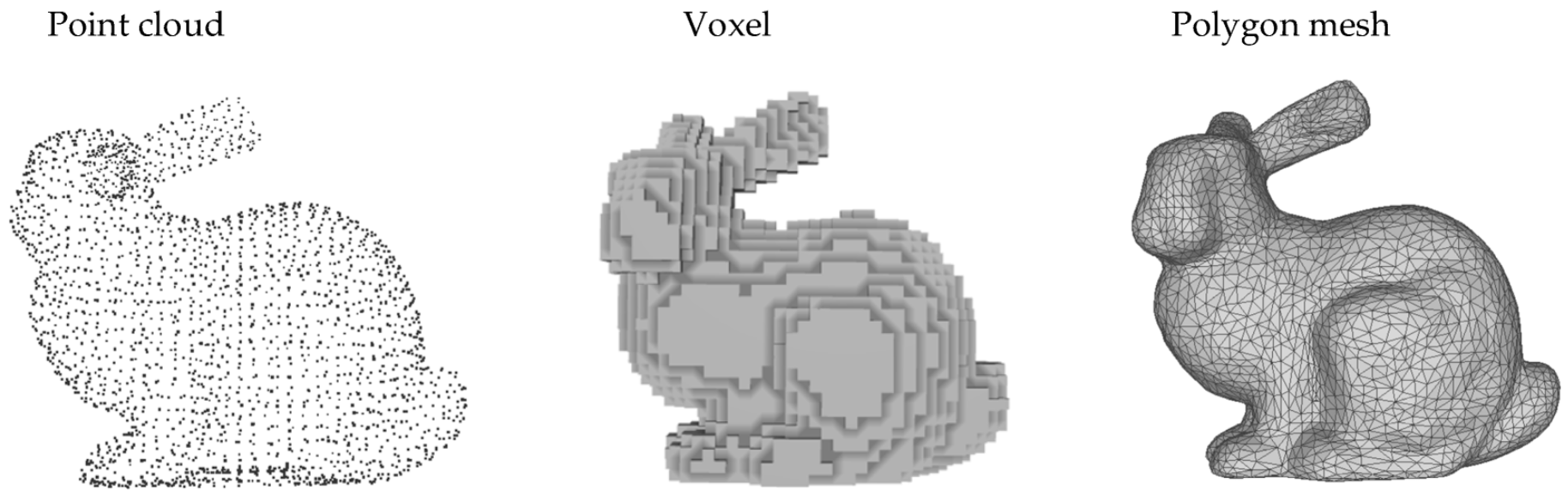}
    \caption[Representations of data in 3D]{Representations of data in 3D. Three-dimensional model of the bunny represented in the $3$ most common representations (from the left): point cloud, voxel grid and polygonal mesh.}
    \label{fig:ch3:3D-representations}
    \footnotesize{Source: \cite{ch3:3D-representations}}
\end{figure}

\noindent
The 3D objects in computer vision can be represented in various ways, depending, among others, on the image acquisition device or specific use case of the task at hand. 
When defining the problem to be solved, we must also consider which representation is most suited to it.
In this work, we will look at the three most common representations of data in 3D: voxel grids, point clouds and meshes (see Fig.~\ref{fig:ch3:3D-representations}). \\

\subsection{Voxel grid}

\noindent
The \textit{voxel grid} is a 3D extension of a 2D \textit{pixel grid}, which is the most common technique of representing images.
The pixel grid is a 2D regular grid built out of square (isotropic) or rectangular (anisotropic) elements called pixels.
The images may come in the form of single or multi-channel grids.
In single-channel grids, each pixel takes a single value, and in multi-channel a tuple of values e.g. RGB, where each pixel is represented with the $3$ element tuple encoding its colour. 
As mentioned earlier, the voxel grid is a natural extension of the pixel grid into 3D, where the smallest element called \textit{voxel} can be either a cube or cuboid. \\

\noindent
When dealing with 3D objects, we often represent them as binary masks with the voxel grid.
The object is placed inside the cuboid bounding box, which is the regular grid of specified voxel density.
The voxels belonging to the foreground (object) take a value of $1$ and to the background of $0$. 
In this manner, we can think of a voxel grid as a construction of an object out of small regular cuboids. \\

\noindent
The main issue of the voxel grid representation of 3D objects lies in the resolution.
Since the grid needs to be regular the regions with the various amount of detail are represented with the same amount of elements.
Thus, when we want to increase the resolution of a certain part of the object, the resolution of the whole grid needs to be increased, resulting in high memory costs (see bunny model labelled as "voxel" in Fig.~\ref{fig:ch3:3D-representations} - the resolution of a surface near the ears is low, the model is "blocky", however for a belly part it is high enough).
Moreover, 3D objects are often a 2D surface embedded in a 3D space and thus the representation of space inside and outside of the object is not required.
The voxel grids are not efficient in this manner, due to the aforementioned constant resolution across the whole grid.
For example, the background of the object is modelled with a large amount of $0$-value voxels which do not provide any information.
For these reasons, we rarely utilized this data representation when dealing with 3D objects, and rather opt to utilize more sparse representations that allow varying resolution across the domain.

\subsection{Mesh}

\noindent
The \textit{mesh} is a graph-based representation of a 3D surface (surface mesh) or volume (volumetric mesh).
We construct the mesh out of nodes, which represent the points embedded in the 3D space, and edges, which represent the connectivity between the nodes to form a 2D surface built out of closed polygons, called faces.
Each node is represented as a tuple of 3D spatial coordinates $x$, $y$ and $z$ in the specified metric space (most commonly we utilize a Euclidean metric space).
The tuples can be further expanded to hold more values per point e.g. colour, normal vector etc. \\

\noindent
Due to its graph nature, the mesh can represent the underlying object surface with varying resolution.
The more detailed areas are densely sampled with nodes, thus conveying more detail, and more simple areas can be sampled sparsely since there is not much detail to covey (see bunny model labelled "polygon mesh" in Fig.~\ref{fig:ch3:3D-representations}).
Moreover, when we deal with surface meshes, the interior and exterior of the object is not modelled at all, making it far more memory-efficient in comparison to the voxel grid.
These properties make a mesh, to be an efficient, detailed and sparse representation of a 3D object. \\

\noindent
However, there are some technical caveats and requirements that must be met when dealing with meshes.
We often require a mesh to be a \textit{manifold}, meaning that it is watertight (the inside and outside are clearly defined and not connected), each edge has exactly two adjacent faces and there are no intersections between the faces.
Keeping these properties is not trivial when some processing techniques are utilized.
Often, fixing techniques in the form of remeshing~\cite{ch1:remeshing} or cleaning of degenerated faces, nodes or edges, are required.
So, even though, the mesh is the far superior technique for representing 3D objects, its irregular nature makes it far more tricky to process than the voxel grid.

\subsection{Point cloud}
\label{sec:3.1.3}

\noindent
The \textit{point cloud} is a lightweight 3D representation that in comparison with the mesh, consists of nodes only without edges.
We can think of a point cloud, as an unordered set of points from some metric space (most commonly a Euclidean one).
As for a mesh representation, nodes can be represented as $n$-dimensional tuples consisting of 3D spatial coordinates and other features of choice. \\

\noindent
Compared with a mesh representation, a point cloud conveys less information due to the lack of connectivity between the points and explicitly defined surface.
To correctly represent the object's surface, more dense point sampling is required than in the case of mesh, where nodes can be further away from each other since the surface is given via connected edges (see bunny labelled "point cloud" in Fig.~\ref{fig:ch3:3D-representations} - without dense sampling in non-detailed areas the surface given implicitly is ambiguous).
However, we shouldn't consider points in a point cloud representation totally isolated.
The relations between the points are given through the metric space and even mesh with the correct surface can be reconstructed from it via some standard techniques such as ball-pivoting algorithm~\cite{ch3:ball-pivoting} or Poisson surface reconstruction~\cite{ch3:poisson-surface}.
Moreover, due to their simple construction, processing point clouds is much easier and more efficient than meshes, especially in the case of neural networks. \\

\noindent
However, designing a neural network architecture suited to the processing of raw point clouds is not a trivial task.
Thus, before introducing the specific deep learning approaches, we will delve into the \textit{Geometric Deep Learning}~\cite{ch3:gdl} which sets up a set of general principles to follow when designing architectures to learning on high-dimensional data. \\

\section[Geometric deep learning]{Geometric deep learning\footnote{In this section, we will only introduce the basic concepts behind the GDL that are important for defining the notion of representation learning on point clouds. All the definitions and equations are based on the work by Bronstein et al.~\cite{ch3:gdl} and for more details the reader is referred to this work.}}
\label{sec:3.2}

\noindent
\textit{Geometric Deep Learning (GDL)} is a set of general principles developed by Bronstein et al.~\cite{ch3:gdl} for designing neural networks that can learn robust and stable representations of high-dimensional data.
The proposed hypothesis behind GDL states that \textit{"learning generic functions in high dimensions is a cursed estimation problem, most tasks of interest are not generic, and come with essential pre-defined regularities arising from the underlying low-dimensionality and structure of the physical world."}~\cite{ch3:gdl}.
The exploitation of such regularities allows for remedying the curse of dimensionality when dealing with large and complex systems. \\

\begin{figure}[ht]
    \begin{subfigure}{.30\textwidth}
      \centering
      \includegraphics[width=\textwidth]{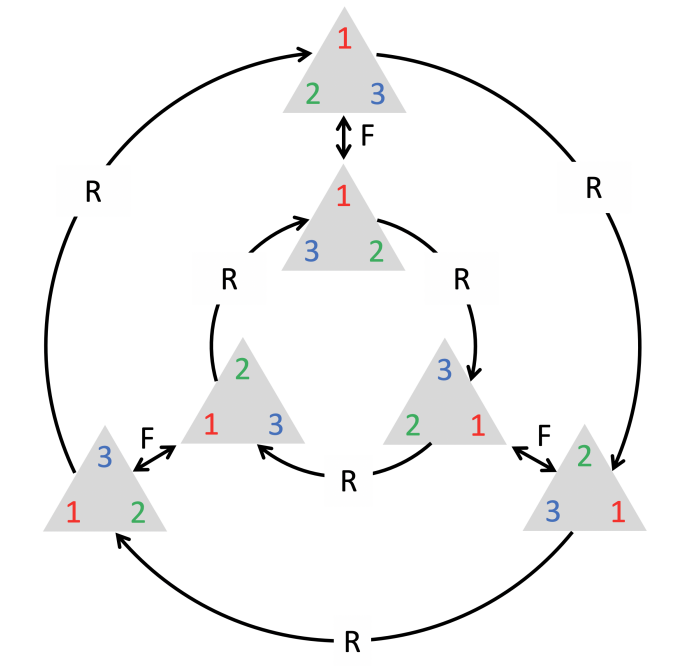}
      \caption{Symmetry}
      \label{fig:ch3:gdl-principles:symmetry}
    \end{subfigure}%
    \begin{subfigure}{.40\textwidth}
      \centering
      \includegraphics[width=\textwidth]{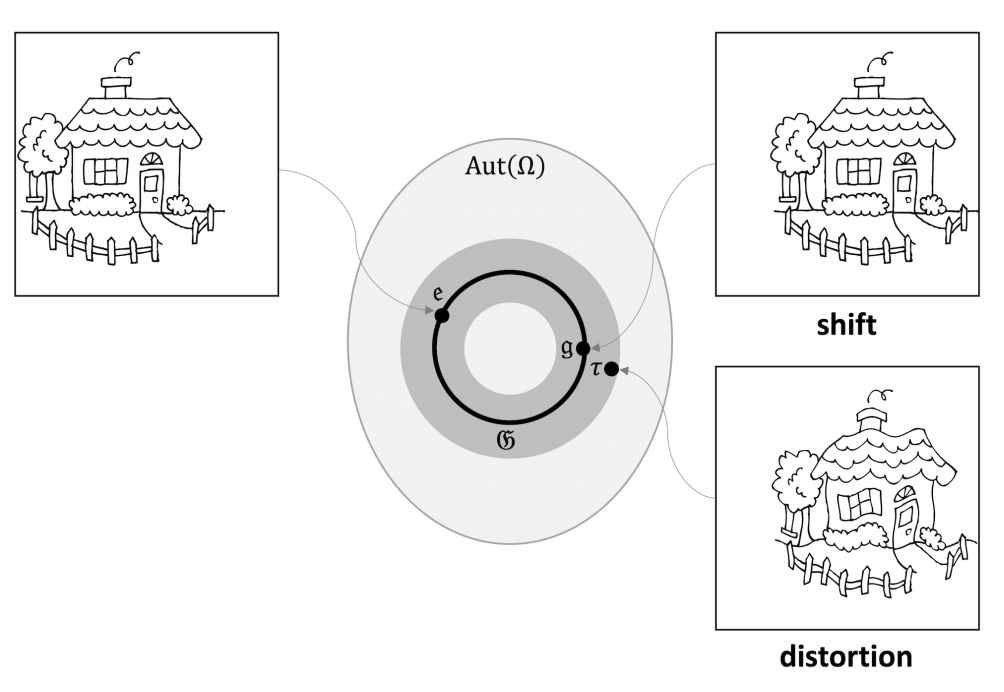}
      \caption{Deformation stability}
      \label{fig:ch3:gdl-principles:distortion}
    \end{subfigure}
    \begin{subfigure}{.30\textwidth}
      \centering
      \includegraphics[width=\textwidth]{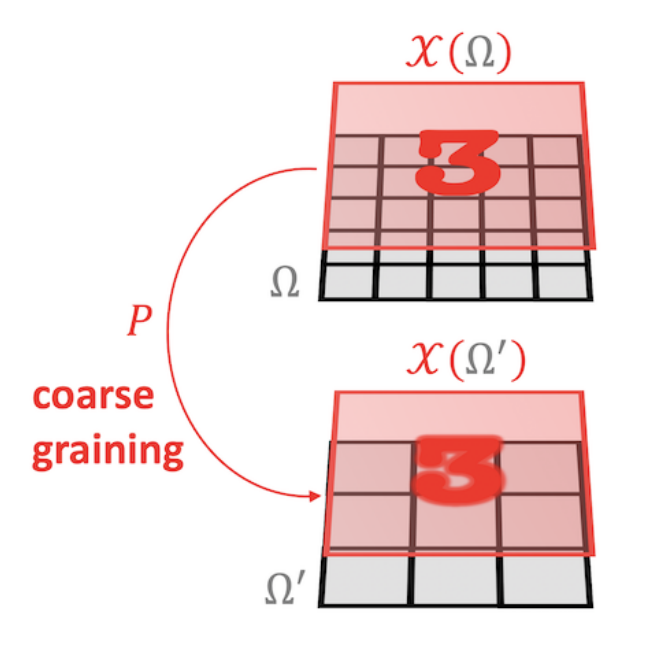}
      \caption{Scale separation}
      \label{fig:ch3:gdl-principles:scale}
    \end{subfigure}

    \centering
    \caption[Principles of Geometric Deep Learning]{Principles of Geometric Deep Learning (GDL). The GDL combines invariance and equivariance to symmetries, stability to deformations of both signals and domain and separation of scale for local-to-global hierarchical learning, to propose a powerful generic blueprint of designing neural networks for learning robust and stable representation on various data domains.}
    \label{fig:ch3:gdl-principles}
    \footnotesize{Source: \cite{ch3:gdl}}
\end{figure}

\noindent
The principles behind representation learning GDL can be explained with three concepts: \textit{symmetry}, \textit{deformation stability} and \textit{scale separation} (see Fig.~\ref{fig:ch3:gdl-principles}). 
Each of these principles addresses different aspects of exploiting geometrical priors for representation learning, and together they built a blueprint of GDL, which is universal with respect to data domains. \\

\subsection{Symmetry}
\label{sec:3.2.1}

\noindent
The \textit{symmetry} of an object is a transformation that leaves its certain property or whole system unchanged~\cite{ch3:gdl} (the Fig.~\ref{fig:ch3:gdl-principles:symmetry} showcases rotational symmetry, the triangle is still a triangle no matter the rotation).
This concept is commonly exploited when dealing with the curse of dimensionality, to radically reduce the dimensionality of the data domain, since it imposes a powerful inductive bias.
For example, when we tackle the classification task on images, the object category is unchanged (or \textit{invariant}) by the translations.
Thus, equipping the classification network with tools that enforce invariance under translation would allow to greatly simplify the problem.
Practically, however, designing such mechanisms is not an easy task and often only some amount of invariance can be achieved.
Coming back to the image classification example, a weaker form of invariance, namely \textit{equivariance} is achieved by utilizing the convolutional layers.
The image representation obtained after the convolutional layer is not the same when the object is shifted, however, the representation of the objects present in the image is the same, since the weights of the convolutional kernel are shared among the whole image.
This is one of the reasons why the linear layers are not well-suited towards image analysis - they do not induce equivariance or any other form of weaker invariance, since the weights are not shared among the whole image. \\

\noindent
Formally, the symmetries are defined by introducing \textit{symmetry groups}, which are \textit{algebraic groups}.
The \textit{algebraic group} is a set $\mathfrak{B}$ with a binary operation $\circ : \mathfrak{B} \times \mathfrak{B} \mapsto \mathfrak{B}$ (denoted also $\mathfrak{g} \circ \mathfrak{h} = \mathfrak{g}\mathfrak{h}$) that satisfies four following axioms~\cite{ch3:gdl}:

\begin{enumerate}
    \item \textbf{Associativity}: $(\forall \mathfrak{g},\mathfrak{h},\mathfrak{l} \in \mathfrak{B})$ $(\mathfrak{g}\mathfrak{h})\mathfrak{l} = \mathfrak{g}(\mathfrak{h}\mathfrak{l})$
    
    \item \textbf{Identity}: $(\exists! \mathfrak{e} \in \mathfrak{B})(\forall \mathfrak{g} \in \mathfrak{B})$ $\mathfrak{e}\mathfrak{g} = \mathfrak{g}\mathfrak{e} = \mathfrak{e}$
    
    \item \textbf{Inverse}: $(\forall \mathfrak{g} \in \mathfrak{B})(\exists! \mathfrak{g}^{-1} \in \mathfrak{B})$ $\mathfrak{g}\mathfrak{g}^{-1} = \mathfrak{g}^{-1}\mathfrak{g} = \mathfrak{e}$
    
    \item \textbf{Closure}: $(\forall \mathfrak{g},\mathfrak{h} \in \mathfrak{B})$ $\mathfrak{g}\mathfrak{h} \in \mathfrak{B}$
\end{enumerate}

\noindent
Having a notion of a symmetry group, the concept of a \textit{group action} on the underlying data domain $\Omega$ and its space of signals $\mathcal{X}(\Omega)$ can be introduced. \\

\noindent
A \textit{group action} of $\mathfrak{B}$ on a set $\Omega$ is defined as a mapping $(\mathfrak{g}, u) \mapsto \mathfrak{g}.u$, where $\mathfrak{g} \in \mathfrak{B}$ and $u,\mathfrak{g}.u  \in \Omega$.
Intuitively, we can think of group action as a process of applying some symmetry transformation $\mathfrak{g}$ to the point $u \in \Omega$ and obtaining another point on the data domain $\Omega$.
Since the data domain $\Omega$ is only an underlying data structure, we associate it with a space of signals $\mathcal{X}(\Omega)$, that defines the mapping between data points and exact values.
For example, when considering a 2D RGB image, the data domain is a 2D pixel grid and the space of signals defines the mapping between pixels and exact RGB values.
The group actions work automatically on the signal space $\mathcal{X}(\Omega)$ as well: $(\mathfrak{g}.x)(u) = x(\mathfrak{g}^{-1}u)$~\cite{ch3:gdl}.
Being equipped with symmetry groups and group actions, we can introduce invariant and equivariant functions which are fundamental building blocks of the GDL blueprint. \\

\noindent
\textit{A function $f: \mathcal{X}(\Omega) \mapsto \mathcal{Y}$ is \textbf{$\mathfrak{B}$-invariant} if $(\forall \mathfrak{g} \in \mathfrak{B}, x \in \mathcal{X}(\Omega))$ $f(\rho(\mathfrak{g})x) = f(x)$, i.e., its output is unaffected by the group action on the input. 
}~\cite{ch3:gdl} \\

\noindent
\textit{A function $f: \mathcal{X}(\Omega) \mapsto \mathcal{X}(\Omega)$ is \textbf{$\mathfrak{B}$-equivariant} if $(\forall \mathfrak{g} \in \mathfrak{B})$ $f(\rho(\mathfrak{g})x) = \rho(\mathfrak{g})f(x)$, i.e., group action on the input affects the output in the same way. 
}~\cite{ch3:gdl} \\

\noindent
where $\rho: \mathfrak{B} \mapsto \mathbb{R}^{n \times n}$ is a \textit{group representation} - a mapping that assigns to each group element $\mathfrak{g}$ a matrix of a size of the feature space on which the group acts. \\

\noindent
Coming back to the example with image classification, we can now view the convolutional operator as a \textit{shift-equivariant} function - the shifting of an image before the convolutional operation gives the same result as shifting it by the same amount after it.
Moreover, when constructing a classification network, we would often utilize some form of global pooling operator to form a $1$-dimensional representation vector.
This global pooling operation, interpreted through the lens of symmetry groups, is a \textit{shift-invariant} function - the pooling is done over the whole image thus it is unchanged by the shifting.
Viewing the construction of commonly utilized layers in deep learning in a more abstract and generic way, as proposed in the GDL blueprint, allows us to see parallels between architectures utilized for various, at first glance completely different, data domains such as graphs, images or 3D shapes.

\subsection{Deformation stability}
\label{sec:3.2.2}

\noindent
The introduced notion of symmetry is a powerful mechanism when we know exactly which transformations are to be considered invariant or equivariant in a given setting.
However, the data examples coming from the real world are often noisy and do not adhere to the strict symmetry definition, thus some weaker form of symmetry needs to be defined.
This weaker notion is to be referred to as \textit{deformation stability} and can be distinguished between two scenarios: stability to signal deformations and to domain deformations~\cite{ch3:gdl}. \\

\noindent
\textbf{Stability to signal deformations:} In this scenario we come from the a priori knowledge that the small deformations of the signal $x$ should not influence the result of the given function $f(x)$.
This idea looks similar to the symmetry problem, however, the deformations, which are defined as small diffeomorphisms $\tau \in \text{Diff
}(\Omega)$~\cite{ch3:gdl}, do not form the group due to not adhering to the closure axiom (see Section 
~\ref{sec:3.2.1}).
The composition of multiple small deformations creates a large deformation, thus the closure can be achieved only for the set of all possible deformations which is in contradiction with our main goal. \\

\noindent
To tackle this problem, the proposition is to introduce some form of a measure $c(\tau)$ which quantifies how far the given $\tau \in \text{Diff}(\Omega)$ is from the given symmetry group $\mathfrak{B} \subset \text{Diff}(\Omega)$.
Obviously $(\forall \tau \in \mathfrak{B})$ $c(\tau) = 0$, so the definition of equivariant and invariant functions can be naturally replaced with a weaker notion of \textit{deformation stability}, given by the following equation~\cite{ch3:gdl}:

\begin{equation}
    ||f(\rho(\tau)x) - f(x)|| \leq Cc(\tau)||x||, \forall x \in \mathcal{X}(\Omega)
\end{equation}

where $C$ is some constant independent of the signal $x$ that quantifies how much deformation we consider to be stable. 
A function $f$ that satisfies the above equation is to be referred to as \textit{geometrically stable}. \\

\noindent
This idea is showcased in Fig.~\ref{fig:ch3:gdl-principles:distortion} with the example of a translation group.
The black ring is a translation symmetry group $\mathfrak{B}$ and elements $\mathfrak{e}$, $\mathfrak{g}$ belong to this group.
The grey ring showcases the set of geometrically stable transformations which are no longer elements of the symmetry group $\mathfrak{B}$, however, are to be considered valid since they are quantified by the given measure $c(\tau)$ to be close enough to the elements of the symmetry group.
This way, the stability to small deformations of the signal can be achieved and incorporated into the neural network design. \\ 

\noindent
\textbf{Stability to domain deformations:} In the second scenario, we are dealing with the deformations of the geometric domain $\Omega$ rather than the signal.
For 2D or 3D images we often talk only about the signal deformations, since the underlying grid is not deformed, however, when dealing with 3D objects such as meshes or point clouds, which we can categorize as forms of spatial graphs, the domain itself is deformed - the coordinates and connectivities between the points can change. \\

\noindent
Formally, for the space of all possible domains $\mathcal{D}$, an appropriate distance metric $d(\Omega, \tilde{\Omega})$, where $\Omega,\tilde{\Omega} \in \mathcal{D}$ and $d(\Omega,\tilde{\Omega}) = 0$ if $\Omega$ and $\tilde{\Omega}$ are equivalent, can be defined.
Again, as for the signal deformation, the measure $d(\Omega,\tilde{\Omega})$ quantifies how large the deformation between the source and obtained domain is, and the notion of stability can be represented with the similarly constructed equation~\cite{ch3:gdl}:

\begin{equation}
    ||f(x,\Omega) - f(\tilde{x},\tilde{\Omega})|| \leq Cd_{\mathcal{D}}(\Omega,\tilde{\Omega})||x||, \forall \Omega,\tilde{\Omega} \in \mathcal{D}, x \in \mathcal{X}(\Omega)
\end{equation}

where $f: \mathcal{X}(\mathcal{D}) \mapsto \mathcal{Y}$ and $\mathcal{X}(\mathcal{D}) = \{(\mathcal{X}(\Omega), \Omega) : \Omega \in \mathcal{D}\}$ is the ensemble of possible input signals defined over a varying domain. 
Again, a function $f$ that satisfies the above equation is considered to be stable to domain deformations. \\

\noindent
The introduction of the notion of deformation stability allows expanding the concept of symmetry to be more flexible when it comes to the data we observe in a real setting.

\subsection{Scale separation}
\label{sec:3.2.3}

\noindent
The symmetry allows us to reduce the data domain by grouping the various data representations that are to be considered invariant or equivariant, and deformation stability strengthens these priors by incorporating invariance to small signal and domain noise.
However, although these concepts address the curse of dimensionality, the problem is not yet overcome, since with the domain growth the number of possible functions that satisfy the aforementioned limitations is still too large.
The key point lies in the incorporation of a multiscale analysis structure that allows building the representation from local to global features. \\

\noindent
This concept is called \textit{scale separation} and it is based on the notion of multiscale coarsening of the data domain $\Omega$ into a hierarchy $\Omega_1,...,\Omega_J$.
The coarsening operation assimilates nearby points $u, u^{'} \in \Omega$ together, based on the given metric over the data domain. 
To process points assimilated in this manner, the so-called \textit{locally-stable} function is utilized.
The global target function $f$, which forms the data representation, often depends on long-range interactions between the features over the whole domain.
By utilizing \textit{locally-stable} functions, the interactions between the features can be separated along the scales, and the final global stable representation can be obtained via local-to-global feature analysis and propagation along the coarse scales~\cite{ch3:gdl}. \\

\noindent
This idea is often incorporated in the design of neural networks in the form of local pooling operators.
Fig.~\ref{fig:ch3:gdl-principles:scale} showcases the classic local pooling operator utilized in convolutional neural networks.
Four adjacent pixels are assimilated together to form one output pixel with the combined signal.
For the images, the representation learning is done hierarchically by utilizing multiple blocks of local feature extraction layers in the form of convolutions and local pooling operators.
For other data domains, the same hierarchical notion can be utilized with exact layers being tailored towards the given data domain.

\subsection{The Geometric Deep Learning blueprint}
\label{sec:3.2.4}
\begin{figure}[ht]
    \centering
    \includegraphics[width=\textwidth]{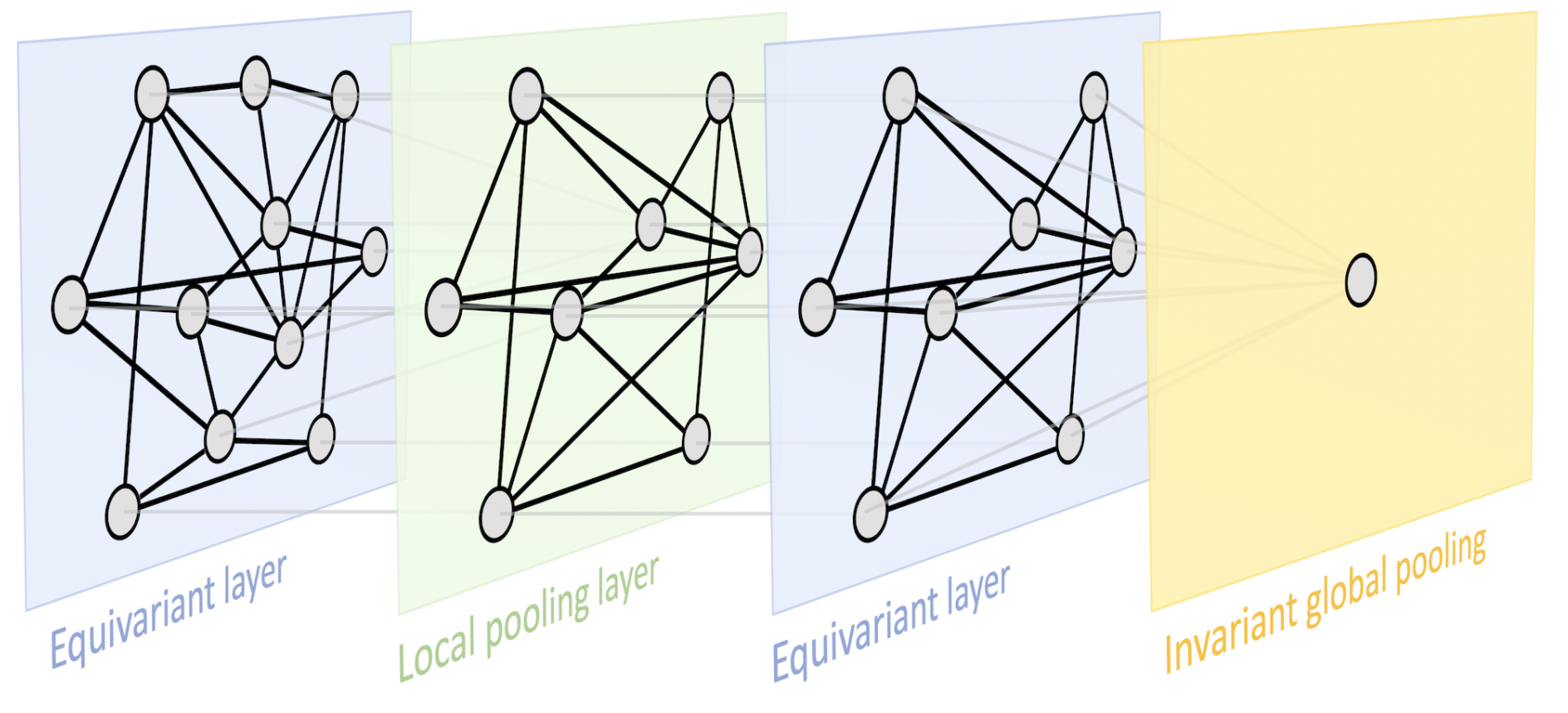}
    \caption[The Geometric Deep Learning blueprint]{Example of the Geometric Deep Learning blueprint applied to learning graph representation. The representation is learned hierarchically with the blocks built out of equivariant and local pooling layers. The final representation is obtained via invariant global pooling.}
    \label{fig:ch3:gdl-blueprint}
    \footnotesize{Source: \cite{ch3:gdl}}
\end{figure}

\noindent
Equipped with the principles of symmetry, geometric stability and scale separation, we can move on to introducing a universal blueprint, proposed by Bronstein et al.~\cite{ch3:gdl}, for learning stable representations of high-dimensional data.
The blueprint does not impose any form of specific deep learning architecture, but rather a set of concepts that must be met.
The principles provide three key building blocks that allow incorporating symmetry invariance, stability to deformations and robust function approximation in the construction of neural networks:

\begin{enumerate}
    \item \textit{Linear $\mathfrak{B}$-equivariant layer:} $B: \mathcal{X}(\Omega, \mathcal{C}) \mapsto \mathcal{X}(\Omega^{'}, \mathcal{C}^{'})$ satisfying $B(\mathfrak{g}.x) = \mathfrak{g}.B(x)$ for all $\mathfrak{g} \in \mathfrak{B}$ and $x \in \mathcal{X}(\Omega, \mathcal{C})$.
    \item \textit{Local pooling (coarsening):} $P: \mathcal{X}(\Omega,\mathcal{C}) \mapsto \mathcal{X}(\Omega^{'},\mathcal{C})$, such that $\Omega^{'} \subseteq \Omega$.
    \item \textit{$\mathfrak{B}$-invariant layer:} $A: \mathcal{X}(\Omega,\mathcal{C}) \mapsto \mathcal{Y}$ satisfying $A(\mathfrak{g}.x) = A(x)$ for all $\mathfrak{g} \in \mathfrak{B}$ and $x \in \mathcal{X}(\Omega,\mathcal{C})$.
\end{enumerate}

where $\Omega$ and $\Omega^{'}$ are domains and $\mathfrak{B}$ a symmetry group over $\Omega$. \\

\noindent
With the usage of the proposed blocks, one may construct a $\mathfrak{B}$-invariant functions that can learn data representation.
The choice of the symmetry group is dependent purely on the data domain we act on and its properties which we want to enforce.
For example, when dealing with images and utilizing Convolutional Neural Networks (CNNs), the data domain $\Omega$ is a regular grid and the symmetry group $\mathfrak{B}$ is translation.
The definition of classic convolution can be further expanded to the concept of \textit{group convolution}~\cite{ch3:group-convs}, which simply put, is a convolution that is invariant or equivariant to the chosen symmetry group.
For example, when we are dealing with images for which we require the invariance to rotations we can utilize Roto-Translation CNNs~\cite{ch3:roto-convs}, which take a group $SE(2)$ (special Euclidean group~\cite{ch3:euclidean-group}) as a symmetry group. 
On the other hand, Fig.~\ref{fig:ch3:gdl-blueprint} showcases the building blocks of a Graph Neural Network (GNN).
In the case of GNNs, we require equivariance to the permutations of the nodes, and thus the symmetry group we deal with is a permutation group $\Sigma_n$~\cite{ch3:permutation-group}. 
The local pooling layer can be realised in many forms, for example with edge pooling~\cite{ch3:edge-pooling}, which downsamples the graph by removing the nodes with edge contraction operation. \\

\noindent
Equipped with the GDL blueprint, we can now move on to introducing the exact architectures utilized for learning on point clouds, which are the main focus of this work, and discussing how they adhere to and realize the proposed principles.

\section{Learning on 3D point clouds}
\label{sec:3.3}

\noindent
We have already described the point cloud data structure (see Section~\ref{sec:3.1.3}) and introduced the GDL blueprint (see Section~\ref{sec:3.2}), hence we can move on to introducing how exactly the learning on 3D point clouds is realised in practice.
To do it, we must first acknowledge the main properties of a point cloud which are of great importance when talking about representation learning~\cite{ch3:pointnet}: \\

\noindent
\textbf{Unordered (symmetry~\ref{sec:3.2.1}):} 
A point cloud is a set of vectors describing the features of individual points. 
Due to this fact, the ordering of the points has no impact on a point cloud itself, hence the representation should be invariant to it as well - the role of a symmetry group is fulfilled by the permutation group $\Sigma_n$.
Thus when constructing the neural networks for consuming point clouds, we should seek to compose them out of permutation-invariant layers~(\ref{sec:3.2.4}). \\  

\noindent
\textbf{Interaction among points (scale separation~\ref{sec:3.2.3}):} 
The points are not isolated, since they come from the metric space (most commonly a Euclidean one), hence the relationships between them should be acknowledged. 
The structure formed by the set of points is meaningful for the analysis of the shape, and both local and global interactions are to be considered when constructing the representation.
We can clearly see the notion of scale separation in this reasoning and thus we would require the local-to-global feature analysis to be present in the neural network design. \\

\noindent
\textbf{Invariance under transformation (symmetry~\ref{sec:3.2.1} \& deformation stability~\ref{sec:3.2.2}):} 
Point clouds are considered to be geometric objects placed in some sort of metric space, as we mentioned previously.
In such spaces, the objects can undergo various types of transformations or deformations, and for some of them, we would like the learned representation to be invariant.
For example, in the tasks of classification or segmentation, the learned representations should not be influenced by the affine transformations of the input.
Moreover, we could also consider some sort of small noise applied to the points, namely domain deformation, to be non-influential in representation construction.
\\

\noindent
With key properties of point clouds, but also challenges in the light of network construction, highlighted, we can now move on to introducing the pioneering work in a raw point cloud analysis in the form of PointNet~\cite{ch3:pointnet} and its successor PointNet++~\cite{ch3:pointnet2}.
Later on, we would also mention some domain-specific adaptations of these general architectures, which utilize prior knowledge about the point cloud structure to build a better more robust representation~\cite{ch3:evg}.

\subsection{PointNet}

\begin{figure}[ht]
    \centering
    \includegraphics[width=\textwidth]{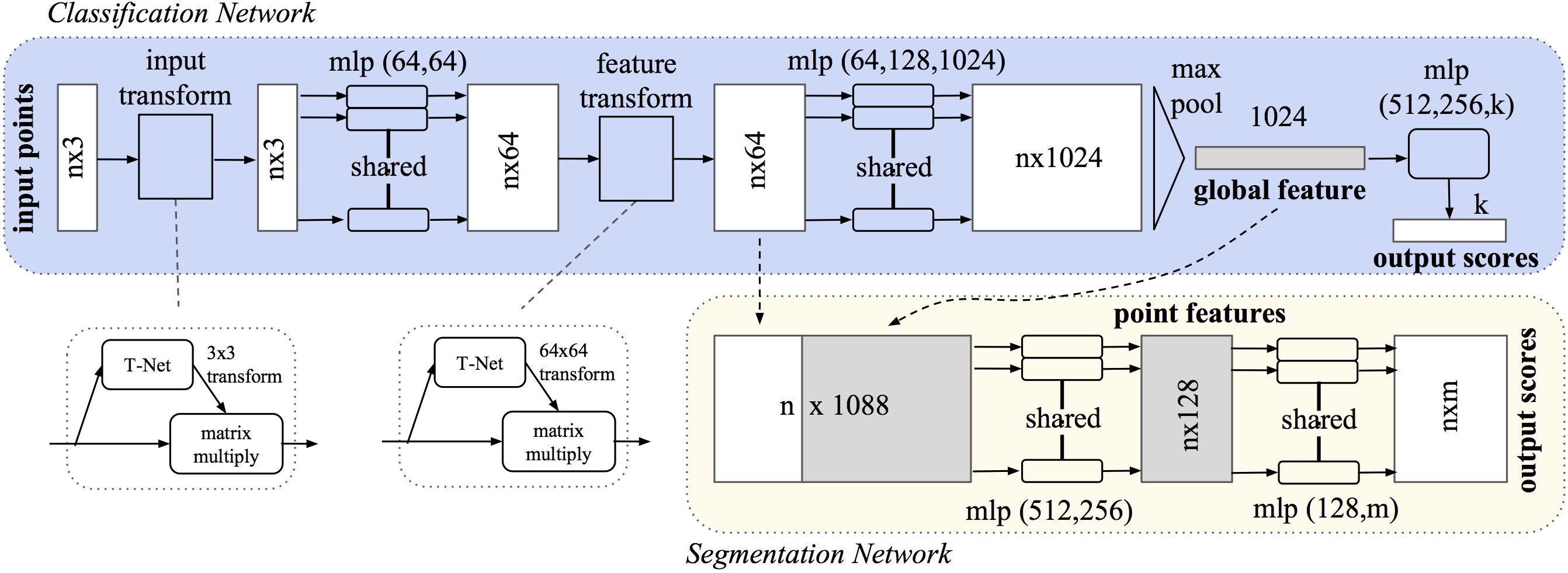}
    \caption[PointNet architecture]{PointNet architecture.}
    \label{fig:ch3:pointnet}
    \footnotesize{Source: \cite{ch3:pointnet}}
\end{figure}

\noindent
The PointNet architecture takes as an input a raw point cloud in the form of a set of $n$ points of $k$ features each (at least $3$ since we need to encode 3D spatial coordinates).
The exact architecture is showcased in Fig.~\ref{fig:ch3:pointnet} in the two configurations: for the classification and segmentation task.
The classification network is built out of the PointNet backbone and classic linear classification head.
Meanwhile, the segmentation configuration utilizes an encoder-decoder scheme, with the same PointNet backbone serving the encoder role, and a simple fully-connected network as a decoder.
The PointNet backbone is composed of three main components. \\

\noindent
\textbf{Shared MLP:} 
The input point cloud takes a shape $n \times k$, as described previously.
The classic MLP approach would be to flatten the $2$-dimensional input into the $1$-dimensional vector of length $nk$.
That, however, is neither a robust nor efficient approach.
First of all, the encoder is fixed to the arbitrally pre-specified number of points and point clouds consisting of fewer or more points are unconsumable without some form of downsampling or upsampling preprocessing.
Secondly, the notion of permutation-invariance is not met, since each vector element is processed with a unique independent neuron, thus the representation is dependent on the point order.
Moreover, the number of parameters in the MLP with $nk$ input neurons grows very fast with the size of a point cloud.
To address these issues, authors propose a shared MLP layer, which can be simply defined as a \textit{single-point encoder}.
The shared MLP takes $k$ input features (features of the single points), and processes them with linear layers, to obtain a single-point representation.
Each point is encoded independently with the same shared MLP, hence the name \textit{shared}, which refers to weight sharing between the point encoders.
Through the lens of the GDL, the shared MLP is a \textit{permutation-equivariant} layer, since the permutation of the points at the input is reflected in the obtained output representation of the points - the representations of points are ordered in the same way as input points. 
As we see in Fig.~\ref{fig:ch3:pointnet}, the shared MLP layer is the main building block of the PointNet with increasing size of representation with the network depth. \\

\noindent
\textbf{T-Net:} 
The second component proposed in the PointNet architecture is a T-Net (transformation net).
It serves the purpose of regressing a transformation matrix that is later on applied to the current point features.
The main idea behind introducing this component is to weakly enforce an invariance under transformations/deformations.
Since, as we mentioned previously (see Section~\ref{sec:3.3}), the input point cloud may come in various orientations or be slightly deformed, we would like to have a mechanism to remedy this problem.
The T-Net is supposed to learn a transformation matrix which mitigates certain transformations implicitly considered to be invariant by the network, and bring all point clouds into some form of a canonical orientation that simplifies and stabilizes the representation learning.
In practice, the T-Net is realised in the form of a small PointNet backbone with a transformation matrix regression head. \\

\noindent
\textbf{Global feature pooling:} 
The point features encoded with the blocks of shared MLPs and T-Nets are at the end run through the global feature pooling layer to obtain a final $1$-dimensional embedding of a given point cloud.
The point cloud before the pooling procedure is of size $n \times z$, where $z$ is the size of a per-point feature vector produced by the preceding shared MLP layer.
The global pooling output vector is then of size $z$ since the pooling is done along the point dimension, and as the pooling function, the \textit{max-pooling} has been proposed by the authors.
By performing the pooling with the \textit{symmetric function} (one whose output is invariant under all permutations of the arguments~\cite{ch3:symmetric-function}) in the form of max-pooling, the obtained point cloud representation is invariant to the permutation of the points at the input.
When considering the GDL blueprint, the global feature pooling is a permutation-invariant layer. \\

\noindent
As we can see, the PointNet architecture follows the principles of representation learning proposed by the GDL blueprint and addresses the notion of points being unordered in the point cloud, by utilizing permutation invariant and equivariant layers.
It also attempts to solve the problem of invariance under transformation, by introducing implicit transformation matrix learning with T-Net.
However, one major aspect of both the GDL and properties of the point cloud is not addressed in any way in the PointNet, namely scale separation / interaction among the points.
The architecture does not propose any form of local-to-global feature analysis and only settles on learning single-point encodings and then performing a global aggregation.
This is a major drawback, which limits the ability of a network to exploit local interactions and structure topology to enhance representation learning.
This aspect of PointNet is to be addressed in the following work, namely PointNet++, which we will introduce in the following section.

\subsection{PointNet++}
\label{sec:3.3.2}

\begin{figure}[ht]
    \centering
    \includegraphics[width=\textwidth]{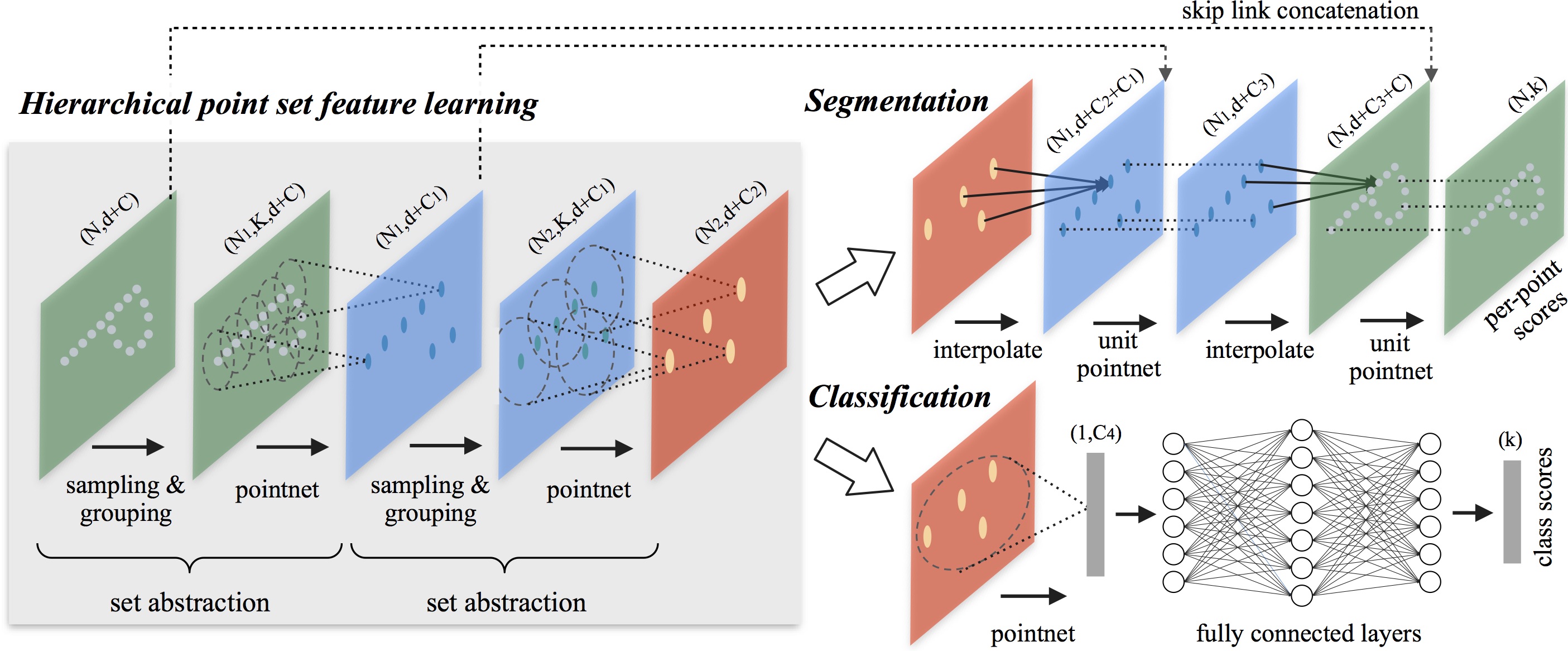}
    \caption[PointNet++ architecture]{PointNet++ architecture.}
    \label{fig:ch3:pointnet2}
    \footnotesize{Source: \cite{ch3:pointnet2}}
\end{figure}

\noindent
The PointNet++ takes the raw point cloud of size $n \times k$ on the input in the same way as PointNet.
The architecture adopts the common encoder-decoder scheme (see Fig.~\ref{fig:ch3:pointnet2}). 
The encoder branch is built out of \textit{set-abstraction} (SA) blocks and for the classification task the decoder is formed out of one PointNet layer which yields a global representation vector, and the fully-connected classification head.
In the segmentation task, however, the decoder is more complex, and it is built out of \textit{feature propagation} (FP) blocks and skip connections between same-level encoder and decoder blocks.
Both SA and FP blocks, realize a missing component of PointNet in the form of exploiting local point interactions. \\

\noindent
\textbf{SA-block:} The block is built out of the three steps: \textit{sampling}, \textit{grouping} and \textit{encoding} (with PointNet).

\begin{enumerate}
    \item \textit{Sampling:} 
    In this step, the subset of an input point cloud $\mathcal{P}$ is sampled to form the representative point cloud $\mathcal{R} \subset \mathcal{P}$.
    Such a subset could be sampled randomly, however, the more densely packed areas would be over-represented this way.
    To avoid this issue, and better catch the global point cloud structure, the authors proposed to utilize a \textit{Farthest Point Sampling} (FPS) algorithm.
    The FPS starts by initializing an empty set of representative points and adds a first point at random.
    Next, the point farthest (in the sense of chosen distance metric) from the whole set is found and added to the set.
    This process is done iteratively until the representative set is of the desired size.

    \item \textit{Grouping:}
    Upon having a representative point cloud $\mathcal{R}$, for each $r \in \mathcal{R}$ the \textit{grouping} step is performed independently to find the set of neighbouring points $\mathcal{N}_r \subset \mathcal{P}$.
    The aim of the grouping procedure is to extract local neighbourhoods of all $r \in \mathcal{R}$, thus it is based on the distance metric in the form of Euclidean distance.
    For each $r \in \mathcal{R}$ a distance to each $p \in \mathcal{P}$ is computed and the points are sorted based on it.
    The neighbourhoods then can be chosen in various ways, but the most common techniques, which were also considered by the authors, are \textit{ball query} and \textit{k-nearest-neighbours (KNN)}.
    The ball query defines a grouping threshold $t$ and considers all points within the shorter distance to be neighbours.
    The $k$-nearest-neighbours, on the other hand, takes simply $k$ points with the shortest distance to the root point.
    The authors decided to utilize the ball query method, and argue that it is preferable due to being isotropic in regards to metric space, and thus the neighbourhoods are of the same size in the sense of distance metric.
    
    \item \textit{Encoding:}
    The neighbourhoods $\mathcal{N}_r$ obtained in the aforementioned matter are later processed with PointNet to obtain neighbourhood embeddings, which become features of representative points.
\end{enumerate}

\noindent
The grouping procedure described above is referred to by the authors as \textit{single-scale grouping (SSG)} and they propose also another scheme in the form of \textit{multi-scale grouping (MSG)}.
In the case of MSG, not one, but multiple thresholds $t_1, t_2, ..., t_k$ are considered, and for each independently, an SSG and encoding with PointNet are performed.
The final features of representative points are a concatenation of encodings of multiple scales.
This scheme is more robust since it allows the network to view a local neighbourhood in different scales and thus have a better view of the point cloud structure. \\

\noindent
The final encoder is built out of multiple SA blocks, which perform both point cloud downsampling and hierarchical feature aggregation from local to global due to the increasing grouping threshold $t$ of the ball query method.
This very process, adopted by PointNet++, is the embodiment of the notion of scale separation proposed by the GDL. \\

\noindent
\textbf{FP-block:} The block is built out of two steps: \textit{interpolation} and \textit{decoding} (with unit PointNet). \\

\begin{enumerate}
    \item \textit{Interpolation:}
    In this step, the features from the representative point cloud $\mathcal{R}$ coming from the previous block are interpolated onto the point cloud from before the sampling stage in the encoder block on the same level.
    The interpolation is done with the KNN and the parameter $k$ is set to $3$ - each point takes features from the $3$ closest points from $R$.
    
    \item \textit{Decoding:} 
    The point cloud with interpolated features is at last processed with the unit PointNet which is the classic PointNet without the global pooling layer at the end.
    Thus the size of the point cloud is not changed and only single-point encodings are performed in this step.
\end{enumerate}

\noindent
As for the encoder branch, the decoder one is also built out of multiple blocks - in this case, the FP blocks.
The number of blocks in each branch is the same, and they are connected together with skip connections to pass the information from different hierarchies. \\

\noindent
As we can see, PointNet++ is a much more complex architecture than PointNet.
The introduction of local-to-global analysis in the form of hierarchical neighborhood aggregation fulfills the role of taking into account interactions between the points in the representation learning and incorporates the notion of the scale separation proposed by the GDL.
Although PointNet++ is a fine generic architecture for learning on point clouds, in some cases we may have prior knowledge about its structure which could be exploited in the network design.
In the next section, we will delve into a few methods that propose such schemes.

\subsection{Domain-specific adaptations}
\label{sec:3.3.3}

\noindent
In some cases, we possess some prior knowledge about some characteristics of the topology of the point clouds that we are working on.
By adopting the classic PointNet++ architecture we do not really utilize this information due to its generic approach.
Since, in this work, we deal with vasculature trees in the form of coronary arteries, we would take a look at two domain-specific adaptations of the PointNet++ tailored toward the task of vessel labeling.
Vessel labeling is the task of assigning a proper vessel segment label to each point in the vascular tree, so we can think of it as a part segmentation task.
The proposed adaptations modify only the grouping procedure in PointNet++ to aggregate the features in a more robust way in light of the available prior knowledge about the data and the task.

\begin{figure}[ht]
    \centering
    \includegraphics[width=\textwidth]{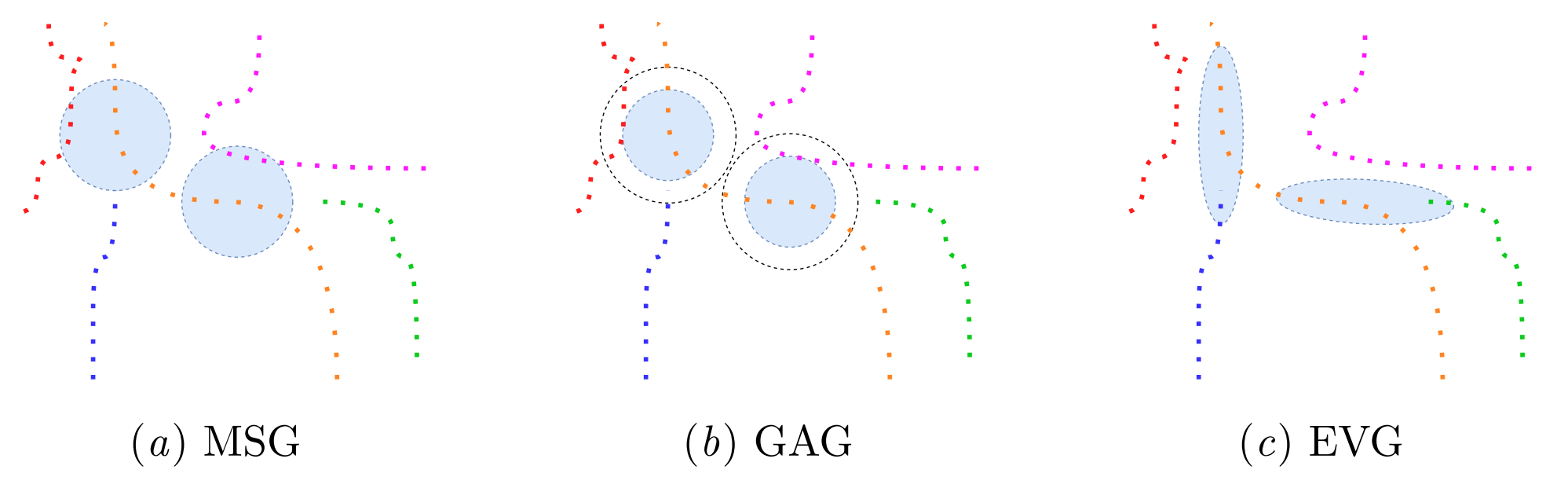}
    \caption[Point grouping strategies]{Point grouping strategies. MSG strategy is the most generic one and does not include any information about the underlying topology of the point cloud. In the GAG scheme, the connectivity between the components is taken into account. The EVG strategy utilizes the underlying topology of the point cloud by performing directed grouping and exploiting the prior knowledge about the vascular tree structure.}
    \label{fig:ch3:groupings}
    \footnotesize{Source: \cite{ch3:evg}}
\end{figure}

\noindent
\textit{Geometry-aware grouping (GAG):}
This grouping scheme has been proposed by He et al.~\cite{ch3:gag} to put more priority on grouping together the points that belong to the same connected component.
The grouping is performed with the ball query as in the MSG scheme, but the distance to the points that belong to the same connected component as the root point is scaled by the parameter $\lambda$ set to $0.25$ by the authors.
Intuitively, this grouping procedure can be interpreted as grouping with two balls queries of different sizes, whereby the smaller one groups all the points and the bigger one only the points from the same component - this idea is portrayed in Fig.~\ref{fig:ch3:groupings}.
This approach exploits prior knowledge in the form of a hypothesis that the connected components are also different segments. \\

\noindent
\textit{Eigenvector grouping (EVG):} 
This grouping scheme, proposed by Rygiel et al.~\cite{ch3:evg}, attempts to exploit the tubular structure of the vessels by performing the grouping along the vector estimating the vessel direction.
The directional vector of the vessel is computed by extracting the top eigenvector of the local root point neighbourhood, hence the method name.
The top eigenvector defines the direction of the most variance in the local point cloud, thus when paired with the tubular nature of the point clouds, allows to perform point grouping along the vessel.
Fig.~\ref{fig:ch3:groupings} showcases how the EVG grouping areas stretch along the vessels' directions, allowing for discrimination between disconnected vessel parts belonging to the same segment, which may happen due to low quality of prior segmentation, and other segments. \\

\noindent
The discussed domain-specific adaptations of PointNet++ showcase how the prior domain knowledge can be incorporated towards the design of the neural network.
Even though these approaches are not specifically designed for the segmentation task, they are still of interest and are worth exploring in the task of hemodynamics estimation which we undertake in this work.
We will discuss more in detail the applicability of these methods to the task at hand in Chapter~\ref{chapter4}.

\section{Summary}

\noindent
In this chapter, we introduced various 3D data representations and discussed the principles of learning robust and stable representation by presenting the Geometric Deep Learning (GDL) blueprint.
Through the lens of the GDL, we introduced how the learning on point clouds does look like, by presenting PointNet and PointNet++ architecture.
At last, the domain-specific adaptions of generic PointNet++ have been showcased for the learning on vascular trees, which are of interest in this work.
In the next chapter, we will move on to presenting the proposed approach to the estimation of FFR in coronary arteries, which is based upon the architectures and principles discussed in this chapter.

	\cleardoublepage

 	\chapter{Proposed approach to the estimation of FFR in coronary arteries}
\thispagestyle{chapterBeginStyle}
\label{chapter4}

\noindent
In this chapter, we will introduce the approach to the estimation of FFR in coronary arteries that is proposed in this work.
We will showcase the whole workflow, by describing the process of data preparation and DNN design.
At last, the training and inference setups will be discussed together with the details and specifics of the implementation.

\section{Proposed approach \& workflow}
\label{sec:4.1}
\begin{figure}[ht]
    \centering
    \includegraphics[width=\textwidth]{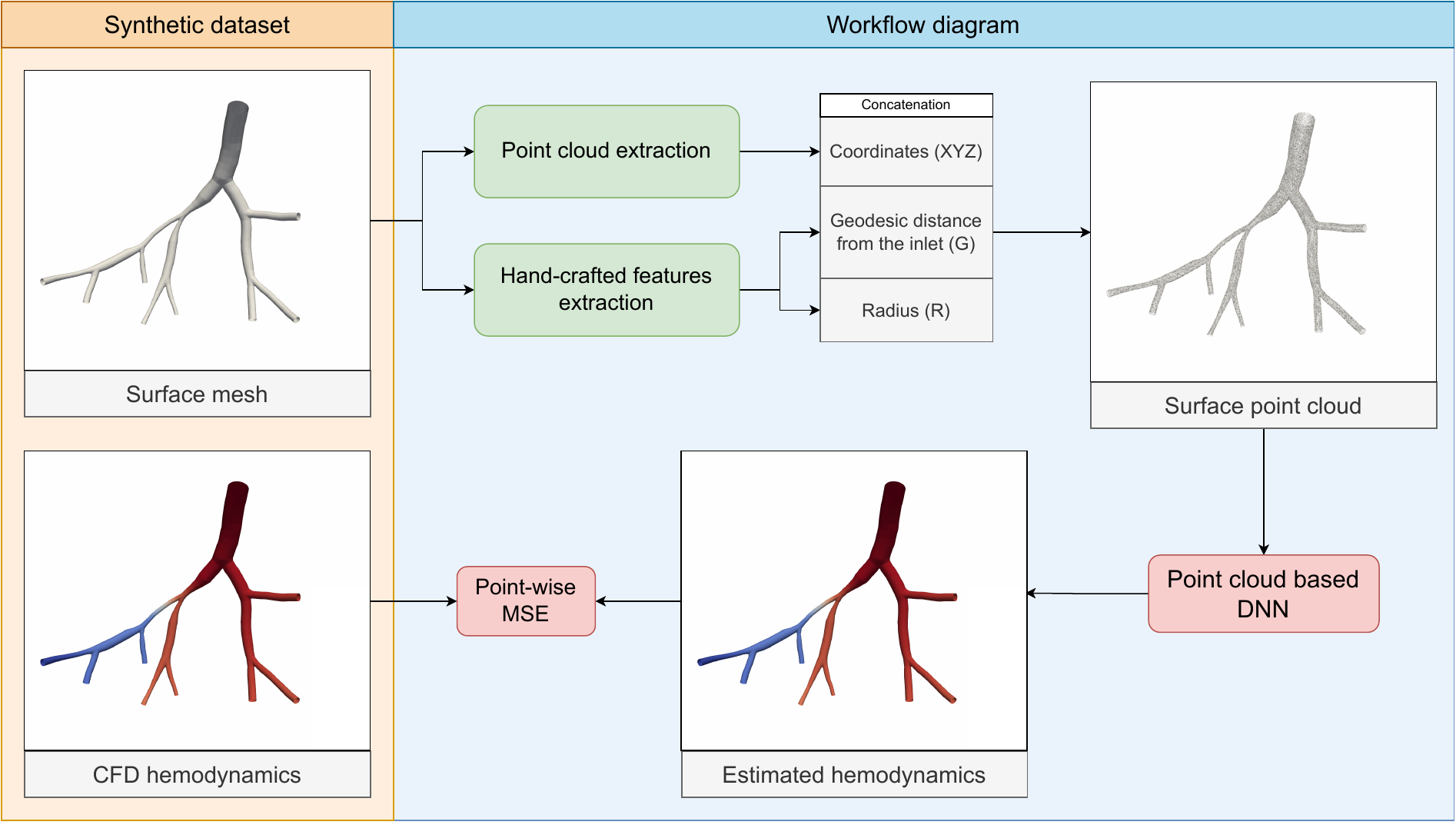}
    \caption[Workflow diagram of the proposed approach]{Workflow diagram of the proposed approach.}
    \label{fig:ch4:workflow-diagram}
\end{figure}

\noindent
As described in Chapter~\ref{chapter2}, there are two common approaches towards learning on geometries for the task of hemodynamics estimation: explicit and implicit feature learning.
In this work, we propose estimating FFR in coronary arteries, mainly focusing on implicit feature learning by representing the input vessel geometry as a 3D point cloud.
However, we also propose to compute some generic hand-crafted features to further enhance the features of the input points.
We argue, that the composition of both implicit and explicit feature learning is the most beneficial approach when dealing with the estimation of hemodynamic features.
The usage of implicit learning allows a DNN to encode the input geometry with respect to the hemodynamic feature at hand.
Moreover, consuming raw point clouds allows for including fine-grained resolution of the surface in the learning procedure.
Such details are almost impossible to be captured with any of the hand-crafted features.
By utilizing hand-crafted features, we equip the network with some additional information that is not trivially extracted from the geometry only.
The design of such features is based on prior knowledge regarding fluid dynamics and the geometry features that do correlate with it.
The chosen features will be described in detail in Section~\ref{sec:4.3.1}. \\

\noindent
Fig.~\ref{fig:ch4:workflow-diagram} showcases the workflow diagram of the proposed approach.
We follow the works by Itu et al.~\cite{ch2:siemens} and Suk et al.~\cite{ch2:gem-wss,ch2:gem-velocity} and construct the dataset of synthetic coronary arteries to both perform the training and evaluation, due to the lack of a representative and large enough database of real-patient geometries.
The geometries of coronary arteries are represented in the form of a surface mesh and for each, the vFFR GT is computed via CFD simulation.
The input mesh is decomposed into the point cloud and hand-crafted features which are later concatenated to form the input features for the points.
The point cloud constructed in this matter is then processed with a point cloud based DNN and as the learning criterion, a simple mean squared error (MSE) between the estimated hemodynamics and CFD-based ones is utilized.
In this work, the estimated hemodynamic is vFFR, however, we do not directly estimate it with the DNN.
Instead, we estimate the so-called \textit{pressure drops}, which inform how much the pressure of the fluid drops along the vessel.
Since the FFR is a ratio of pressures, it can be directly computed by prescribing the patient-specific pressure at the vessel inlet and computing the following pressures with the usage of pressure drops (this process will be further discussed in Section~\ref{sec:4.2.2}).
This is a high-level description of our approach.
In the next sections, we will describe each workflow step in detail.\\

\noindent
Most importantly our proposed approach allows us to drastically limit the time required for vFFR estimation in comparison to CFD simulation.
The estimation of vFFR with CFD engine (one utilized in this work, more information about will be provided in Section~\ref{sec:4.2.2}) takes up to \textbf{2h} for single geometry, meanwhile, our proposed approach requires only around \textbf{15s} for single geometry.

\section{Data preparation}

\noindent
As we briefly mentioned previously (see Section~\ref{sec:4.1}), there aren't any available large enough databases of real-patient geometries of coronary arteries.
This fact renders the training on only real-patient unfeasible and thus the incorporation of synthetic data is required.
For this work, we generated a dataset of $1700$ synthetic coronary artery geometries and labels in the form of pressure drop maps obtained via CFD simulations.

\subsection{Generation of synthetic coronary artery geometries}
\noindent
In this work, we utilize an undisclosed in-house synthetic coronary artery generator developed by Patryk Rygiel at Hemolens Diagnostics sp. z o.o.~\footnote{\url{https://hemolens.eu/}}.
We cannot share much detail about the method, due to the information being company know-how.
However, we can disclose that the method is derived from the work by Itu et al.~\cite{ch2:siemens}, and discuss the set of generic steps necessary to obtain biologically proper synthetic coronary artery models. \\

\begin{figure}[ht]
    \centering
    \includegraphics[width=\textwidth]{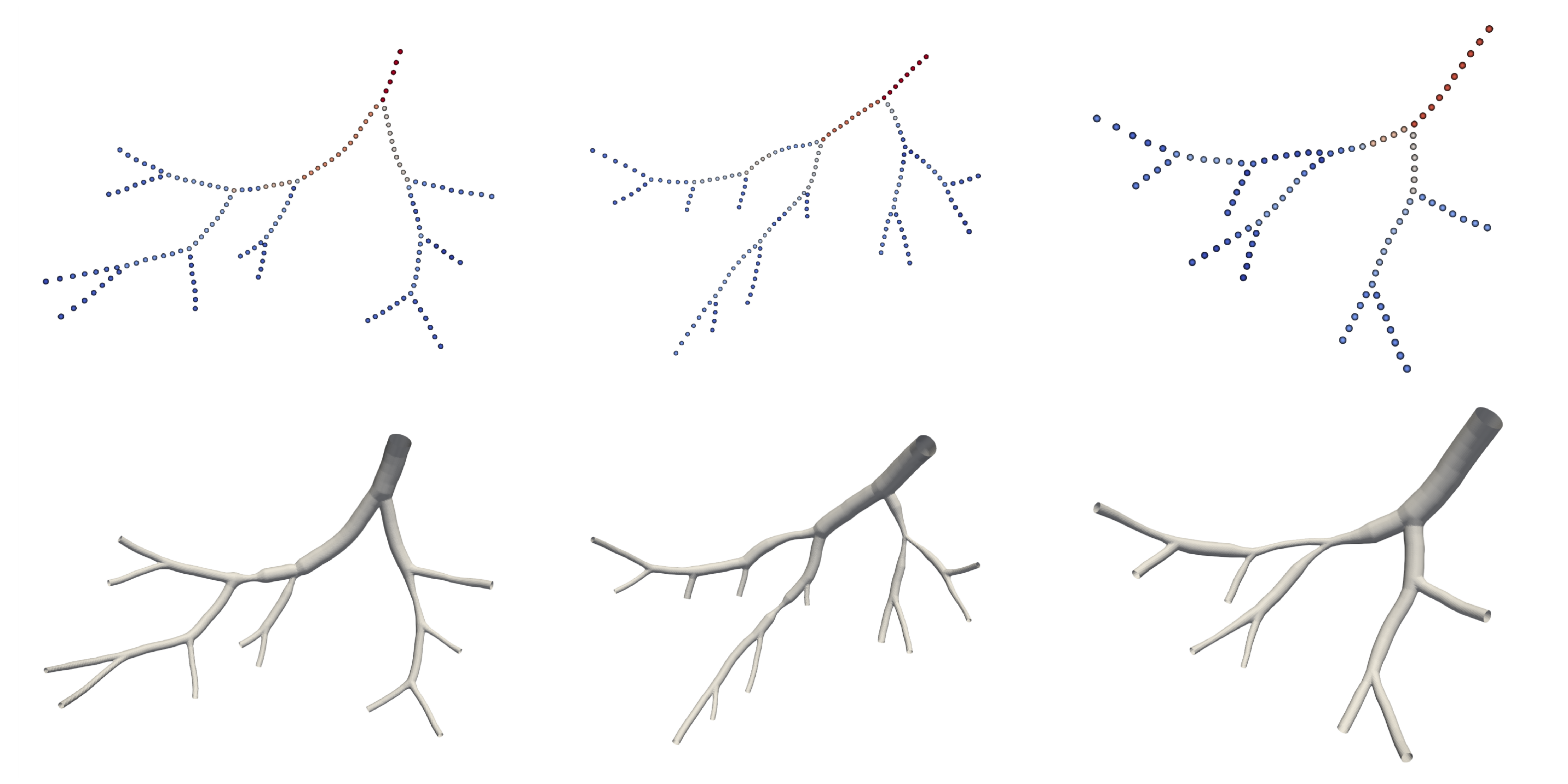}
    \caption[Samples from the synthetic artery generator]{Samples from the synthetic artery generator. The first row showcases generated centerline graph with colour describing the radii at the nodes (from red (large) to blue (small)). In the second row, the surface meshes created based on the centerlines are presented.}
    \label{fig:ch4:synthetic-samples}
\end{figure}

\noindent
The generation of synthetic vessels can be split into two general steps: \textit{vessel centerline} and \textit{vessel surface} generation. \\

\noindent
\textbf{Vessel centerline generation:}
The centerline is a weighted shortest path traced between two extremal points~\cite{ch4:vmtk-centerlines} - in the case of the vessels as the points we consider a vessel inlet and outlets.
The path needs to be set within the vessel, and at each point be at approximately equal distance from all the surface points in the perpendicular plane to the centerline at this point.
Intuitively, we think of a centerline as a line that describes the vessel skeleton which is represented as a graph.
In the most common cases, the centerline is extracted from the vessel geometry, as the process of vessel analysis.
However, during the generation process, we, first of all, generate a biologically proper centerline and later generate a vessel surface around it.
For the synthetic vessel to be topologically correct and biologically relevant, the centerline is generated based on the anatomical atlases and studies~\cite{ch2:siemens,ch2:coronary-anatomy} which provide intervals of various coronary artery geometrical features e.g. radii, tapering or bifurcation angles.
Since the atlases provide only the details about the healthy coronary trees, we must further enhance the synthetic data with the pathological cases.
Such cases are constructed by generating stenotic segments along the centerlines according to the Coronary Artery Disease - Reporting and Data System (CAD-RADS)~\cite{ch4:cad-rads} stenoses percentage intervals. 
The first row of Fig.~\ref{fig:ch4:synthetic-samples} showcases examples of generated centerlines, where the colours of nodes denote the radius of the vessel (from the largest (red) to the smallest (blue)). \\

\noindent
\textbf{Vessel surface generation:}
In the second step, based on the generated centerline, a 3D surface mesh is created. 
The construction of the surface mesh can be easily done with the usage of tools available in the Vascular Modelling Toolkit (VMTK) library~\cite{ch4:vmtk}, which is the common choice when dealing with the generation of synthetic vasculature.
The second row of Fig.~\ref{fig:ch4:synthetic-samples} showcases examples of surface meshes constructed based on the centerlines (row above) generated in the previous step. \\

\noindent
Each synthetic vessel in the dataset was generated in the described two-step process with the topology, geometrical features and stenotic areas sampled randomly based on the provided value ranges.
The generation of one sample took approximately $2.30$min.

\subsection{CFD simulations}
\label{sec:4.2.2}

\noindent
As we already mentioned in Sections~\ref{sec:1.2.2}, \ref{sec:2.1} and \ref{sec:2.2}, the most common way of obtaining the ground truth labels for hemodynamics estimation is to run a CFD simulation.
We take the same approach in this work and utilize a CFD engine by Kosior et al.~\cite{ch1:hemolens-cfd-patent} to run \textit{stationary} blood flow simulations.
By the \textit{stationary} we consider the simulations when the velocity, pressure and fluid properties remain constant at any point in the flow field~\cite{ch4:stationary-cfd} - basically, they do not change in time.
In this work, we treat the CFD engine as a black box and do not delve into the specifics of the algorithm, for the details the reader is referred to the cited patent~\cite{ch1:hemolens-cfd-patent}. \\

\begin{figure}[ht]
    \centering
    \includegraphics[width=\textwidth]{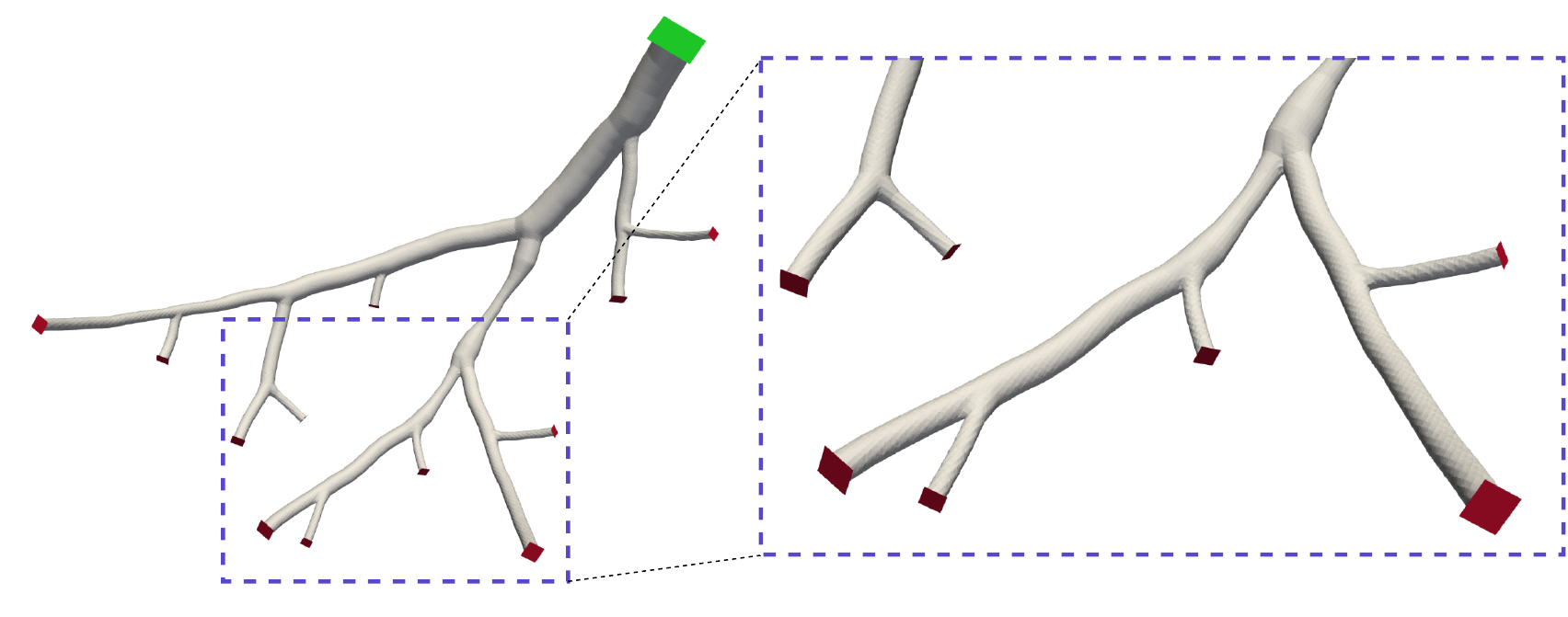}
    \caption[Placement of inlets and outlets for the CFD simulation]{Placement of inlets and outlets for the CFD simulation. The green plane denotes the inlet through which the fluid comes into the system. The red planes denote the outlets through which the fluid leaves the system.}
    \label{fig:ch4:cfd-inlets-outlets}
\end{figure}

\noindent
The input to the CFD engine is a vessel geometry represented as a surface mesh, inlet and outlet planes which describe respectively the place where the fluid comes into the system and where it leaves it (see Fig.~\ref{fig:ch4:cfd-inlets-outlets}), and a single number representing the prescribed flow of the blood at the vessel inlet given in ml/s (referred to as inflow).
The inflow value, simply means how much blood is coming into the system, and since the simulation is stationary this value is constant across the time.
For each vessel, we run three individual CFD simulations with inflow values from a biologically relevant range: $3$, $5$ and $7$ ml/s, to simulate a patient under rest, mild exercise and high-intensity exercise conditions~\cite{ch4:cfd-conditions}. 
One simulation takes approximately $2$ hours on the off-the-shelf processor with $16$ processes.

\begin{figure}[ht]
    \begin{subfigure}{.3\textwidth}
      \centering
      \includegraphics[width=\textwidth]{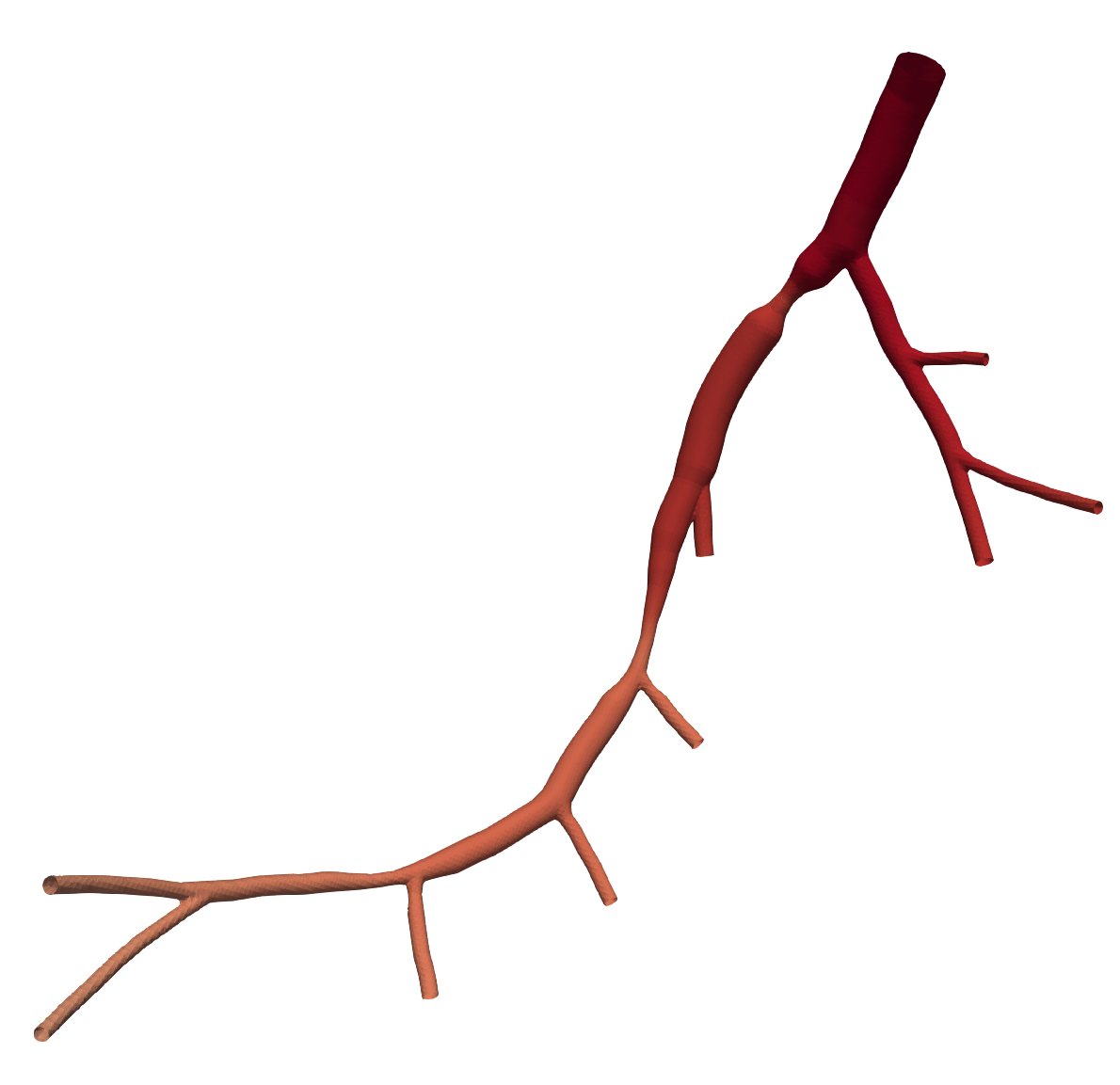}
      \caption{Inflow: $3$ ml/s}
      \label{fig:ch4:cfd-simulation:300}
    \end{subfigure}%
    \begin{subfigure}{.3\textwidth}
      \centering
      \includegraphics[width=\textwidth]{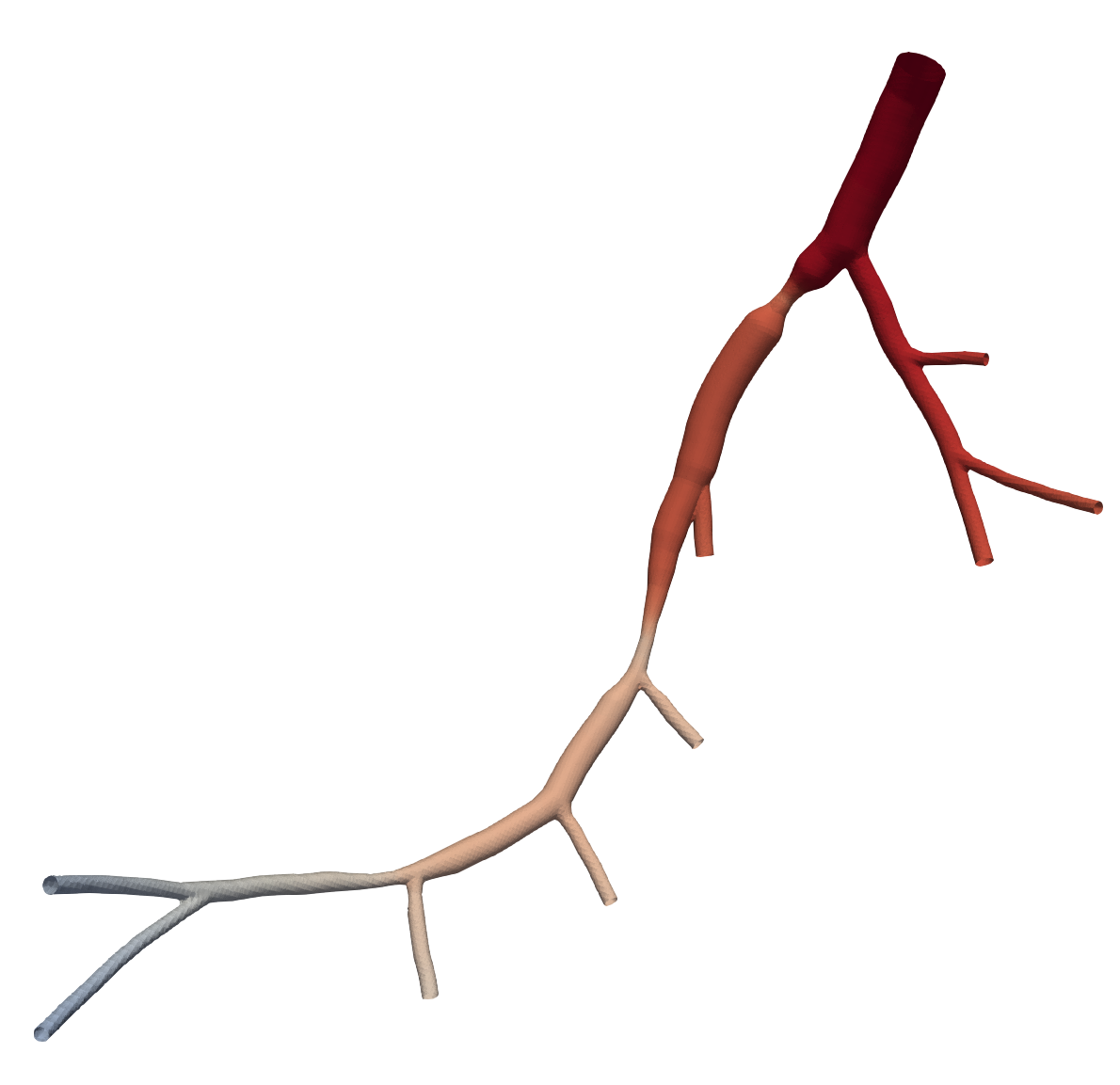}
      \caption{Inflow: $5$ ml/s}
      \label{fig:ch4:cfd-simulation:500}
    \end{subfigure}%
    \begin{subfigure}{.3\textwidth}
      \centering
      \includegraphics[width=\textwidth]{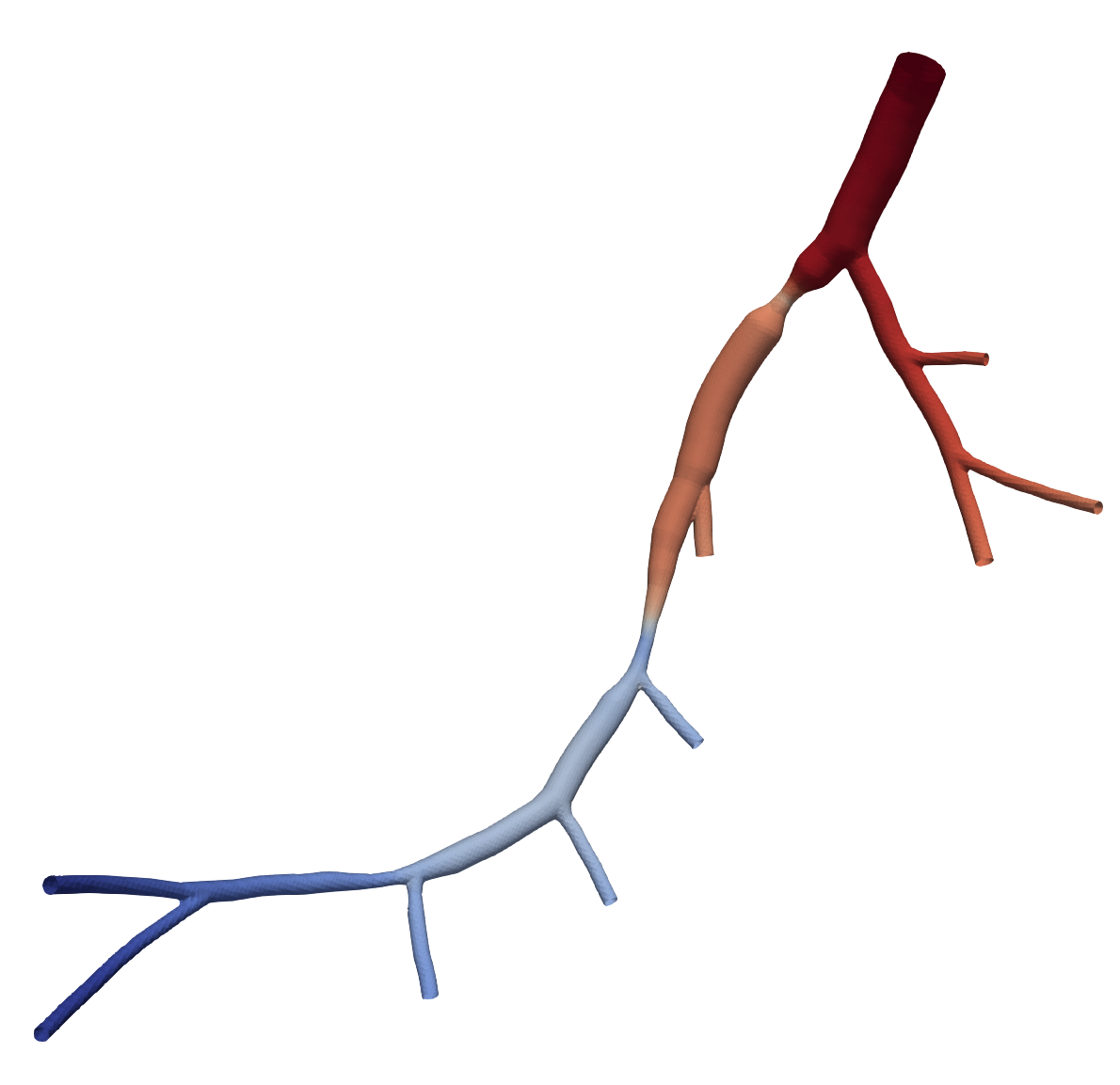}
      \caption{Inflow: $7$ ml/s}
      \label{fig:ch4:cfd-simulation:700}
    \end{subfigure}%
    \begin{subfigure}{.1\textwidth}
      \centering
      \includegraphics[width=\textwidth]{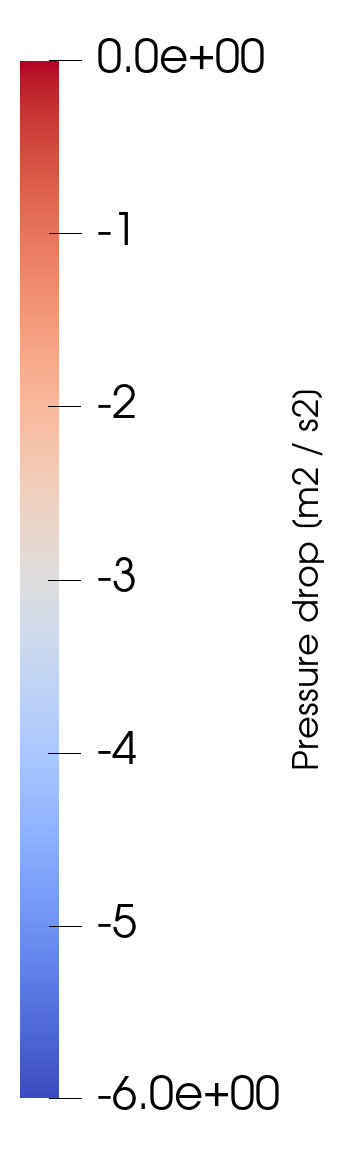}
      \label{fig:ch4:cfd-simulation:legend}
    \end{subfigure}%

    \centering
    \caption[CFD simulation for various inflow values]{CFD simulation for various inflow values. The larger the amount of blood coming through the inlet, the larger drops of the pressure at the stenotic segments. Intuitively, there is a cap on the amount of blood that can go through the stenotic segment in a set amount of time. Thus, the more blood there is in the system the more of it gathers before the stenosis, hence the higher pressure is induced and the drop after the stenosis is larger.}
    \label{fig:ch4:cfd-simulation}
\end{figure}

\noindent
As we already mentioned in Section~\ref{sec:4.1}, as the labels for a deep learning method, we extract \textit{pressure drops} for each surface point from the CFD simulation.
A \textit{pressure drop} informs how much of the pressure of the fluid is lost as it travels along the vessel.
The value of the pressure drop at the vessel inlet is $0$ and the higher the drop the more negative it gets (Fig.~\ref{fig:ch4:cfd-simulation} showcases how the values of pressure drops are distributed along the vessel geometry and how they are impacted by the stenotic segments for various inflow values). 
To calculate the vFFR based on the pressure drops, the patient-specific pressure at the inlet needs to be prescribed.
The value of the pressure for each point is computed by summing the inlet pressure with the point's pressure drop.
Then the target vFFR at the given point is simply a ratio of the pressure at this point to the inlet pressure. 
The AI-based vFFR estimation is computed in the same way but based on the pressure drops obtained from an AI model.

\section{Estimation of vFFR with point cloud based neural networks}

\noindent
In this work, as already mentioned in Section~\ref{sec:4.1}, we propose to utilize a hybrid approach to learning both on implicit and explicit features.
The explicit features are provided as the input point features and the implicit ones are extracted by the point cloud based neural network.
The process of point cloud construction and specifics of both utilized deep learning architectures and training regimes are provided in the following sections.

\subsection{Point cloud construction}
\label{sec:4.3.1}

\noindent
As highlighted in Fig.~\ref{fig:ch4:workflow-diagram}, the construction of the point cloud is done based on the input mesh representing the vessel surface.
The mesh is constructed out of nodes with 3D coordinates and edges which convey the connectivity of the surface.
In the \textit{point cloud extraction} step (see Fig.~\ref{fig:ch4:workflow-diagram}) the edges of the mesh are stripped and only the nodes with 3D coordinates (denoted XYZ) are kept.
In the \textit{hand-crafted features extraction} step (see Fig.~\ref{fig:ch4:workflow-diagram}), the two features are extracted from the surface mesh: \textit{geodesic distance from the inlet} (denoted G) and \textit{radius} (denoted R). \\

\begin{figure}[ht]
    \begin{subfigure}{.5\textwidth}
      \centering
      \includegraphics[width=\textwidth]{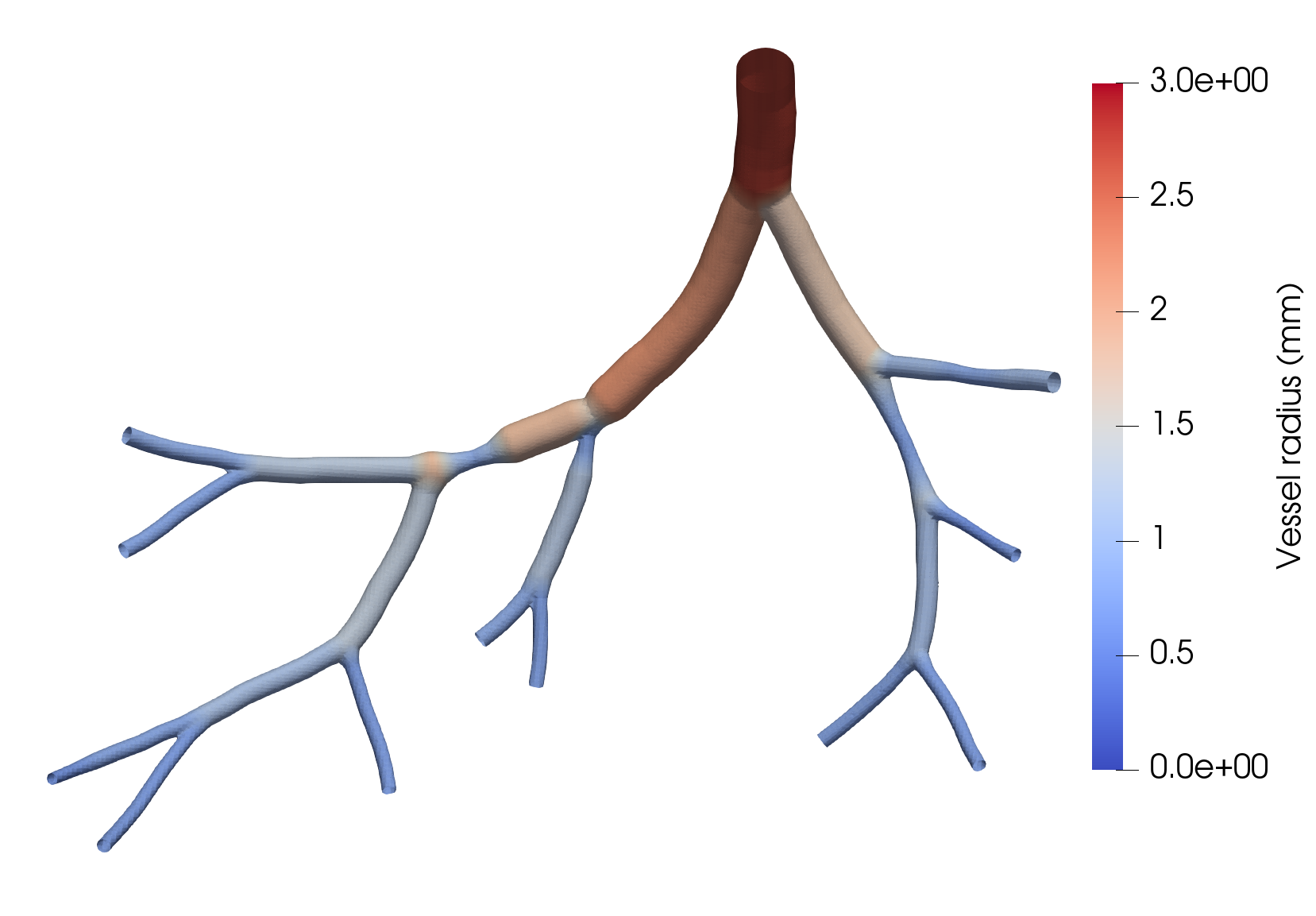}
      \caption{Vessel radius}
      \label{fig:ch4:hand-crafted-features:radius}
    \end{subfigure}%
    \begin{subfigure}{.5\textwidth}
      \centering
      \includegraphics[width=\textwidth]{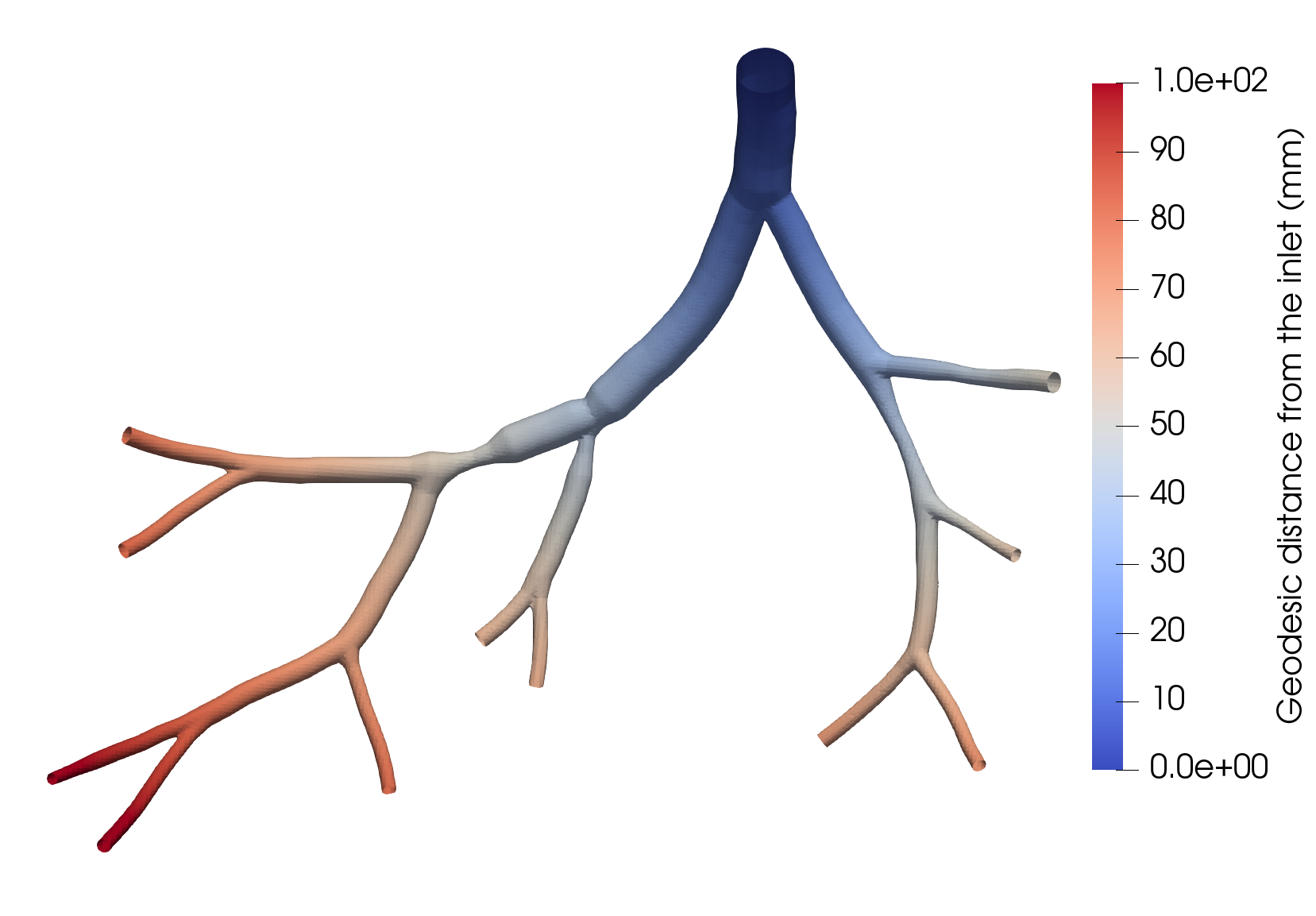}
      \caption{Geodesic distance from the inlet}
      \label{fig:ch4:hand-crafted-features:geodesics}
    \end{subfigure}%

    \centering
    \caption[Additional hand-crafted features]{Additional hand-crafted features. (a) The vessel radius feature informs how wide is the vessel at the given point. (b) The geodesic distance from the inlet feature informs how far along the vessel surface the given point is from the inlet plane.}
    \label{fig:ch4:hand-crafted-features}
\end{figure}

\noindent
\textit{Geodesic distance from the inlet:}
In the coronary tree, the direction of the blood flow is set - the fluid moves from the inlet to the outlets.
This information is crucial when estimating the hemodynamic features since the behaviour of the system is totally different when the fluid comes from a different side.
Thus, to incorporate the information about the blood flow direction, in the form of a hand-crafted feature, we equip each surface point with its geodesic distance along the surface mesh from the inlet plane. 
The geodesic distance along the surface mesh is simply the shortest path between the given pair of points in the graph representing the surface mesh.
This feature allows us to determine how far the given point is from the vessel inlet, and it also is continuous, thus even in the local vessel segment, the direction of the blood flow is provided.
Fig.~\ref{fig:ch4:hand-crafted-features:geodesics} showcases an example of the coronary tree with the colours denoting values of the geodesic feature given in mm. \\

\noindent
\textit{Radius:}
The second feature that we provide to the network explicitly is the vessel radius.
For each surface point, its shortest Euclidean distance to the centerline is computed and denoted as the radius feature.
This feature simply informs how wide the vessel is at the given point.
One could argue that such a feature should be easily learnt implicitly by the neural network given the vessel geometry.
Although, obviously, such a feature is learnt in some form implicitly, the fine-grained precision of it might be lacking.
When incorporating radius explicitly as the input feature, we provide the network with an exact value of the narrowing which allows for a better estimation of the pressure drops in these key points.
Fig.~\ref{fig:ch4:hand-crafted-features:radius} showcases an example of the coronary tree with the colours denoting values of the radius features given in mm. \\

\noindent
When extracted, both coordinates (XYZ) and hand-crafted features (R, G), are concatenated together to form the $5$-element tuple (X, Y, Z, R, G) representing a single point.
The point cloud constructed in such a manner is then provided as input to the point cloud based DNN.

\subsection{Architectures}

\begin{figure}[ht]
    \centering
    \includegraphics[width=\textwidth]{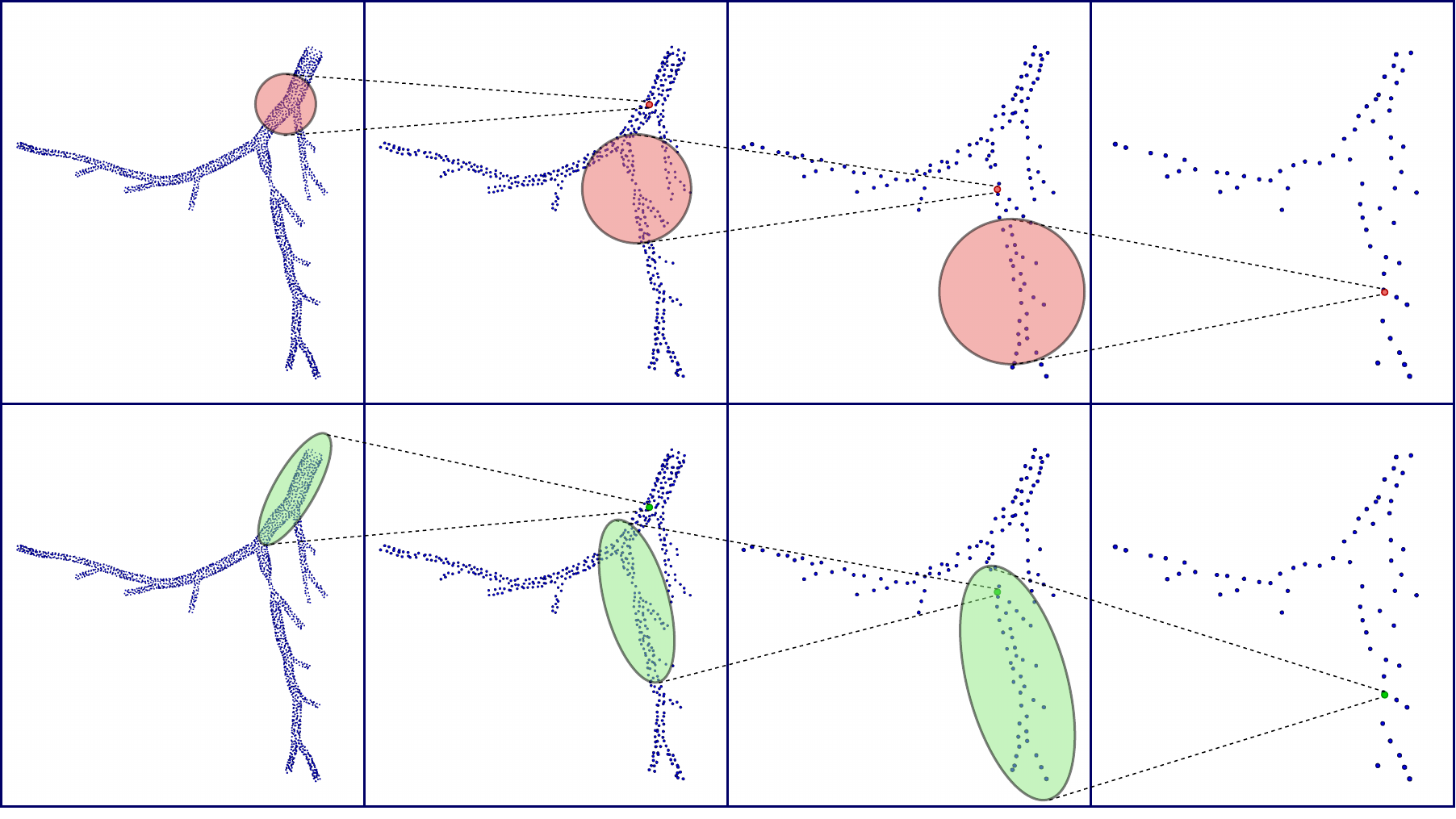}
    \caption[Grouping methods for the vFFR estimation]{Grouping methods for the vFFR estimation. The figure showcases the process of hierarchical point cloud encoding in PointNet++. In the first row, the MSG method is utilized - the points are grouped based on the Euclidean distance. In the second row, the EVG method is utilized - the points are grouped along locally dominant eigenvectors, which estimate the local vessel direction.}
    \label{fig:ch4:grouping-arteries}
\end{figure}

\noindent
In this work, as the point cloud neural network, we utilize PointNet++ architecture (for the detailed description please refer to Section~\ref{sec:3.3}).
We tested two different configurations of the network, namely usage of standard grouping technique multi-scale grouping (MSG) and eigenvector grouping (EVG), described in Section~\ref{sec:3.3.3}.
The geometry-aware grouping (GAG) which has also been described in Section~\ref{sec:3.3.3} is not applicable to the problem at hand.
It is mainly tailored towards the analysis of point clouds with disconnected components.
Meanwhile, in our task of hemodynamics estimation, the vessel is a single non-disconnected component, which renders GAG to be non-different from MSG.
The difference between MSG and EVG in the process of hierarchical point cloud encoding is showcased in Fig~\ref{fig:ch4:grouping-arteries}.
When utilizing Euclidean distance to group (MSG), in a single neighbourhood multiple branches might be present.
Meanwhile, with EVG the neighbourhood mostly is stretched along a single branch. \\

\begin{figure}[ht]
    \centering
    \includegraphics[width=\textwidth]{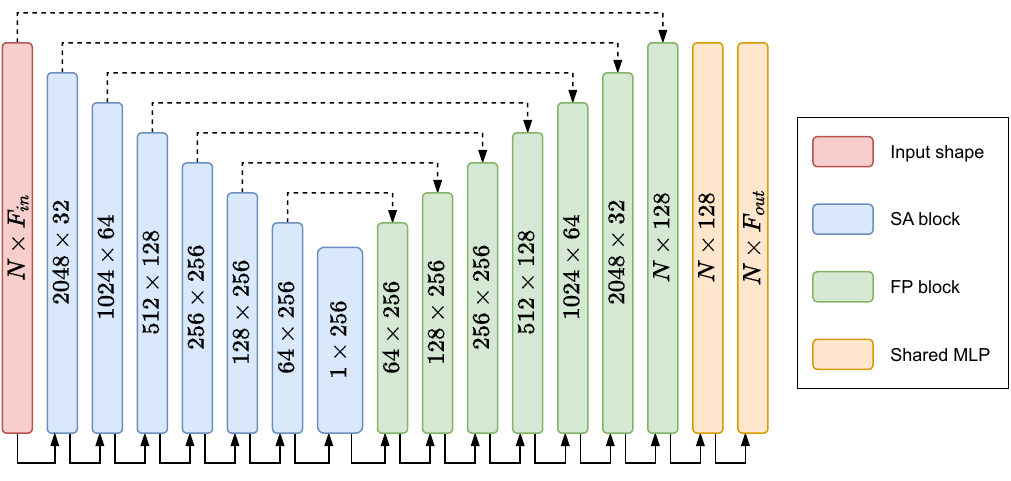}
    \caption[PointNet++ architecture details]{PointNet++ architecture details. We employ the general architecture which consists of $7$ pairs of SA and FP blocks, and a regression head built out of $2$ shared MLP layers.}
    \label{fig:ch4:architecture-details}
\end{figure}

\noindent
For both grouping methods, we employ the same general architecture setup showcased in Fig.~\ref{fig:ch4:architecture-details}.
The network is built out of $7$ SA blocks which hierarchically perform feature extraction and downsampling for the point cloud to be finally encoded as a 1D feature vector of size $256$.
The decoder branch is built out of $7$ FP blocks, and topped with two shared MLP layers to output the desired number of features per point $F_{out}$.
In our case, $F_{out} = 1$ since we only estimate scalar pressure drop values, and $F_{in} = 5$, due to the usage of 3D coordinates (XYZ) and two hand-crafted features (R, G) described in detail in Section~\ref{sec:4.3.1}.
The information extracted from the following SA blocks is passed to decoder ones to perform feature interpolation and upsampling.
For MSG and EVG configuration, only the grouping procedures in SA blocks are changed.

\noindent
The EVG configuration utilizes MSG grouping in the first block and only EVG in the following ones.
The neighbourhoods in the first block are small and do not span the vessel along its direction that much, thus there's no need for directional grouping.
For both MSG and EVG configurations, we employ the multi-scale grouping scheme (see Section~\ref{sec:3.3.2}) with $2$ scales and denote hyperparameters for these scales with respective subscripts $1$ or $2$.
The MSG method is configured with a single hyperparameter, namely grouping radius $r$.
The EVG one takes additional $2$ parameters: $k$ - number of nearest neighbours defining the prior neighbourhood to compute the eigenvector of, and $L$ - a length of the grouping vector.
For the EVG method, we set the following parameters to be constant across the encoder blocks: $L_1 = 0.001$, $L_2 = 0.002$, $k_1 = 64$ and $k_2 = 64$.
The radii $r_1, r_2$ change between the blocks and are the same for both MSG and EVG - the exact values are provided in Table~\ref{tab:ch4:grouping-radii}.

\setlength{\tabcolsep}{0.5em}
{\renewcommand{\arraystretch}{1.2}%
    \begin{table}[H]
    \centering
    \caption[Short Heading]{Values of the grouping radii in the MSG and EVG PointNet++ configurations. SA@n denotes the following SA-blocks in the encoder branch, and the $ALL$ marker denotes that all the remaining points are grouped together.}
    \begin{tabular}{l|llllllll}
    \hline
    \toprule 
        \textbf{Radius} & \multicolumn{7}{c}{\textbf{Encoder}}\\
        & \textbf{SA@1} & \textbf{SA@2} & \textbf{SA@3} & \textbf{SA@4} & \textbf{SA@5} & \textbf{SA@6} & \textbf{SA@7}\\
        
        \midrule
        $r_1$ & $0.04$ & $0.04$ & $0.08$ & $0.16$ & $0.24$ & $0.6$ & $ALL$\\
        $r_2$ & $0.08$ & $0.08$ & $0.16$ & $0.24$ & $0.32$ & $0.8$ & $ALL$\\
        
    \hline
    \end{tabular}
    \label{tab:ch4:grouping-radii}
    \end{table}
}

\subsection{Training \& Inference setup}
\label{sec:4.3.2}

\noindent
We split the generated synthetic dataset of $1700$ samples into $3$ subsets train, validation and test with $1500$, $100$, and $100$ samples respectively.
We utilize the same split in each experiment. \\

\noindent
\textbf{Preprocessing:}
During the preprocessing stage, before either the training or inference pipeline, the point cloud and hand-crafted features are extracted based on the provided surface mesh.
The obtained point cloud $\mathcal{P}$ is centred around the origin of Euclidean space and resized to fit the cube centred around the origin as well, with the edge of length $1$: $\mathcal{P}_{resized} \in [-0.5, 0.5]^3$ (values for the radii in Table~\ref{tab:ch4:grouping-radii} are provided in this resized metric space).
For the sake of time efficiency, the preprocessed data is cached on the drive. \\

\noindent
\textbf{Training setup:} 
During training, in each iteration, the input point cloud is subsampled to have exactly $20,000$ points.
The resolution of the original mesh is about $200-400k$ nodes, which is far too memory and time-exhausting.
Subsampling to about $5-10\%$ of the original resolution is a good tradeoff between the efficiency and quality of the results.
As the loss function, we utilize a mean squared error (MSE) which is calculated point-wise between pressure drop labels and estimation from the model.
The models are trained for $500$ epochs with the batch size of $8$ and for the optimizer we employ \texttt{Adam} with a constant learning rate of $0.001$. 
For augmentations, we utilize random rotations of any angle in all planes. \\

\noindent
\textbf{Inference setup:}
For the inference to be performed on the whole point cloud, not only a subsampled one, as done in the training pipeline, we adopt the scheme proposed by Balsiger et al.~\cite{ch4:inference-scheme}.
The input point cloud is split randomly into chunks of $20,000$ points, and for each one, a different inference is run.
For the results to be more comprehensive, this procedure is performed $10$ times and final per-point estimations are an average along all the runs.

\subsection{Implementation}

\noindent
The models are implemented in Python with the usage of PyTorch~\cite{ch4:pytorch}, PyTorch Geometric (PyG)~\cite{ch4:pytorch-geometric} and pytorch-lightning~\cite{ch4:pytorch-lightning} libraries.
The CUDA kernel supporting the EVG grouping scheme was developed to be compatible with the PyG library and it is provided in the attached CD~\ref{plytaCD}.

\section{Summary}

\noindent
In this chapter, we introduced and described in detail our proposed approach to the estimation of vFFR in the coronary arteries.
We discussed the reasoning behind combining explicit and implicit feature learning to perform hybrid and robust surface encoding.
The process of the synthetic dataset generation was described, and the methods of obtaining the geometry and CFD labels were discussed.
Finally, the employed architecture configurations of PointNet++ were introduced with the explicitly stated hyperparameters.
At last, we described the training and inference pipelines and mentioned the implementation details.
In the next chapter, we will describe the performed evaluations, of the proposed approach, together with the results and the discussion of limitations and future works. 
	\cleardoublepage

 	\chapter{Experiments \& results}
\thispagestyle{chapterBeginStyle}
\label{chapter5}

\noindent
In this chapter, we describe the performed evaluations of the proposed approach and report obtained results.
The experimental settings are tailored towards evaluating the generalization of the method for the various blood flow characteristics and stenosis grades.
Moreover, since the method is to be used in the clinical setting to qualify the patient for the invasive treatment, we showcase its clinical viability by performing classification under the FFR clinical threshold of $0.8$. 
At last, we showcase the qualitative comparisons between the AI and CFD-based simulation results and discuss common issues and limitations.

\section{Metrics}

\noindent
The problem of vFFR estimation is a per-point regression task, thus to evaluate the models' performance we utilize three common regression metrics: 

\begin{equation}
    \text{Mean Absolute Error (MAE)} = \frac{1}{n} \sum_{i=1}^n |y_i - \hat{y}_i|
\end{equation}

\begin{equation}
    \text{Normalized Mean Absolute Error (NMAE)} = \frac{1}{ny_{max}} \sum_{i=1}^n |y_i - \hat{y}_i|
\end{equation}

\begin{equation}
    \text{R$^2$ score (R2)} = 1 - \frac{\sum_{i=1}^n (y_i - \hat{y}_i)^2}{\sum_{i=1}^n (y_i - \overline{y})^2}, \;\;\;\;\ \overline{y} = \frac{1}{n} \sum_{i=1}^{n} y_i
\end{equation}

where $y$ are ground truth labels, $\hat{y}$ network predictions. The $y_{max}$ denotes a maximum value label over the whole test dataset, which serves as a normalization term. \\

\noindent
In this work, we also utilize well-known classification metrics, to perform the evaluation of the clinical viability of the proposed approach: accuracy, precision, recall and F1 score.

\section{Quantitative results}
\noindent
We perform a wide series of quantitative evaluations to test the robustness, generalization and clinical applicability of the proposed approach.
As mentioned in Section~\ref{sec:4.2.2}, we generated a $3$ set of ground truth labels for various values of blood inflow from the biologically relevant range: $3$, $5$ and $7$ ml/s.
The evaluation of network performance on these various settings of boundary conditions is relevant in the light of real-life applications.
All experiments provided in the following sections are evaluated on the same test set of $100$ geometries and for all $3$ inflow settings.
The experiments' results are reported for PointNet++ in two configurations, namely MSG and EVG, described previously in detail in Section~\ref{sec:4.3.2}.

\subsection{Evaluation under various blood flow characteristics}

\noindent
First of all, we perform the evaluation of the estimation of the pressure drops for various inflow values.
The results are provided in Table~\ref{tab:pressure-inflow-table} and we report results for mean, median and $75$th percentile for both NMAE and MAE metrics.
As we can see, the results between the MSG and EVG methods for all inflow $Q_{in}$ values are quite similar across the board.
The percentile error reported by NMAE is almost identical across the $Q_{in}$ values, meaning that the proposed approach has good generalization capabilities and can perform on the same level for datasets of different boundary conditions of blood flow simulation.
Of course, when looking at MAE, the error rises, however since this is the absolute error and values of pressure drops are much larger for the higher inflow (see Fig.~\ref{fig:ch4:cfd-simulation} for the explanation of this phenomena), such behaviour is expected.
Thus we consider the NMAE metric to be a more comprehensive one for the problem at hand. \\

\setlength{\tabcolsep}{0.5em}
{\renewcommand{\arraystretch}{1.5}%
    \begin{table}[!h]
    \small
    \caption[Evaluation of pressure drops estimation]{Evaluation of pressure drops estimation. We report mean, median and $75$th percentile of NMAE and MAE metrics for inflow $Q_{in}$ values of $3$, $5$ and $7$ ml/s. We evaluate two PointNet++ configurations, namely MSG and EVG, and denote them accordingly. The NMAE metric is a percentage-based metric and MAE is provided in kinematic pressure units $m^2/s^2$.}
    \centering
    \begin{adjustbox}{width=.7\textwidth}
    \begin{tabular}{ll|lll|lll}
    \hline
    \toprule         
        & & \multicolumn{3}{c}{\textbf{NMAE (\%)}} & \multicolumn{3}{c}{\textbf{MAE ($m^2 / s^2$})}\\
        \textbf{Model} & \textbf{$Q_{in}$ $(ml/s)$} & \textbf{mean} & \textbf{median} & \textbf{75th} & \textbf{mean} & \textbf{median} & \textbf{75th}\\
        
        \midrule
        \textbf{MSG} & 3 & 2.09 & 0.79 & \textbf{1.44} & 0.29 & 0.11 & \textbf{0.20}\\
        \textbf{EVG} & 3 & \textbf{1.85} & \textbf{0.74} & 1.62 & \textbf{0.26} & \textbf{0.10} & 0.23\\
        
        \midrule
        \textbf{MSG} & 5 & \textbf{1.40} & \textbf{0.60} & \textbf{1.54} & \textbf{0.45} & \textbf{0.19} & \textbf{0.49}\\
        \textbf{EVG} & 5 & 1.83 & 0.66 & 1.59 & 0.58 & 0.21 & 0.51\\
        
        \midrule
        \textbf{MSG} & 7 & \textbf{1.63} & \textbf{0.50} & \textbf{1.55} & \textbf{0.90} & \textbf{0.28} & \textbf{0.86}\\
        \textbf{EVG} & 7 & 1.80 & 0.66 & 1.58 & 1.00 & 0.36 & 0.88\\
        
    \hline
    \end{tabular}
    \end{adjustbox}
    \label{tab:pressure-inflow-table}
    \end{table}
}
\setlength{\tabcolsep}{0.5em}
{\renewcommand{\arraystretch}{1.5}%
    \begin{table}[!h]
    \small
    \caption[Evaluation of vFFR estimation]{Evaluation of vFFr estimation. We report mean, median and $75$th percentile of MAE metric for inflow $Q_{in}$ values of $3$, $5$ and $7$ ml/s, and input pressure $p_{in}$ of $80$, $100$ and $120$ mmHg. We evaluate two PointNet++ configurations, namely MSG and EVG, and denote them accordingly.}
    \centering
    \begin{adjustbox}{width=.8\textwidth}
    \begin{tabular}{ll|lll|lll|lll}
    \hline
    \toprule 

        & & \multicolumn{9}{c}{\textbf{MAE (10e2)}}\\
        & & \multicolumn{3}{c}{$p_{in} = 80$ $mmHg$} & \multicolumn{3}{c}{$p_{in} = 100$ $mmHg$} & \multicolumn{3}{c}{$p_{in} = 120$ $mmHg$}\\
        
        \textbf{Model} & \textbf{$Q_{in}$ $(ml/s)$} & \textbf{mean} & \textbf{median} & \textbf{75th} & \textbf{mean} & \textbf{median} & \textbf{75th} & \textbf{mean} & \textbf{median} & \textbf{75th}\\
        
        \midrule
        \textbf{MSG} & 3 & 2.95 & 1.11 & \textbf{2.03} & 2.27 & 0.85 & \textbf{1.56} & 1.97 & 0.74 & \textbf{1.35}\\
        \textbf{EVG} & 3 & \textbf{2.61} & \textbf{1.04} & 2.28 & \textbf{2.01} & \textbf{0.80} & 1.75 & \textbf{1.74} & \textbf{0.69} & 1.52\\
        \hline
        
        \midrule
        \textbf{MSG} & 5 & \textbf{4.45} & \textbf{1.91} & \textbf{4.89} & \textbf{3.42} & \textbf{1.47} & \textbf{3.76} & \textbf{2.97} & \textbf{1.27} & \textbf{3.26}\\
        \textbf{EVG} & 5 & 5.85 & 2.11 & 5.08 & 4.50 & 1.62 & 3.91 & 3.90 & 1.41 & 3.39\\
        \hline
        
        \midrule
        \textbf{MSG} & 7 & \textbf{9.04} & \textbf{2.76} & \textbf{8.61} & \textbf{6.95} & \textbf{2.12} & \textbf{6.62} & \textbf{6.03} & \textbf{1.84} & \textbf{5.74}\\
        \textbf{EVG} & 7 & 10.02 & 3.65 & 8.78 & 7.71 & 2.81 & 6.75 & 6.68 & 2.43 & 5.85\\
        
    \hline
    \end{tabular}
    \end{adjustbox}
    \label{tab:ffr-inflow-table}
    \end{table}
}

\noindent
The evaluation of clinical applicability is hard to be done based on the estimation of pressure drops since they serve only as an intermediate step in the calculation of relevant biomarker FFR.
Thus in Table~\ref{tab:ffr-inflow-table}, we report the quantitative results obtained for the vFFR estimation for various inflow and input pressure values prescribed for vFFR calculation (see Section~\ref{sec:4.2.2} for the in detail description of obtaining vFFR from pressure drops).
The vFFR takes values from between $0$ and $1$, and we report the MAE metric and all the values in the table are provided in the scientific notation of $10e2$ for the readability sake.
We see that the results between the MSG and EVG configurations do not vary much, however, for larger inflows the MSG seems to behave a bit better.
Across the board, the mean MAE is larger than the median and the $75$th percentile.
This is due to the fact, that there are a few very pathological cases in the test dataset, for which the pressure drops are very large, and the estimations of vFFR have quite a large error - we will discuss this issue more in detail in Section~\ref{sec:5.4}).
However, across the board, the absolute error of estimated vFFR is kept under $0.1$ and for more than half of the dataset, the error is almost negligible - below $0.03$.
Obviously, for larger input pressures the error gets smaller since the pressure drops are relatively smaller.
These are promising results which convey that the proposed approach can be easily trained for different boundary conditions of underlying fluid simulation without a noticeable drop in percentage performance.

\subsection{Evaluation of stenosis grade impact}
\begin{figure}[ht]
    \centering
    \includegraphics[width=\textwidth]{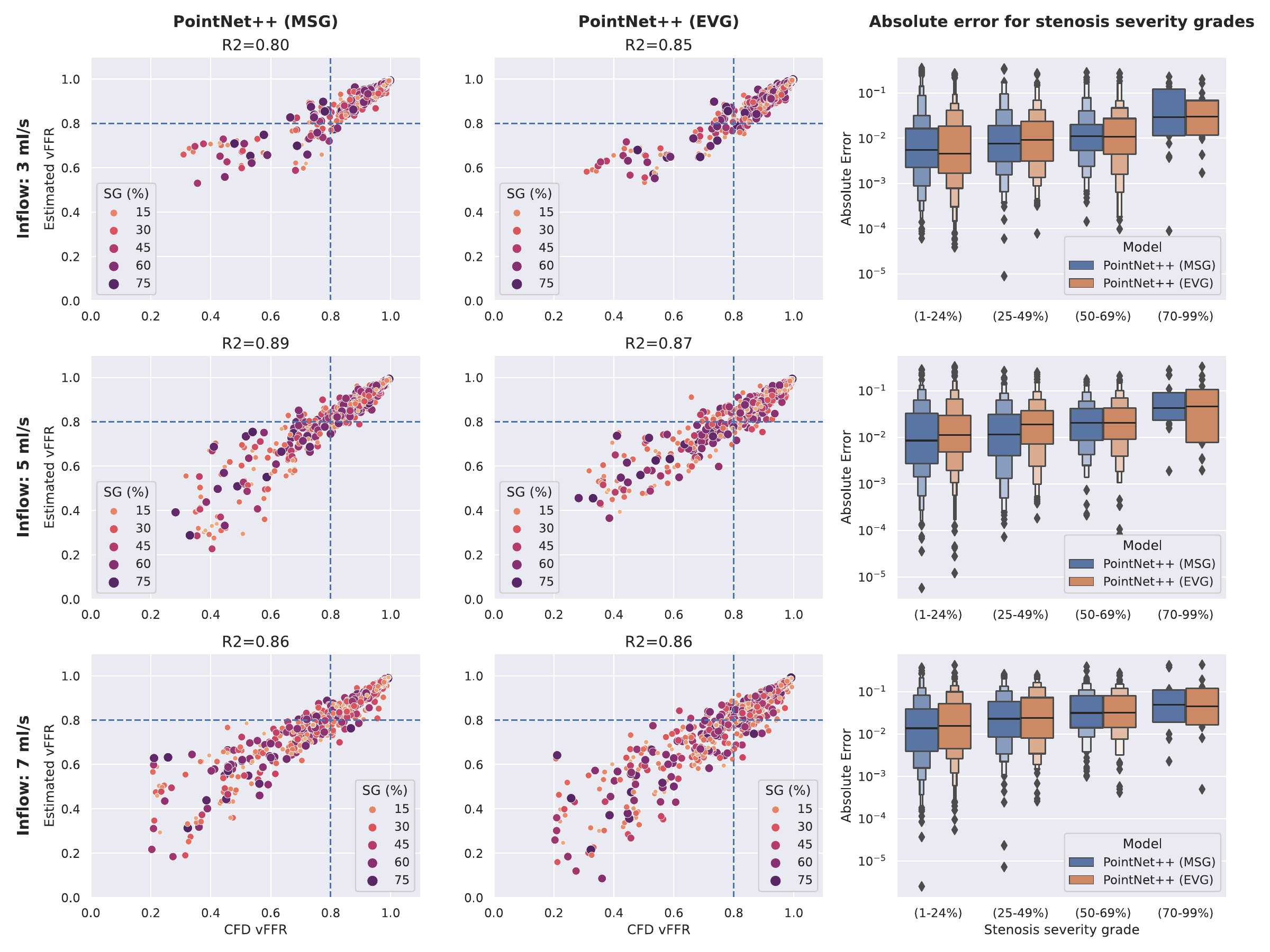}
    \caption[Evaluation of stenosis grade impact]{Evaluation of stenosis grade (SG) impact. The scatter plots showcase the correlation between AI and CFD-based vFFR estimations. The blue dashed line denotes the clinical FFR threshold of $0.8$. The boxplots showcase the absolute estimation errors grouped based on the CAD-RADS scale of stenoses' severity grades.}
    \label{fig:ch5:cad-rads}
\end{figure}

\noindent
In the second experiment scenario, we want to evaluate the method's correlation with the CFD-based FFR and its performance with respect to the stenosis grade.
We extract the stenotic segments from the test set (about $700$ samples) and label them with their percentage severity grade (SG).
This time we report results only for the average input pressure $p_{in}$ of $100$ mmHg.
In Fig.~\ref{fig:ch5:cad-rads}, we showcase the correlation plots between AI and CFD-based vFFR estimations for MSG and EVG model configurations and various inflow values.
The stenoses' severity grades (SG) are marked with markers of varying size and colour, and the clinically accepted FFR threshold of $0.8$ is marked with the blue dashed line.
Again, both model configurations seem to behave quite similarly, and the R2 score is kept above $0.8$ across the board.
For the inflow of $5$ ml/s we report the highest R2 scores of $0.89$ (MSG) and $0.87$ (EVG), which showcases that the method has very high correlations with CFD-obtained vFFR.
Moreover, as we can see the correlation in the decision margin around the FFR clinical threshold of $0.8$ is of good quality (more on the clinical viability will be discussed in Section~\ref{sec:5.2.3}).
The errors on the stenoses well outside the decision range are much less impactful, since whether the patient's FFR is estimated as $0.3$ or $0.5$, the clinical decision regarding the treatment is the same.
In the third column, we showcase the boxplots of the absolute errors with respect to stenoses' severity grades.
The severity grades are grouped based on the CAD-RADS scale~\cite{ch4:cad-rads}.
We report, as expected, that the errors grow larger with the stenoses severity grade.
For all the grades, models and inflow configurations, the error is kept under $0.1$ despite a few outliers.
For smaller severity grades, the mean error is somewhere around $0.01-0.02$.
These results prove that the method is robust towards the estimation of the impact of various stenoses' grades without much loss in performance.

\subsection{Evaluation under clinical viability}
\label{sec:5.2.3}

\noindent
In this experiment, we evaluate the clinical viability of the proposed approach.
Following the well-known Latin phrase \textit{"Primum non nocere"} (\textit{eng. first, do no harm}), which serves as the fundamental principle in healthcare, we need to evaluate whether the proposed approach is safe for clinical use, or at least does not cause any unnecessary risk for the patient. \\

\noindent
To do that, we report the classification results under the clinically accepted FFR threshold of $0.8$, of whether to perform the treatment or not.
The results for various inflow and model configurations are showcased in Table~\ref{tab:clinical-viability-table}.
As we may see, the results are quite high across the board, not dropping below $90\%$.
The precision scores seem to be a bit higher for the EVG method, with all the other metrics being in favour of MSG.
For higher-intensity exercise conditions, reflected with higher inflow values, we see a drop in the performance of the proposed approach.
These are more extreme blood flow characteristics than usual, however, when dealing with patients with suspected CAD, such conditions are not that uncommon.
Although this aspect does not limit the clinical viability of the approach that much, it should be reported and known to the medical specialist utilizing the system for the treatment decision process. \\

\noindent
In this work, we utilize only synthetic data for both training and testing of the proposed approach.
Obviously, for proper and safe usage in the clinical routine, the approach would need to be trained and evaluated on real patient geometries.

\setlength{\tabcolsep}{0.5em}
{\renewcommand{\arraystretch}{1.5}%
    \begin{table}[!h]
    \small
    \caption[Evaluation under clinical viability]{Evaluation under clinical viability. We perform the classification of whether to perform the treatment or not, by thresholding the vFFR based on the clinically accepted threshold of $0.8$.}
    \centering
    \begin{adjustbox}{width=.8\textwidth}
    \begin{tabular}{ll|llll}
    \hline
    \toprule 
        & \multicolumn{1}{c}{$Q_{in}$ $(ml/s)$} & \textbf{Accuracy (\%)} & \textbf{F1-score (\%)} & \textbf{Precision (\%)} & \textbf{Recall (\%)} \\

        \midrule
        \textbf{MSG} & 3 & \textbf{95.66} & \textbf{97.57} & 95.93 & \textbf{99.27} \\
        \textbf{EVG} & 3 & 95.02 & 97.19 & \textbf{96.42} & 97.96 \\

        \midrule
        \textbf{MSG} & 5 & \textbf{94.11} & \textbf{96.12} & \textbf{95.45} & \textbf{96.80} \\
        \textbf{EVG} & 5 & 92.64 & 95.15 & 94.56 & 95.74 \\

        \midrule
        \textbf{MSG} & 7 & \textbf{92.37} & \textbf{93.91} & 91.83 & \textbf{96.07} \\
        \textbf{EVG} & 7 & 91.10 & 92.63 & \textbf{93.84} & 91.46 \\
        
    \hline
    \end{tabular}
    \end{adjustbox}
    \label{tab:clinical-viability-table}
    \end{table}
}

\section{Qualitative results}
\begin{figure}[h!]
    \centering
    \includegraphics[width=\textwidth]{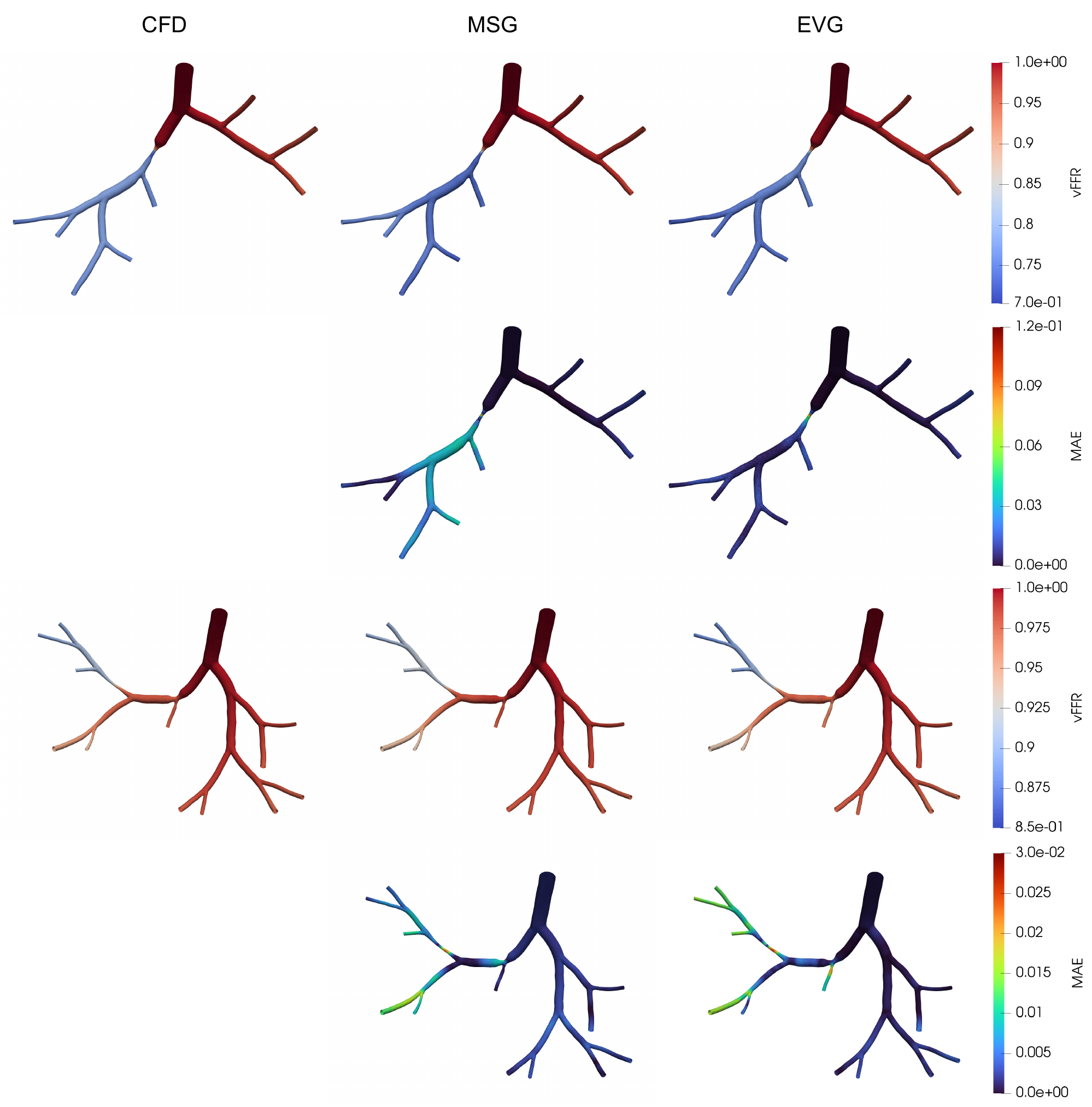}
    \caption[Qualitative results of vFFR estimation.]{Qualitative results of vFFR estimation for $Q_{in} = 3$ ml/s and $p_{in} = 100$ mmHg. We visualize vessel geometries with estimations of vFFR and MAE for both tested PointNet++ configurations - MSG and EVG. For each row, the individual legend scale has been chosen to better visualize the intricacies between the methods.}
    \label{fig:ch5:qualitative-results-good}
\end{figure}
\begin{figure}[h!]
    \centering
    \includegraphics[width=\textwidth]{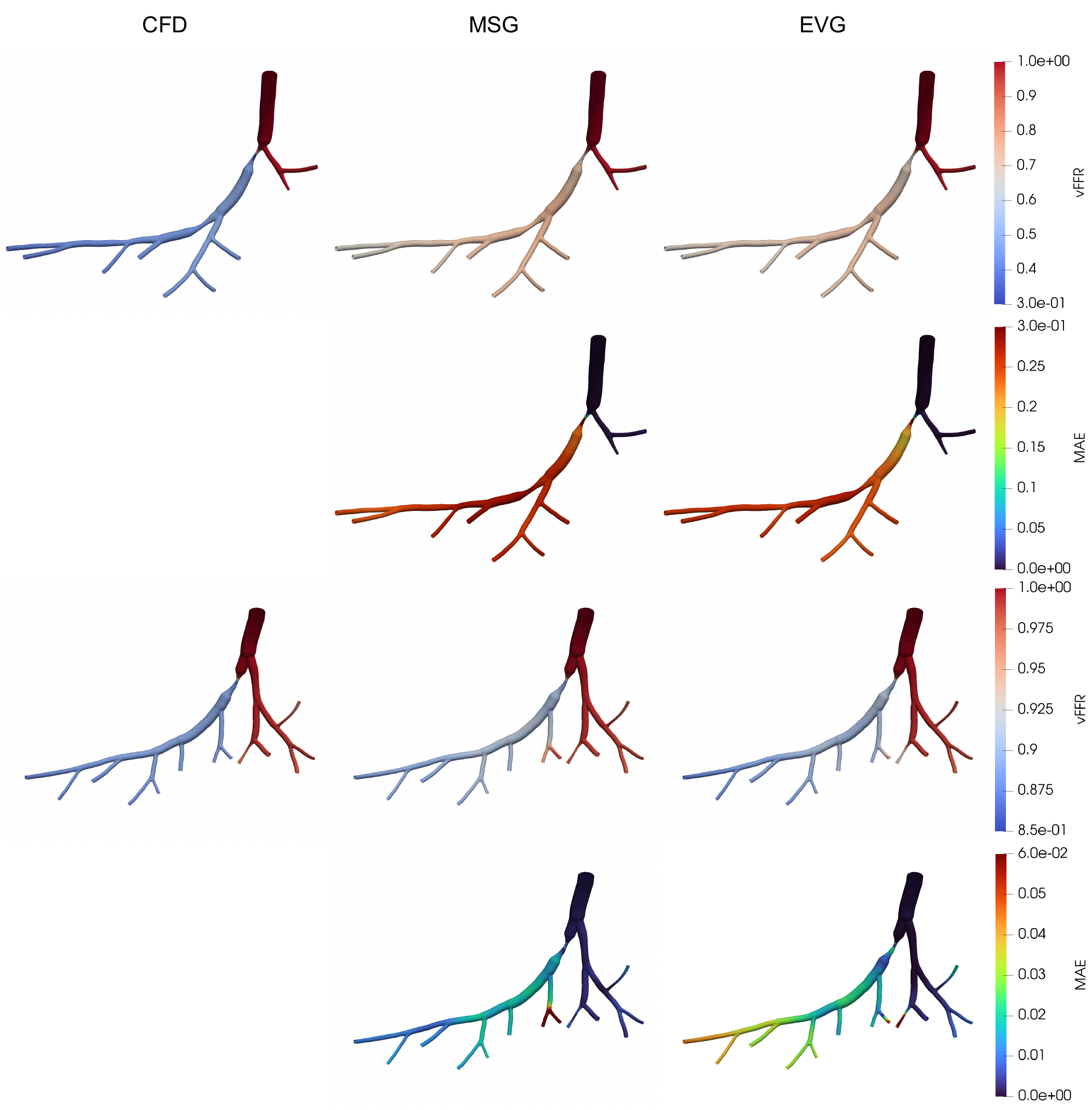}
    \caption[Issues with vFFR estimation.]{Issues with vFFR estimation for $Q_{in} = 3$ ml/s and $p_{in} = 100$ mmHg. We visualize vessel geometries with estimations of vFFR and MAE for both tested PointNet++ configurations - MSG and EVG. For each row, the individual legend scale has been chosen to visualize the intricacies between the methods better. The first sample showcases the problem with the underestimation of the vFFR after large stenosis. The second sample showcases the problem of "spilling" estimations of vFFR between two close branches.}
    \label{fig:ch5:qualitative-results-bad}
\end{figure}

\noindent
In Fig.~\ref{fig:ch5:qualitative-results-good} we showcase a few qualitative comparisons between ground truth labels obtained from CFD and from our proposed approach (in both PointNet++ configurations, namely MSG and EVG) and plot both vFFR predictions and MAE.
All the rows utilize individual legend scales for better visualization of the intricacies of the predicted vFFR.

\noindent
For the first sample (rows $1$ and $2$) we see a very early, large $~60\%$ narrowing of the vessel lumen on the leftmost branch.
This stenosis is an important one from a clinical point of view since it is close to the decision threshold of $0.8$.
The CFD-obtained vFFR reports vFFR of $~0.75$, and both MSG and EVG configurations yield similar results.
For EVG and error is almost negligible, as seen on the MAE plot, and for MSG the error is in the range of $0.03$.
But most importantly, for both methods, the predicted vFFR is under $0.8$, which is crucial in a clinical setting.
The largest error is reported in the very middle of the stenosis, however, FFR is never estimated in such places but rather behind the stenosis where the blood dynamics are more stable. \\

\noindent
The second sample (rows $3$ and $4$) has $2$ medium-size narrowings on the leftmost branch, the first one of size $~30\%$, and the second one of size $~50\%$.
In comparison to the first sample we discussed, the stenoses present here are not causing vFFR to drop below $0.8$ - it is kept quite high almost above $0.9$ for the whole artery.
We see that EVG is a bit better at estimating the first narrowing, however, it lacks a bit of precision on the small branches for which the reported error is about $0.01$.
Both methods, however, correctly estimate the importance of the narrowings of the blood dynamics.
For both showcased examples we see that the error on non-stenotic segments is almost negligible, meaning that the network is robust enough to correctly grasp the notion of fluid dynamics in standard vessel geometries. \\

\section{Problems \& issues}
\label{sec:5.4}

\noindent
The proposed approach achieves promising results in all evaluation scenarios.
It generalizes well for various blood flow characteristics, achieves good correlation with CFD-estimated vFFR and is clinically viable under the assumption that the synthetic data can be replaced with real geometries without a large loss in estimation quality.
However, there are some common problems and issues that can be observed in a few samples - these problems are showcased in Fig.~\ref{fig:ch5:qualitative-results-bad}. \\

\noindent
One common issue, when estimating very large narrowings ($>70\%$) is underestimation - the network cannot correctly assess the impact of the stenosis.
The example showcasing this problem is plotted in the first and second row of Fig.~\ref{fig:ch5:qualitative-results-bad}.
The network spots the stenosis and considers it significant since the estimated vFFR is clearly below $0.8$.
However, it considers vFFR to be higher after the stenosis by almost $0.2$ than it is in reality. \\

\noindent
The second problem appears when there are two very close, in the Euclidean sense, branches in the vessel tree.
In such cases, the estimated vFFR from one branch may "spill" onto another, causing the vFFR to rise, which is biologically impossible and simply incorrect.
The example showcasing this issue is plotted in the third and fourth row of Fig.~\ref{fig:ch5:qualitative-results-bad}.
A little branch in the middle of the tree for both MSG and EVG has increased vFFR more than the prior vessel segment.
This is caused by the close presence of another branch on the right side with drastically different vFFR.
The network has problems with such areas because points are considered neighbours based on the Euclidean distance and not proximity along the manifold, and the features are hierarchically aggregated in this manner. \\

\section{Future works}

\noindent
This work serves as a proof-of-concept of utilizing point cloud based approaches to vFFR estimation.
We plan on developing this approach further and thus outline below a few ideas that are to be addressed in the future:

\begin{itemize}
    \item Collection of representative real-patient database to perform training and validation on.
    \item Incorporating information about the boundary conditions of the underlying CFD simulation into the neural network, to be able to train one model which can be parametrized in the inference setting.
    \item Enhancing the proposed approach with PINNs (physics-informed neural networks) to introduce a physics prior that can help with the problems mentioned in Section~\ref{sec:5.4}. 
\end{itemize}

\section{Summary}

\noindent
In this chapter, we reported the results of the proposed model evaluation for various experiment scenarios.
We showcased that the method can be easily trained for the various underlying CFD simulations without loss of performance and that it achieves promising quantitive results.
We evaluated the impact of the stenosis severity grade on the model prediction and calculated the correlation with CFD simulation.
To prove the clinical viability of the model, the classification of whether to perform the treatment or not based on its predictions has been performed.
We showcase some qualitative examples by plotting geometries with vFFR and highlight issues and problems with the estimations of vFFR with the proposed approach.
At last we touched on future works, and describe what might be next steps of developing and further enhancing the proposed approach in this work.

	\cleardoublepage
	
	\chapter{Summary}
\thispagestyle{chapterBeginStyle}

\noindent
To conclude, in this work we propose a novel approach to estimating vFFR in coronary arteries with the usage of point cloud representation.
We propose a hybrid approach of utilizing both implicit and explicit feature learning to encode the vessel geometry in a robust matter.
To demonstrate the applicability of the proposed approach in the clinical practice we construct and run a series of evaluation scenarios.
We prove that the proposed approach is a compelling replacement for commonly utilized CFD-based methods in this task, due to the achieved high correlation of vFFR and drastically decreased inference time from hours to seconds.
This study serves as a promising proof-of-concept which we plan on developing further on. \\

\noindent
During my master's studies, my main areas of research have been applications of deep learning in cardiovascular imaging and in the analysis of 3D shapes, in particular 3D point clouds.
My work "Coronary Ostia Localization Using Residual U-Net with Heatmap Matching and 3D DSNT"~\cite{ch5:ostia} dealing with the localization of landmarks of interest in 3D medical images in form of CCTA, was published at Machine Learning in Medical Imaging (MLMI) 2022 Workshop held with MICCAI 2022.
This master thesis utilized parts of research previously conducted during my bachelor thesis titled "Artefact removal from 3D coronary artery segmentation from CT via point cloud neural network." which was concluded with a paper "Eigenvector Grouping for Point Cloud Vessel Labeling"~\cite{ch3:evg} at Geometric Deep Learning in Medical Imaging (GeoMedIA) 2022 Workshop endorsed by MICCAI 2022.
A part of this work with a follow-up proposing centerline-guided surface embedding~\cite{ch3:centerlinepointnet} was published at MICCAI 2023 conference.

	\cleardoublepage


	\pagestyle{bibliographyStyle}
	\bibliographystyle{abbrv}
	\bibliography{literatura}
	\thispagestyle{chapterBeginStyle}
        \addcontentsline{toc}{chapter}{Bibliography}

	\cleardoublepage
	
	
	\appendix
	\pagestyle{appendixStyle}
	
	\chapter{Additional qualitative results}
\thispagestyle{chapterBeginStyle}
\label{supplementary1}

\noindent
In this supplementary, we include additional qualitative results of vFFR estimation.

\begin{figure}[h!]
    \centering
    \includegraphics[width=\textwidth]{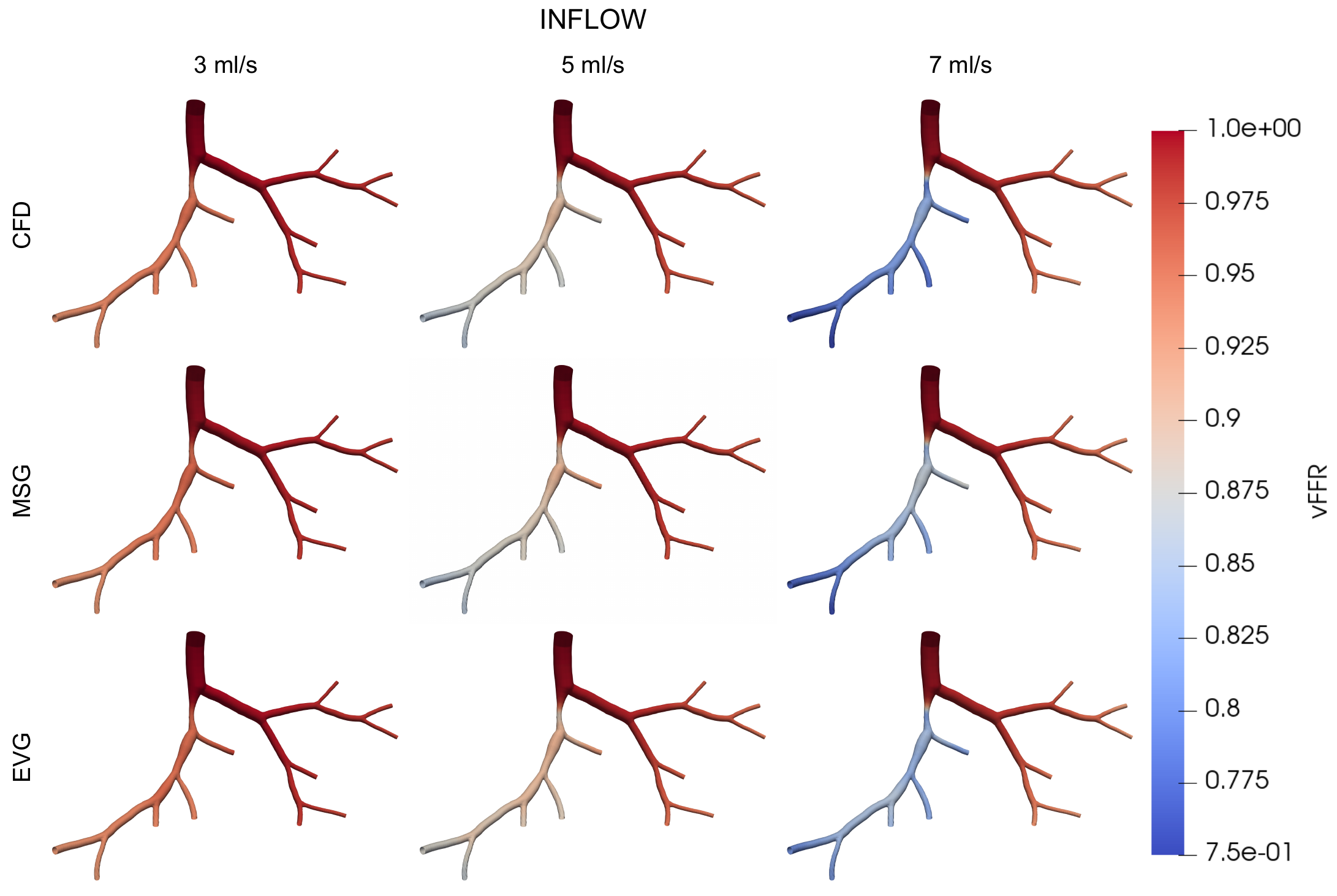}
    \caption[Qualitative results of vFFR estimation for various inflow values]{Qualitative results of vFFR estimation for various inflow values. We can see that the proposed approach can be easily adapted for various sets of boundary conditions e.g. inflow, of underlying CFD simulation. The absolute errors do become larger, when inflow gets larger, however, the percentage errors stay the same - see Table.~\ref{tab:ffr-inflow-table}. } 
    \label{fig:supp:inlow-results}
\end{figure}
\begin{figure}[h!]
    \centering
    \includegraphics[width=\textwidth]{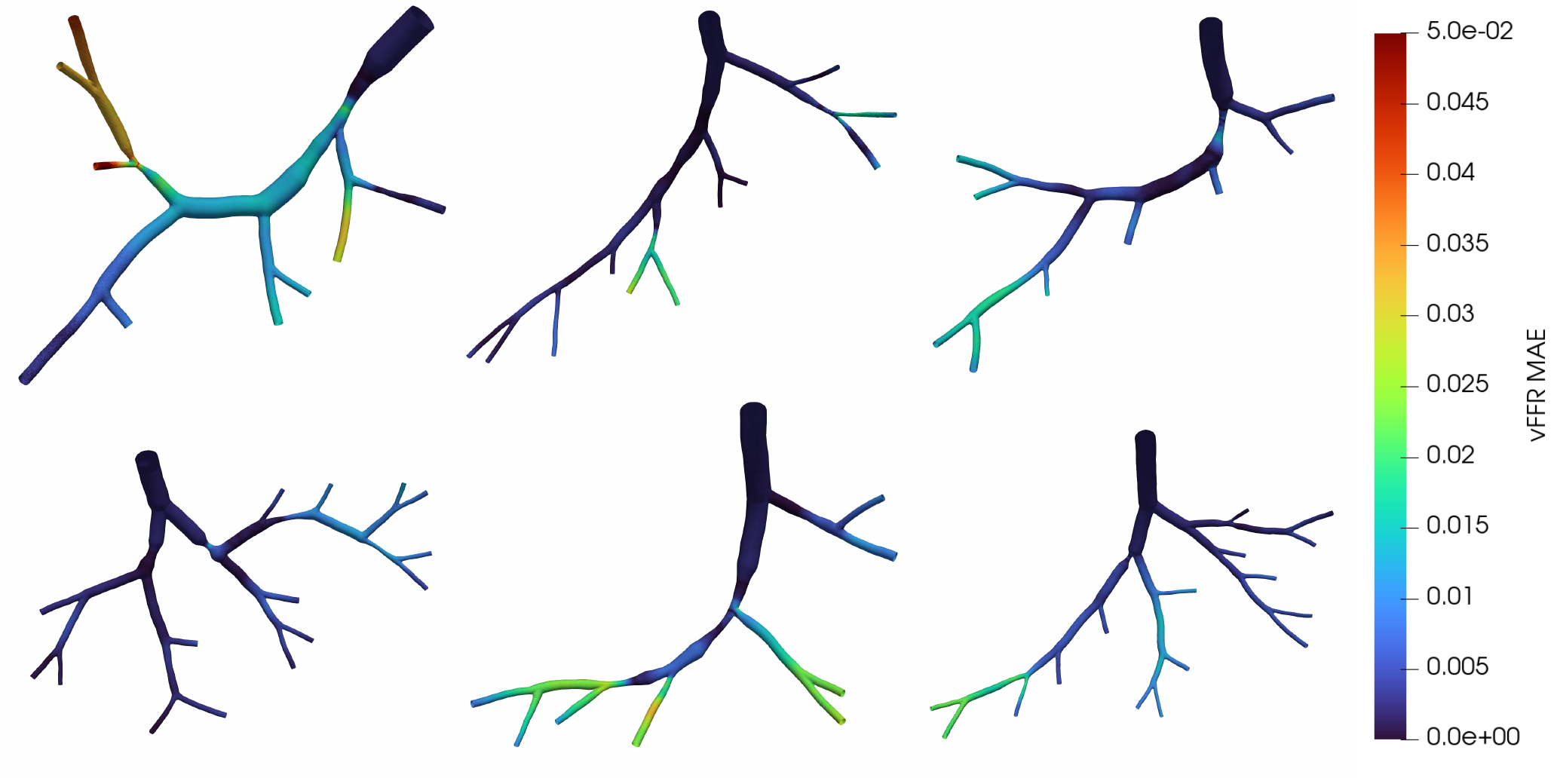}
    \caption[Additional MAE results for MSG model.]{Additional MAE results for MSG model. We see that the highest errors appear in the stenotic areas or after them. The errors tend to be propagated after under or over-estimated narrowing impact.} 
    \label{fig:supp:msg-mae-results}
\end{figure}
\begin{figure}[h!]
    \centering
    \includegraphics[width=\textwidth]{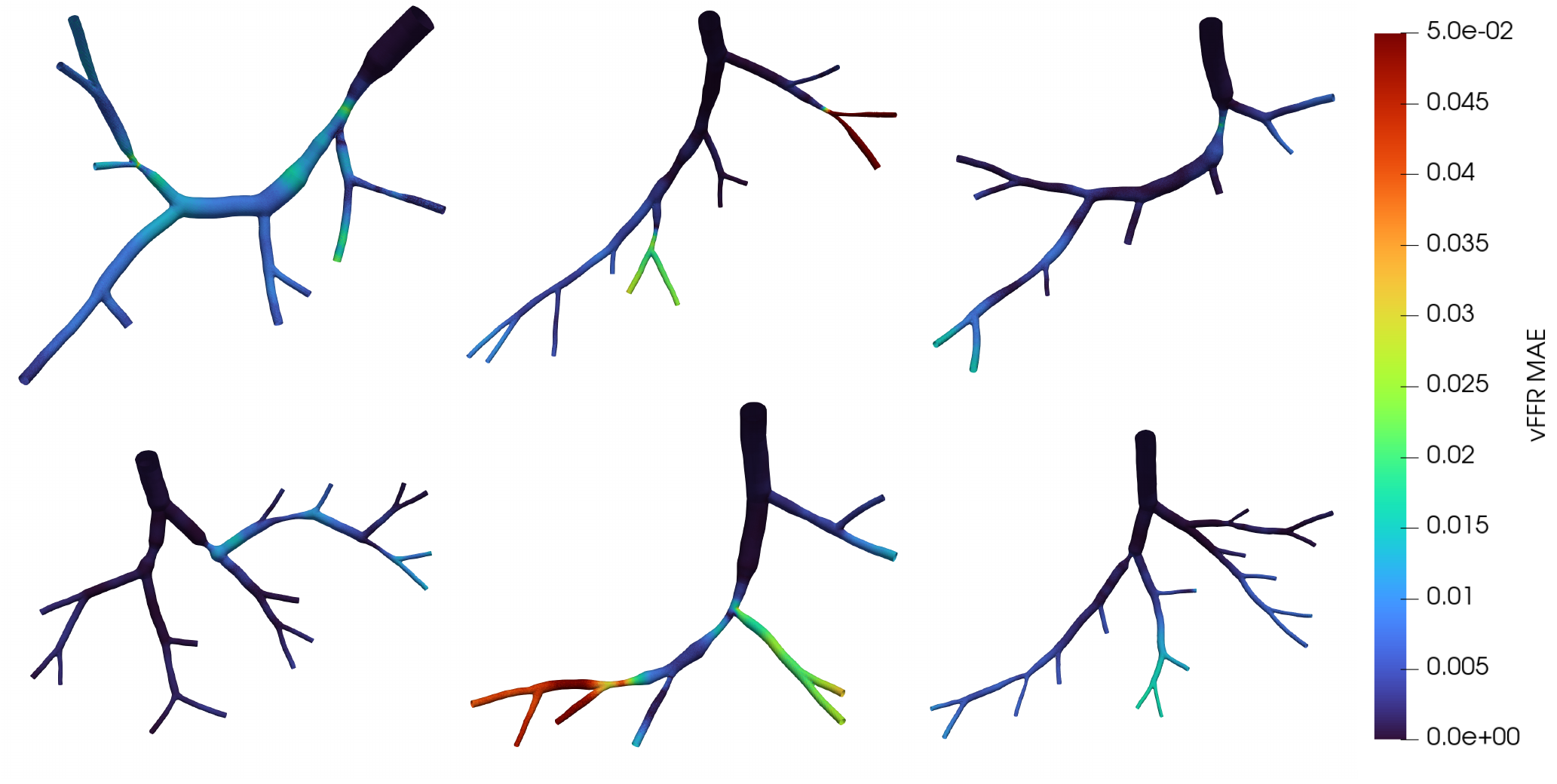}
    \caption[Additional MAE results for EVG model.]{Additional MAE results for EVG model. We report the same issues for EVG as well - there are no significant differences between both methods.} 
    \label{fig:supp:evg-mae-results}
\end{figure}
	\cleardoublepage

\end{document}